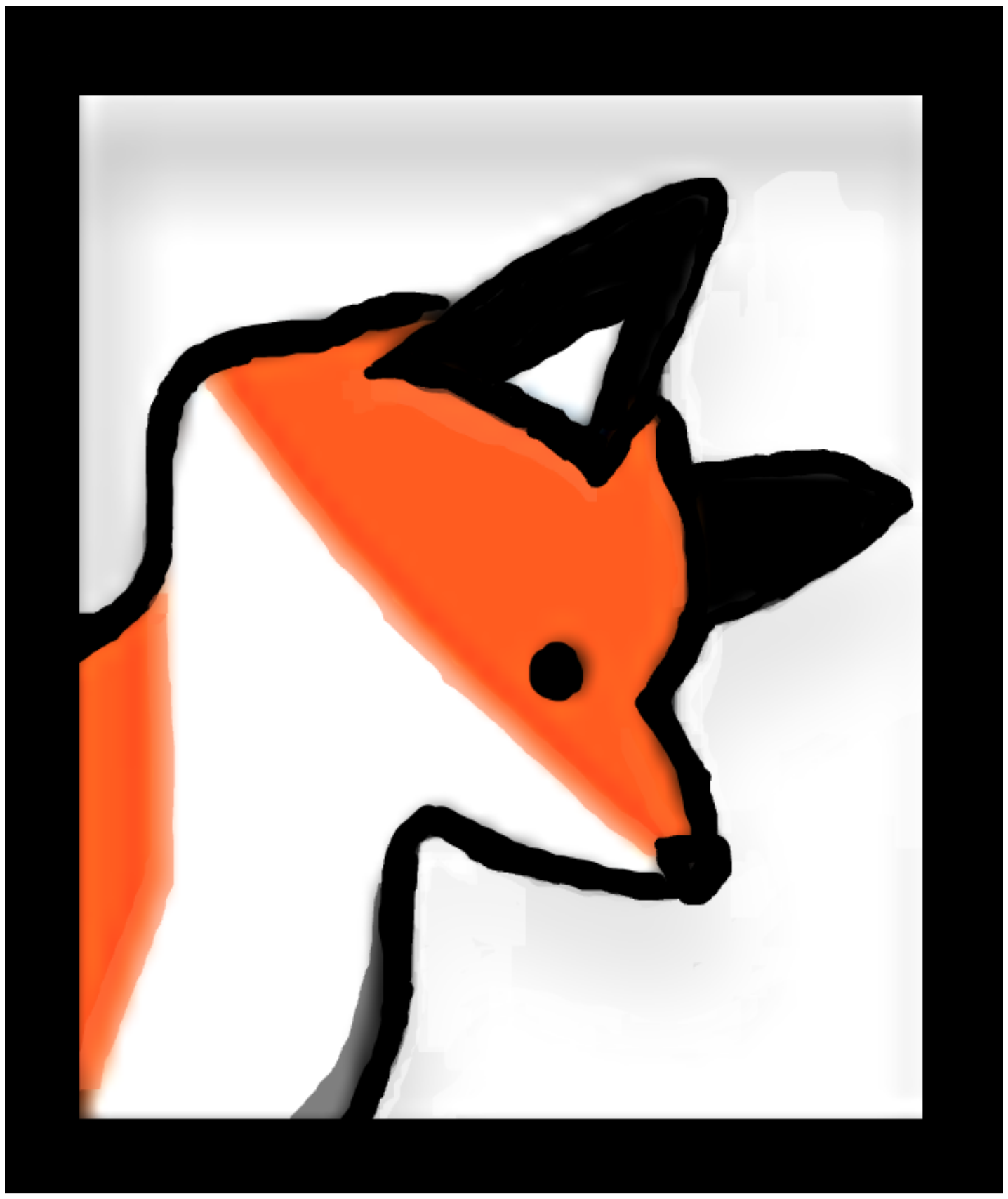

# The Quick Red Fox gets the best Data Driven Classroom Interviews:

## A manual for an interview app and its associated methodology

Version 1.0

[1]Jaclyn Ocumpaugh, [2]Luc Paquette, [3,4]Ryan Baker, [3]Amanda Barany, [2]Jeff Ginger, [3]Neithan Casano, [3]Andres Zambrano, [3]Xiner Liu, [3]Zhanlan Wei, [2]Yiqui Zhou, [2]Sophie Liu, [5]Stephen Hutt, [6]Alexis Andres, [3]Nidhi Nasiar, [3]Camille Giordano, [7]Martin van Velsen, [7]Michael Mogessie,

[1]University of Houston, [2]University of Illinois Urbana Champaign
[3]University of Pennsylvania, [4]Adelaide University, [5]University of Minnesota, [6]Learning Data Insights, & [7]Carnegie Mellon University

## Acknowledgements:

**We gratefully acknowledge funders who have supported the development of this methodology and the associate tech, including NSF DUE 1561676 and NSF DRL 2301173. We are also grateful to collaborators who worked with us during the development, including Chelsea Porter, H Chad Lane, Gautam Biswas, Anabil Munshi, Hayden Stec, Bruce McClaren, Ram Rajendran, and Hyeongjo Kim. Thank you also to the numerous collaborators over the years who have helped us to get to a point where this technology is both necessary and possible. We would not have gotten here without you.**

# TABLE OF CONTENTS

# List of Tables & Figures

**Tables:**

**Figures:**

# Chapter 1: Introduction to DDCIs, QRF, and the Goals of this Manual

## 1.1 Intro to Data Driven Classroom Interviews (DDCIs)

Data Driven Classroom Interviews (DDCIs) are an interviewing technique that is facilitated by recent technological developments in the learning analytics community. DDCIs are short, targeted interviews that allow researchers to contextualize students' interactions with a digital learning environment (e.g., intelligent tutoring systems or educational games) while minimizing the amount of time that the researcher interrupts that learning experience, and focusing researcher time on the events they most want to focus on (Baker et al., 2023).

The goal of DDCIs may differ somewhat from one research project to the next, but they are used to gather the kind of qualitative data that would help you to contextualize students' activities within a learning software. The method was invented in order to conduct these interviews in real classrooms, as opposed to in laboratory settings, so that researchers can conduct broader sampling across a wider range of students while also ensuring ecological validity (i.e., a student's experience in a paid activity in a lab is often very different than their experience in a classroom with their peers). Interviews range in length, but rarely exceed 10 minutes. They occur immediately after an action of interest has occurred, meaning the interviewer is able to both observe the student's affective engagement and ask questions about the student's experience before the student has had time to self-reflect, forget, reprocess, or resolve any experiences related to the event.

## 1.2 Overview of the Quick Red Fox (QRF) App

DDCIs are facilitated by a research tool called the Quick Red Fox (QRF)—an open-source server-client Android app that optimizes researcher time by directing interviewers to users that have just displayed an interesting behavior (previously defined by the research team). As Figure 1.1 shows, QRF integrates with existing student modeling technologies (e.g., behavior-sensing, affect-sensing, detection of self-regulated learning) to alert researchers to key moments in a learner's experience. QRF 'listens' for events (e.g., a detector recognizing a behavior, or a sequence of those behaviors) and identifies moments of interest, prioritizes them, and directs interviewers accordingly, allowing the interviewer to record DDCIs directly in the app along with relevant metadata (e.g. participant ID, time stamps, trigger information etc.). In addition to pushing triggers to the mobile phone, it also pushes them to a dashboard that allows fellow researchers to monitor the triggering events in real time.

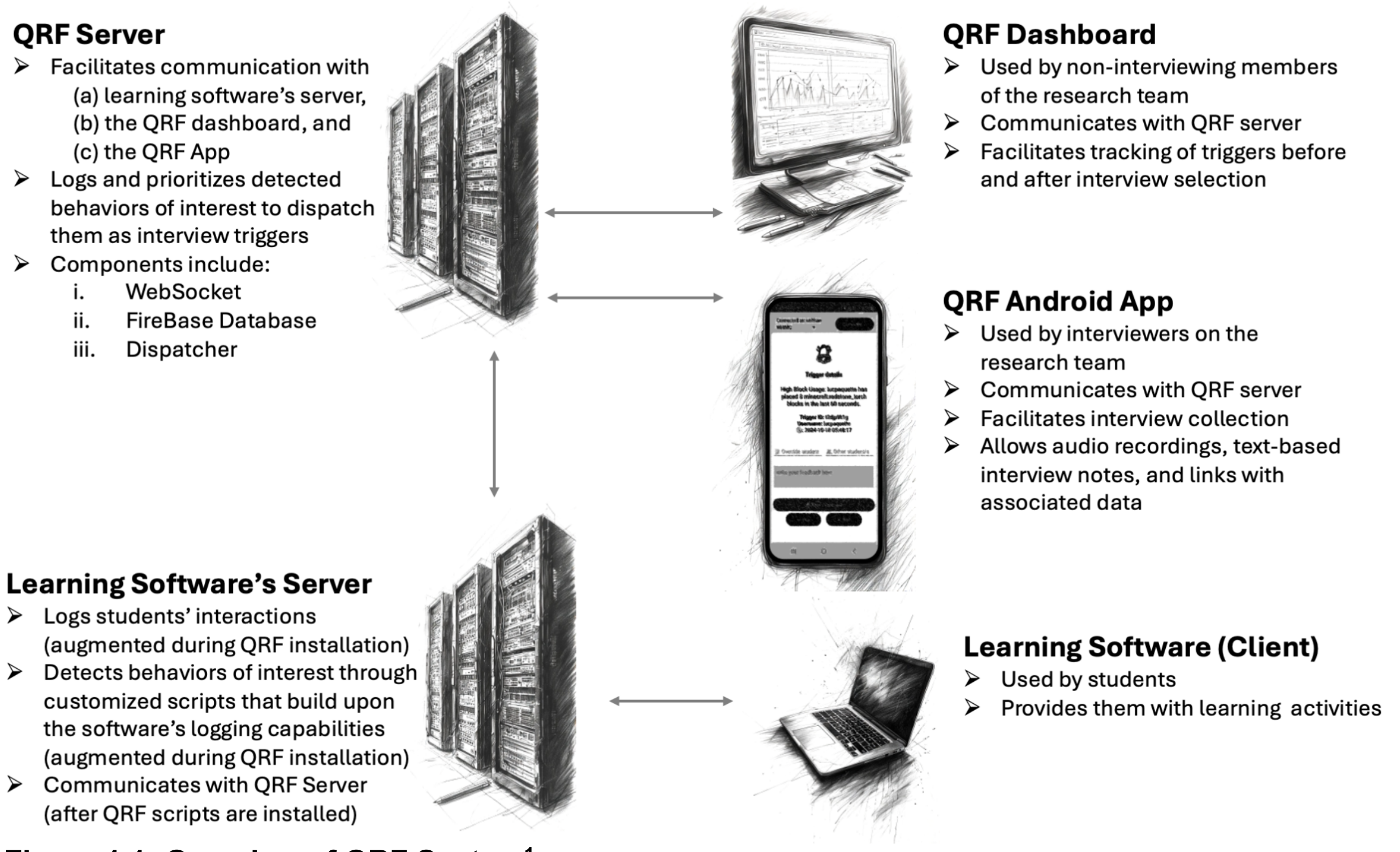

**Figure 1.1: Overview of QRF System**[1]

## 1.3 Overview of the DDCI Process

DDCIs are designed to allow researchers to interview students at specific, relevant times (after a specific event) in their learning experience, but what those specific events are will vary by both the learning system and the current research question. This means that DDCI research requires significant planning. Researchers must have access to the back-end of the learning system so that they can integrate the QRF server, which "listens for" events of interest, prioritizes them (according to predefined criteria), and then pushes them to the App.

Although QRF is open-source, it can only detect what is currently logged and/or modeled in the learning system. For example, when DDCIs were used to study epistemic emotions in Betty's Brain (Baker et al., 2024; Bosch et al., 2021; Andres et al., 2022; Hutt et al., 2022), researchers first had to develop models of these emotions that could detect them in real time from students' interactions with Betty's Brain (Jiang et al., 2018) so that these could be used to trigger each interview. However, QRF need not be triggered based on complex, machine-learned models. For example, in research studying the development of STEM interest, researchers triggered on student interactions with non-player characters, visits to points of interest, and other interactions that were common in the WHIMC learning system (see more detailed discussion in **Section 3.3**). Oftentimes, it is possible to augment what data is being logged by an online learning system, but you will need to check with your developers as to what is feasible.

In general, your process will be to integrate the QRF app and server within the learning software and to develop triggers for the interviews you would like to conduct as you work on developing interview questions that align to your research goals. You will also develop a priority list, which ensures that the most urgent triggers are pushed to the phone first. When you get to the field, those triggers will alert an interviewer using the QRF app, who can then approach the specific student within moments of an event

---

[1] Image credit to [Perchance.com](https://perchance.com) for generating the line drawings used to illustrate this system

of interest being triggered. These interviews are often short (e.g., <5 minutes), allowing you to capture highly contextual data related to the student's experience *in situ*, ensuring ecological validity that might not be possible with other methods. You will then harvest those interviews from the phone, alongside any associated metadata needed for your analysis.

## 1.4 Goals for this Manual

Conducting DDCIs involves pinpointing specific events or interactions for targeted interviews. This manual is designed to get you from the early stages of setting-up the QRF app to the analysis of DDCI data. We have attempted to organize this manual in as linear a fashion as possible, but the reality is that preparing for DDCIs is an inherently iterative process.

As you may already have gathered from the overview in this chapter, the process of connecting with students across repeated interviews requires careful planning and practice. Researchers must identify appropriate times to conduct interviews (designing triggers for the QRF app), must design these triggers (and the prioritization rules for their use) in a way that does not interrupt a given student too frequently, and then must execute these interviews in a way that builds trust with students while also asking interview questions that reflect their own research goals.

In early chapters, we cover how to **integrate** the QRF app and its associated server into an existing digital learning system (Section 2.5), implement **event detection** to trigger each interview (Chapter 3), and **prioritize triggers** so that the interviewer gets the most important and/or rare events first (Section 3.4). We will also cover infrastructure related to the QRF app (Section 2.5), **the dashboard** (Section 2.4), and **recording equipment** (Section 2.3). Once this infrastructure has been integrated and tested, you are ready to go into the field.

We then cover interview strategies. As the DDCI method makes use of short, frequent, and in situ interviews, this includes aligning interview questions with your research questions (Section 4.2.1), but also prepping for the interviews to be able to change as the learners get to know you better (Section 4.4.4). This also includes the **guidelines for conducting interviews** (Section 4.3), as well as features of the app, including the skip and override functions (Section 4.4) and the note-taking functions. The guidelines provided have been developed based on research on conducting interviews, with a special focus on conducting interviews with children. However, we believe the use of QRF to conduct DDCIs is flexible enough to adapt to a range of interview contexts, and we have provided an overview of **related research** that could be useful for those efforts (Section 4.7).

Finally, we also cover **data extraction** (Section 5.2), which includes both the metadata and the interview recordings themselves. We also discuss **transcription** choices (Section 5.3) and an overview of the kinds of **coding** (Section 5.4) and **analyses** (Section 5.4-5.8) that might be useful for exploring these data. This overview is far from comprehensive, but the goal is to preview the various types of analyses that may be important in order to inform (a) the selection and prioritization of triggers and (b) the preparation of interview questions and strategies that will ensure that you maximize these opportunities. In other words, this manual is organized in a linear fashion, but customizing QRF to conduct DDCIs requires an iterative design process to calibrate across multiple needs. As such, readers may not want to wait until they've gotten to the later stages of this research process to read about those parts of the DDCI method.

# Chapter 2: Intro to the QRF App and Associated Tech

## 2.1 Technological Overview

In this chapter, we will cover the technology necessary to implement Data Driven Classroom Interviews (DDCIs), which can be used to investigate the digital learning system of your choice. This includes the Quick Red Fox (QRF) app (Section 2.2), which the interviewer will use to conduct interviews, as well as a researcher-facing dashboard that allows the interviewers' support team to help monitor triggers as they are detected by the server (Section 2.4), and the backend server (Section 2.5). For the ease of reading, this chapter deals exclusively with the tech involved with the app and its server, and not with the actual development and prioritization of interview triggers, which will be covered in Chapter 3.

## 2.2 The QRF App and its Interface

The QRF App is an open-source application developed by a team of researchers at multiple universities, as part of two research grants. It was developed for the Android Platform and is available at https://github.com/pcla-code/QRF for free download for research purposes. We only ask that you cite this manual when you conduct research using the app so that we can report that information back to our funding agencies.

QRF, which enables you to conduct qualitative research fast,[2] is designed both to display and log the relevant metadata each time an interview is triggered. As Figure 2.1 shows, the app shows **connectivity information** at the top, including the name of the interviewer (in this case, "neithan"), the name of the learning software that the app has connected to (in this case, "WHIMC"), and confirmation that the app is "Connected" (shown in the blue button at the top). This connection button can be toggled as needed.

The app then provides **trigger details**. In the example shown here, which occurred while we were developing triggers for the *What-if Hypothetical Implementations for MineCraft* (aka **WHIMC**) learning system (Lane et al., 2022), the trigger itself is related to building (block-placing activities) by *lucpaquette*[3]. In addition, we see a unique trigger ID and a time stamp, which allow the interviewer to communicate with the rest of the research team and to monitor how recently the trigger event occurred.

Below the trigger details, you will find **three text boxes**. These allow you to (a) **override which student you are interviewing** (top, left text box), and/or **add the usernames of any additional students** who spontaneously join the interview (top, right text box). They also allow you to **take notes** about the interview or any other observations you might have about the student. These details will be saved alongside the associated metadata for each interview.

[2] The research team believes that "qualitative research fast" was originally the reason that we called the app QRF, but someone (who shall remain nameless) could not remember this, and the lab began to call the app the Quick Red Fox. Foxes are cute, and they make for better logos, so the name stuck.

[3] Note that in this test case, our co-developer Luc Paquette was playing the part of a student.

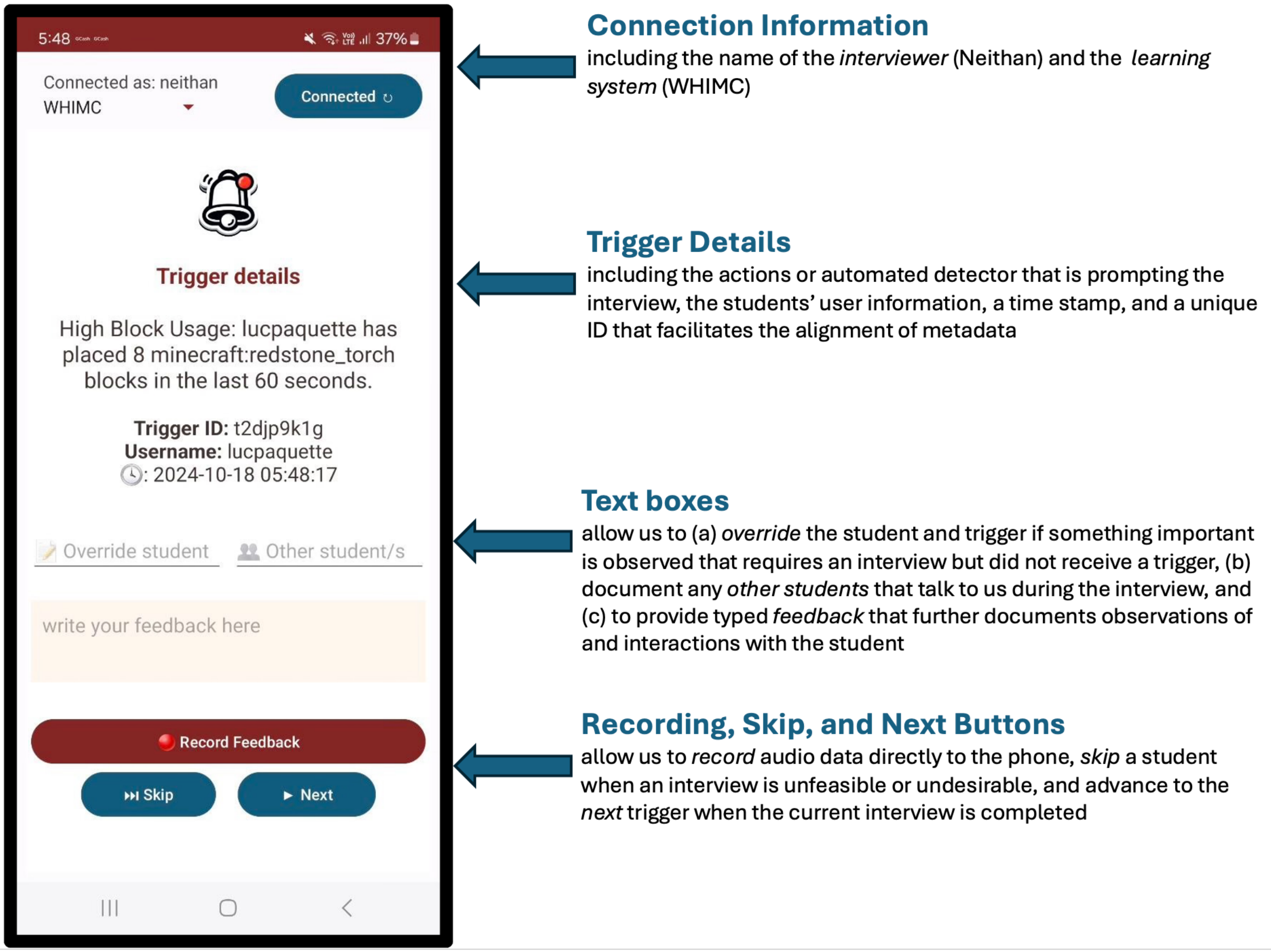

**Figure 2.1: The QRF Interface**

Finally, below the text boxes, you will see three function buttons, including the **record button**. This is the button you will use to start and stop each interview. You can also restart the interview, as necessary. If you do so, additional interview files will be generated and associated with this trigger. On the bottom right, you will see a **next button** that allows you to advance to the next trigger when you are done with an interview. You will also see a **skip button**, which allows you to move on to the next trigger if, for any reason, you do not want to conduct an interview with that student at that particular moment.[4]

## 2.3 Additional Recording Equipment Needs

### 2.3.1 Potential Problems with Cell Phone Microphones

Although many modern phones have excellent microphones, they may not capture the entire speech signal, and the directionality of their microphones is not optimal for noisy classroom environments. We also know that some phones use multiple microphones to pick up sound from more than one direction, which is useful for the average consumer, who may move the phone further or closer away from themselves while on speaker phone and who may also use it for conference calls. However, in a loud classroom, you want to have more control over which direction the microphone is picking voice data from.

---

[4] At the time of this publication, we are working to modify the skip button functionality so that we can track whether or not a student was skipped because (a) the prioritization algorithm was off, (b) the student was upset, (c) the student was unavailable, etc. These design changes have not been fully implemented yet.

In addition, there are gender and age differences in the speech signal that may make them work unequally on some populations. For adult men, most of the speech signal occurs under 2500 Hz, while for women it may go as high as 3000 Hz and for children that number can be even higher (e.g., 3500 Hz; Lee et al., 1999). Many phones do not include specifications on their microphones, so their exact frequency ranges are unknown, but historically there have been issues with landlines cutting out frequencies above 2000 Hz (under the necessary frequency for accurately capturing women's voices). Even if the phone's internal microphone captures these upper frequencies, some phones may apply compression algorithms that are more likely to distort higher frequencies.

### 2.3.2 External Microphone Recommendations

Our previous studies have shown that an external microphone is an important investment. We strongly recommend purchasing an external, unidirectional microphone that you can connect to your phone for conducting interviews. In our case, we consulted with a local music store, who recommended microphones that boosted the signal in the ranges of speech that we wanted (including those that were in the range of younger children). They also helped us to find the necessary secondary equipment for connecting those microphones to the phones (e.g., a USB Pre). Because technology changes quickly, we do not recommend a specific product. Instead, we recommend you check these specifications and/or consult with someone who understands the available products before making any purchases.

In addition to improving the accuracy of our transcriptions, we have also found that a unidirectional mic is important for ensuring privacy. Since many have a very short distance range, you will pick up fewer voices incidentally, and those that you do pick up will be less likely to be audible enough to accurately transcribe (helping to ensure that a student disclosing PII in the background is not incidentally recorded). This does mean that you may need to move the microphone back and forth between the interviewer and the student being interviewed. However, if you introduce the microphone to the students before the interviews start (see **Section 4.3.1**), they usually acclimate to this quickly.

## 2.4 The QRF Dashboard

The newest versions of the QRF app (since version 2.0) now feature a dashboard that allows researchers who are not conducting interviews to monitor the triggering events as they are detected in real time. For privacy reasons, you will need to set up your own dashboard. **Appendix 2** provides step-by-step instructions for this process.

The dashboard provides a range of information about each trigger, some of which is updated as interviews are completed. These items include:

A. The unique trigger ID
B. The timestamp
C. The student
D. The triggering event
E. The interviewer
F. Any feedback/notes taken by the interviewer
G. Whether or not the student/trigger was overridden
H. Whether or not an additional student participated in the interview
I. Information about the recording

These items can be seen in Figure 2.2.

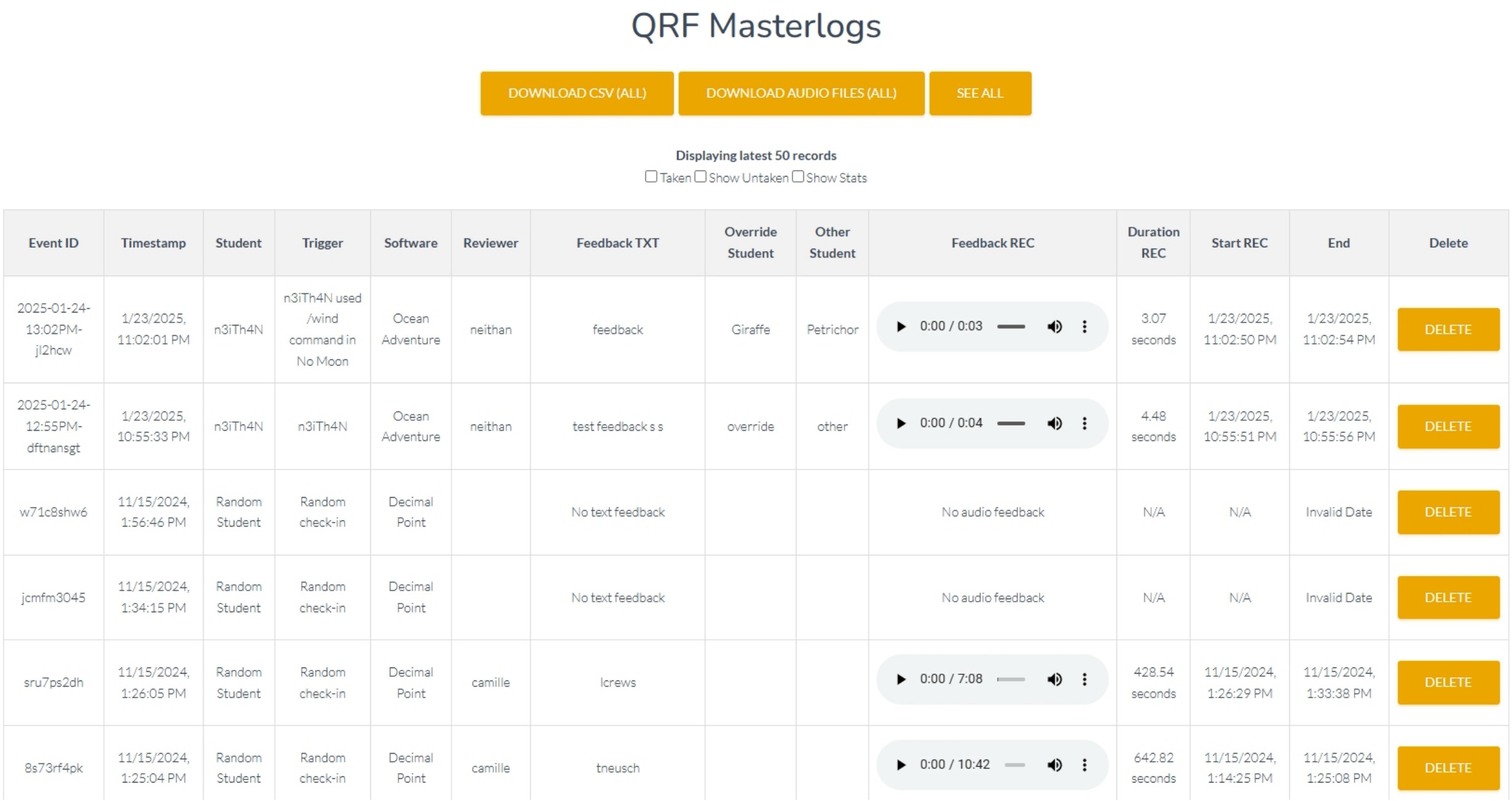

QRF Masterlogs

DOWNLOAD CSV (ALL) | DOWNLOAD AUDIO FILES (ALL) | SEE ALL

Displaying latest 50 records

☐ Taken ☐ Show Untaken ☐ Show Stats

| Event ID | Timestamp | Student | Trigger | Software | Reviewer | Feedback TXT | Override Student | Other Student | Feedback REC | Duration REC | Start REC | End | Delete |
|---|---|---|---|---|---|---|---|---|---|---|---|---|---|
| 2025-01-24-13:02PM-jl2hcw | 1/23/2025, 11:02:01 PM | n3iTh4N | n3iTh4N used /wind command in No Moon | Ocean Adventure | neithan | feedback | Giraffe | Petrichor | 0:00 / 0:03 | 3.07 seconds | 1/23/2025, 11:02:50 PM | 1/23/2025, 11:02:54 PM | DELETE |
| 2025-01-24-12:55PM-dftnansgt | 1/23/2025, 10:55:33 PM | n3iTh4N | n3iTh4N | Ocean Adventure | neithan | test feedback s s | override | other | 0:00 / 0:04 | 4.48 seconds | 1/23/2025, 10:55:51 PM | 1/23/2025, 10:55:56 PM | DELETE |
| w71c8shw6 | 11/15/2024, 1:56:46 PM | Random Student | Random check-in | Decimal Point | | No text feedback | | | No audio feedback | N/A | N/A | Invalid Date | DELETE |
| jcmfm3045 | 11/15/2024, 1:34:15 PM | Random Student | Random check-in | Decimal Point | | No text feedback | | | No audio feedback | N/A | N/A | Invalid Date | DELETE |
| sru7ps2dh | 11/15/2024, 1:26:05 PM | Random Student | Random check-in | Decimal Point | camille | lcrews | | | 0:00 / 7:08 | 428.54 seconds | 11/15/2024, 1:26:29 PM | 11/15/2024, 1:33:38 PM | DELETE |
| 8s73rf4pk | 11/15/2024, 1:25:04 PM | Random Student | Random check-in | Decimal Point | camille | tneusch | | | 0:00 / 10:42 | 642.82 seconds | 11/15/2024, 1:14:25 PM | 11/15/2024, 1:25:08 PM | DELETE |

**Figure 2.2: Example QRF Dashboard from WHIMC Project**

## 2.5 The QRF Server

In addition to the Android app, QRF makes use of several other components. This includes the **QRF Dispatcher**, a server that routes triggers from the learning software (client) to the **Firebase Database**, where all triggers are stored in JSON files. The QRF Dispatcher also organizes (prioritizes) triggers and transmits them (as appropriate) to the **QRF Android App**. These communications are transmitted via WebSocket. The **QRF Dashboard**, connects to the Firebase Database to provide a real-time overview of stored triggers and system activity. Setup and Installation instructions can be found in **Appendix 2.** Note that for privacy and security reasons, you will need to set up your own QRF Dashboard, Dispatcher, and Firebase Database.

## 2.6 Notes about Firewalls

Researchers should be aware that school firewalls and cybersecurity can be tricky. Many are quite restrictive. Therefore, even if all connections are considered secure, you may need the school IT team to permit all relevant URLs on student machines and open ports required for both the learning environment[5] as well as the QRF dispatcher and associated infrastructure. Failure to test your system beforehand in the actual classroom environment may lead to connection errors or data transfer blocks that cannot be easily resolved, especially while learners are present and require your attention. If you will be relying upon the school's web services,[6] it is advisable to contact the school IT department early to notify them about the QRF server, and, when possible, arrange for virtual and/or onsite tests ahead of time to discover and troubleshoot problems.

---

[5] For example, when using QRF with WHIMC it was discovered that schools may block the Xbox authentication client on Windows because it is used by many kinds of games, including Minecraft.

[6] The QRF team has had success employing University-supplied VPN clients to bypass some firewall issues, employing phone hotspots for instructor machines or running a local server version of the learning environment.

# CHAPTER 3: CHOOSING AND DEVELOPING INTERVIEW TRIGGERS

## 3.1 Overview of QRF Triggers

Data Driven Classroom Interviews are facilitated by **QRF Triggers**, or triggering events—which must be identified and/or developed before research can occur. These events could be very simple. For example, in research with the WHIMC learning software, we triggered interviews whenever a student had been moving without stopping during the last minute or if the student had been exploring alone (far from any other student) for more than 30 seconds, both behaviors that might be related with students' interest development. However, triggers can also include more complex constructs. For example, when studying Betty's Brain, we developed BROMP-based, machine-learned detectors of epistemic emotions (Jiang et al., 2018). We then used those detectors, which were capable of running in real time, to trigger interviews when particular sequences of emotions (e.g., boredom to confusion) occurred (Baker et al., 2021; Andres et al., 2022; Hutt et al., 2021). The triggers you choose for your project will depend on your own research questions and the available data/modeling capabilities within the learning software you are studying. This chapter provides an overview of that process.

## 3.2 Selecting Triggers

There are multiple considerations when designing triggers for interviews. First, researchers must become thoroughly familiar with the software or system in question, identifying which key actions or events in the game are most relevant to the research questions. It is also important to consider which theoretical frameworks are of interest to the research goals/questions and how to operationalize key components of those theoretical models using the available data.

For example, the first studies using QRF sought to better understand the relationships between self-regulated learning strategies and student emotions (Andres et al., 2022a; Andres et al., 2022b; Hutt et al., 2022; Baker et al., 2024). This work was grounded in D'Mello and Graesser's (2012) model of affective dynamics, which describes expected sequences of affective states and how they might impact learning outcomes. These sequences served as triggers for interviews and were detected by affect detectors already embedded in the system, which assessed students' affective states every 20 seconds. In this project, interviews were triggered by D'Mello and Graesser's *facilitative cycle*, in which students eventually return to engaged concentration after experiencing confusion, and their *inhibitory cycle,* in which students fall into boredom if they fail to regulate their confusion and frustration. However, interviews were also triggered during transitions that were not predicted by D'Mello and Graesser's model—such as when students return to concentration after experiencing boredom—so that interviewers could investigate what might motivate these more frequent transitions.

For other learning environments, such as educational games—especially open-world games—the breadth of possible player actions can make it more challenging to operationalize a theoretical model. Researchers must identify which specific student behaviors might represent core constructs of the theory. For instance, in the WHIMC project, the focus is on analyzing students' interest development in terms of Hidi and Renninger's (2010) four-stage model of interest development. Considering that social behaviors can trigger and maintain this interest, identifying when students explore virtual worlds alongside or following other players could serve as triggers for interviews. Similarly, taking time to observe, explore and reflect on the exploration might also be an indicator of maintained situational interest. Therefore, a trigger based on the student moving continuously (without pauses) could be important to understand whether the student is potentially disengaged from the learning efforts or is simply demonstrating another form of engagement and interest not initially anticipated.

Historical data can help researchers identify potentially relevant triggers. If this data is available, researchers can identify correlations between specific in-game actions and learning outcomes or personal traits pertinent to the research questions. For instance, in WHIMC, historical data on students' science and Minecraft interest was used to identify in-game actions that correlated with these subscales. Through this correlation analysis, we identified which in-game tools (e.g., pressure or temperature measurement tools) were tied to the interest scales and defined triggers based on their use. Notably, such correlation-focused exploration is intended to reveal moments where an interview might yield deeper insight into student behaviors rather than to confirm a formal hypothesis about action-outcome associations. Consequently, even if certain correlations are not statistically significant (as with the two movement-related triggers previously mentioned), actions that theory or past studies suggest might warrant further investigation are still worth including as triggers.

When designing triggers, it is also important to consider how often they would fire. Some actions may correlate with relevant outcomes, but triggering an interview every time a frequent action occurs can overwhelm both students and interviewers. For instance, in Minecraft-related environments such as WHIMC, the number of blocks placed may be correlated with interest; however, even un-interested students might place hundreds—or thousands—of blocks, making it impractical to interview them after every block. Instead, researchers should define thresholds that represent outlier behavior—such as unusually high or low block usage—using confidence intervals from historical data or other criteria to define those behaviors that need for further investigation. If historical data is available, it can be useful to simulate how frequently these triggers would prompt interviews and then adjust thresholds accordingly.

## 3.3 Triggers Components

Each trigger will be presented to the interviewer alongside several pieces of information, including (1) a unique ID, (2) a date and time stamp, and (3) the student's user ID. When writing code for triggers, all of this information should be accounted for, but there will also be several other considerations. Specifically, you will want to write the code in a way that

A. accounts for how frequently the event is expected to occur (or not occur)
B. accounts for how long an interview might take
C. accounts for how frequently a given student should be interviewed
D. accounts for how long you want the trigger to stay valid
E. accounts for any other variables/classifications within your data
F. prevents the trigger from immediately refiring

As **Section 3.4** (below) discusses, the complexity of designing these triggers will depend in part on how frequently they are likely to be available to an interviewer.

For example, in the Betty's Brain project, most of the triggers were related to BROMP-based affect detectors, which make predictions at 20 second intervals. Triggers were based on sequences of these affective states that were aligned with our research questions, and these affect labels arrive in a steady stream, virtually guaranteeing that we'd be able to have available affect-related triggers at any time that an interviewer used the QRF app to signal that they were ready for their next interview. That is, even as we prioritized highly infrequent triggers for theoretical reasons, we took into account how frequently those important triggers occurred. We planned for interviews that would last approximately 3-minutes each, which was relatively easy since the prediction intervals for the affect detectors were so frequent (i.e., every 20 seconds). This allowed us to ensure that an interviewer was not waiting for twenty minutes between triggers to start the next interview.

In contrast, the WHIMC project contained far more triggers that were highly variable in their distribution across time. As Table 3.1 shows, this includes the *Antisocial* trigger, which fires when a student has been far from other players for more than a specific amount of time (e.g., more than 40 blocks away for more than 120 seconds). This trigger, like several others already has a time frame built into it (i.e., more than 120 seconds), but others are more global. For example, *Used Tool Gravity Multiple Times*, keeps a running total of how many times the gravity tool has been used by a given student, but does not worry about how much time passed between usage. Depending on student gameplay strategies, the first trigger could potentially occur every two minutes, but the other might not happen for days. As we will discuss in **Section 3.4** (below), you should account for these kinds of differences in the customization of the QRF prioritization queue, but you may also want to adjust the time bracket for some features to ensure that an interviewer is missing important opportunities to talk to students.

**Table 3.1: Trigger Examples from the WHIMC Project. Note: [x] is used to denote variables that could be easily changed, such as the number of seconds that a student had been doing something or the number of blocks they had used/destroyed. Categorical variables are shown with [y-VARIABLE].**

| Name | Description |
|---|---|
| *Antisocial* | has been more than [x] blocks (distance) from other students for more than [x] sec. |
| *NPC Astrozoologist* | has been interacting with NPC Astrozoologist for more than [x] sec. |
| *NPC ElectricalEngineer* | has been interacting with NPC ElectricalEngineer for more than [x] sec. |
| *NPC Mysterious Mynoan* | has been interacting with NPC Mysterious Mynoan for more than [x] sec. |
| *Ignoring Mysterious Mynoan* | has been within [x] distance of NPC Mysterious Mynoan for more than [x] sec but has not interacted with them |
| *POI Pond* | has been within POI Pond for more than [x] seconds |
| *POI Underground Complex* | has been within POI Underground complex for more than [x] seconds |
| *Destroyed Blue Ice Blocks* | has destroyed [x] minecraft:blue_ice blocks in the last in [x] minutes |
| *Destroyed End Rod Blocks* | has destroyed [x] minecraft:end_rod blocks in the last in [x] minutes |
| *Placed RedStone* | Has placed more than [x] minecraft:red_stone blocks in the last in [x] minutes |
| *High tool use* | Uses more than [x] tools in [x] minutes |
| *Few Stops* | has fewer than [x] stops in [x] minutes |
| *Used Tool Gravity* | Used tool gravity in the last in [x] minutes |
| *Used Tool Gravity Multiple Times* | used tool gravity [x>1] times |
| *Appropriate Tool near POI* | Used [y-tool] near corresponding [y-POI] |
| *Appropriate Tool near NPC* | Used [y-tool] near corresponding [y-NPCI] |
| *High Tool Comparison* | Used [x] tools in this world, where X+3>y, and y=the highest number of tools used by anybody else in this world |
| *Visits multiple NPCs* | Has visited [x] NPCs in the last [x] minutes |
| *Visit unmarked POI* | The student is near [y-region], where that region is an area not marked by a minimap marker that has science relevance, for [x] minutes |

Moreover, the WHIMC learning environment has some notable design differences from Betty's Brain, including in how it allows students to access the various parts of the game. In Betty's Brain, the students were allowed to move completely freely from one section of the learning system to the next. For example, the students could choose when to add or subtract concepts from the causal map they were constructing and when to navigate to pages that allowed them to read about the learning domain. WHIMC players are also given a lot of freedom, but they are manually transported to several different simulated planets over the course of the week by instructors. On Monday, for example, they might spend 20 minutes learning the basics of making observations and measuring science-relevant characteristics on the moon and then the following day visit simulations of earth with a colder sun and earth with two moons for only 15 minutes each. This means that any triggers related to specific points of interest (POI) would only occur within a relatively short window of the students' total gameplay experience. In this case, we opted to have triggers that were both highly specific and far more generic. For example, the *NPC Mysterious Mynoan* trigger fires when a student has been interacting with a specific non-player character (NPC Mysterious Mynoan) for more than [x] seconds, and *Appropriate Tool near NPC* fires when a student uses a tool near an NPC who would have suggested its use. The latter trigger may require more complicated code, but it ensures that the trigger is more likely to occur across contexts.

**Table 3.2. Hypothetical Timeline of Triggering Events (for four students) and Interviewer Activities**

| Time | Bear | Caterpillar | Dragon | Tardigrade | Interviewer |
|---|---|---|---|---|---|
| 3:05:05 | *Placed RedStone* | | | *Destroyed Blue Ice Blocks* | |
| 3:05:10 | | *Placed RedStone* | | | |
| 3:05:15 | *Placed RedStone* | | *Placed RedStone* | *Destroyed Blue Ice Blocks* | |
| 3:05:20 | *Used Tool Gravity* | | | | |
| 3:05:25 | *Placed RedStone* | *NPC Mysterious Mynoan* | | | |
| 3:05:30 | | | *Destroyed Blue Ice Blocks* | | Stops recording interview |
| 3:05:35 | *Placed RedStone* | | | | |
| 3:05:40 | | *Destroyed Blue Ice Blocks* | | *POI Pond* | |
| 3:05:45 | *Placed RedStone* | | | | |
| 3:05:50 | *Used Tool Gravity Multiple Times* | | | | |
| 3:05:55 | | *Destroyed Blue Ice Blocks* | | *Destroyed Blue Ice Blocks* | |
| 3:06:00 | *Placed RedStone* | | | | Requests Trigger |
| 3:06:05 | | *Placed RedStone* | | *Destroyed Blue Ice Blocks* | |
| 3:06:10 | *Placed RedStone* | | | | |
| 3:06:15 | *Used Tool Gravity Multiple Times* | *Destroyed Blue Ice Blocks* | | *Destroyed Blue Ice Blocks* | Starts recording interview |
| 3:06:20 | | | | | |
| 3:06:25 | *Placed RedStone* | *Visits multiple NPCs* | | *Destroyed Blue Ice Blocks* | |
| 3:06:30 | | | | | |
| 3:06:35 | *Placed RedStone* | | | | |
| 3:06:40 | *Used Tool Gravity Multiple Times* | | | *POI Pond* | |
| 3:06:45 | | | | | |
| 3:06:50 | *Placed RedStone* | | | *Destroyed Blue Ice Blocks* | |
| 3:06:55 | | | | | |
| 3:07:00 | *Placed RedStone* | | *POI Underground Complex* | *Destroyed Blue Ice Blocks* | |
| 3:07:05 | | | | | |
| 3:07:10 | | *Visit unmarked POI* | | *Destroyed Blue Ice Blocks* | Stops recording interview |
| 3:07:15 | | | | | |
| 3:07:20 | *Destroyed Blue Ice Blocks* | | | *Destroyed Blue Ice Blocks* | |
| 3:07:25 | | | | | |
| 3:07:30 | | | | | Requests Trigger |
| 3:07:35 | | *Appropriate Tool near NPC* | | | |
| 3:07:40 | *Appropriate Tool near NPC* | | | *POI Pond* | |
| 3:07:45 | | | *Few Stops* | | |
| 3:07:50 | | *Visits multiple NPCs* | | *Destroyed Blue Ice Blocks* | |
| 3:07:55 | | | | | |
| 3:08:00 | | *Placed RedStone* | | *Destroyed Blue Ice Blocks* | Starts recording interview |
| 3:08:05 | | | | | |
| 3:08:10 | | *Appropriate Tool near NPC* | | *Destroyed Blue Ice Blocks* | |
| 3:08:15 | | | | | |
| 3:08:20 | | | | *Destroyed Blue Ice Blocks* | |
| 3:08:25 | | *Visits multiple NPCs* | | | |
| 3:08:30 | | | | | |
| 3:08:35 | *Used Tool Gravity Multiple Times* | | | | |
| 3:08:40 | | | | *POI Pond* | Stops recording interview |
| 3:08:45 | *Few Stops* | *Appropriate Tool near NPC* | | | Requests Trigger |
| 3:08:50 | | | | | |
| 3:08:55 | | *Destroyed Blue Ice Blocks* | | | |

It is important to ensure that (a) there is a constant flow of triggers to the interviewers and (b) no particular trigger sits in the queue so long that interviewing the student is no longer relevant. Table 3.2 provides a hypothetical example of how frequently a given set of triggers might fire for each student within the client server (e.g., the WHIMC server) alongside the hypothetical interviewer's actions. In this example, yellow highlights are used to contrast the time scale at which actions are logged within the client server (in this example, every row is logged at 5 second intervals) and the time scale at which the QRF server might be programmed to ping the client server (e.g., shown with yellow rows in this table). It is possible to adjust how frequently the QRF server polls the client server to queue new triggers, but in this hypothetical example, that event is happening every 15 seconds. Using historical data to better understand such distributions can help you to make timing decisions about how frequently each trigger might fire and how long the trigger should remain active.

There is more than one way to adjust how frequently an interviewer might receive any given trigger. The first way to do so would be to adjust how individual triggers are operationalized. For example, you can increase the frequency that a trigger applies by adjusting the time window in within its code (i.e., to make it fire more frequently, the *Antisocial* trigger could be coded to apply when the student has been far away from other players for only 30 seconds instead of every two minutes). As will be discussed in **Section 3.5**, the code for individual triggers also includes **timestamps**, **state flags**, and **dataframes**, each of which might be used to prevent a trigger from firing repeatedly. However, as discussed in **Section 3.4**, you can also throttle how often a trigger is fired by adjusting the QRF prioritization algorithm.

## 3.4 Prioritizing Triggers

Aligning potential QRF triggers to the theoretical frameworks driving your research questions is a primary goal. QRF is specifically designed for understanding how a student interprets their experience at key points in their online learning activities. However, these qualitative concerns must be balanced with quantitative concerns in order to make your time in the field effective. Once you have chosen your triggers, you will need to provide QRF with a ranking of their prioritizations so that it knows which to send you in the event that multiple triggers are available at the same time.

### 3.4.1 Principles for Prioritization

Researchers can develop numerous valuable triggers to guide interviews. However, this can result in multiple students being triggered simultaneously—an impractical situation in groups of more than ten students when only one or two interviewers are available. One possible solution is to reduce the number of triggers or design triggers that fire less frequently. However, this carries the risk of interviewers remaining inactive during multiple moments or missing events that, while not initially considered high priority, could yield valuable insights—especially when no other triggers are firing. Therefore, a more effective strategy is to define a diverse set of triggers and assign each a priority level based on its importance and expected frequency. The following prioritization principles should guide this selection.

**Principle 1:** Prioritize triggers that align with your research questions over those that are more distant from your research questions.

**Principle 2:** Prioritize triggering events that occur rarely over triggering events that occur more frequently.

**Principle 3:** Prioritize triggering events that occurred more recently over those that occurred longer ago.

**Principle 4:** Prioritize triggering events with shorter expiration times over those with longer expiration times.

**Principle 5:** Stagger triggers so that they do not all fire at once.

**Principle 6:** Ensure that interviews are distributed across students.

**Principle 7:** Ensure that any given student is not interrupted too frequently.

**Principle 8:** Always include a random trigger (labeled as such) as the last possible trigger in your prioritization list.

Triggers tied directly to the core research question should receive a higher priority, as stated in **Principle 1**. For instance, in Betty's Brain, the D'Mello and Graesser's (2012) *facilitative* and *inhibitory* cycles of epistemic emotions were prioritized because they were closely aligned with the study's central research question and theoretical framework on which the project was grounded (Andres et al., 2022). These rare cycles were selected so that we could talk to students about their self-regulation processes when one of these cycles had just occurred, but so were cycles that had been hypothesized not to be important by the theoretical model. For example, there was no theoretical explanation for why a student would cycle from boredom to engaged concentration, and so interviewing at those times offered the potential to determine why such events might occur.

Theoretical alignment should be balanced with **Principle 2**, which states that you should prioritize rare events over those that are more common. For example, we consider two triggers from the WHIMC system that are related to student interest levels. The first is a rare indicator of student disengagement: a student who plays the game for 20 minutes without making a scientific observation. By definition, this trigger would be rare because it could only fire after twenty minutes. The second is a rare indicator of high knowledge and/or heightened interest: the use of in-game tools that measure magnetic and tectonic fields. Given the low frequency of these two events, these triggers should be prioritized over more common events (of the same level of importance) that could be captured at multiple points during the learning experience.

In fact, **Principle 2** (prioritize rare triggers) should sometimes supersede other principles, including **Principle 3**, which states that triggers that occurred more recently should be prioritized over those that occurred longer ago, and **Principle 4**, which states that triggers with shorter expiration times should be prioritized over those with longer expiration times. **Principle 2** could be implemented simply by changing the rank of your project's triggers in QRF's prioritization queue, but it could also be implemented by changing the settings on the expiration of your triggers so that rare events stay in the queue for a longer period of time. For example, by ensuring that a trigger for an exceptionally rare event has a five-minute expiration and that triggers for more common events have a two-minute expiration, you can ensure that the QRF server is more likely to present the rarest trigger to the interviewers.

In parallel, **Principle 5** states that you should stagger the timing of triggers that might be likely to fire simultaneously (or within a short amount of time from one another). For example, it might be important to interview a student who has *not* looked at the reading material in a learning system in the last 5 minutes. This might be operationalized by starting a timer at the moment that a student closes the last page with reading material, which would keep the trigger from firing across multiple students at one time. However, starting your timer from the last time that the student has closed their reading material would exclude students who had never opened any pages with reading materials. To capture those students, you would also need to include triggers for "student has never opened the reading material" and set them to fire periodically throughout the learning experience (e.g., after 5, 10, and 15 minutes). However, this operationalization will result in the same event triggering for several students in a classroom at roughly the same time. Likewise, if you have several triggers that are operationalized by the same time interval (e.g., not reading every five minutes, not entering an answer every five minutes, not asking for a hint every ten minutes), then an inactive student will set off three triggers at exactly the same time, effectively causing you to only interview on the trigger with the highest prioritization ranking in the QRF dispatcher.

**Principles 6 and 7** have to do with how frequently you are sampling within your population. **Principle 6** states that you should ensure that triggers sampling from all students. For example, in the WHIMC project, we balanced the triggers associated with high interest (e.g., the use of redstone blocks) with triggers that were associated with low or mid-range interest development. This ensures that we are able to ask questions that could help to improve the experience for those students, while also ensuring that we didn't draw conclusions about the differences in these kinds of experiences without having sufficient evidence about how they might contrast with one another.

In parallel, **Principle 7** states that a student should not be interviewed so frequently that we fully disrupt their learning process. Although the interviews may sometimes help support students' metacognition, it is important to give students a break in between interviews so that they are able to work independently before talking to you again. This ensures greater ecological validity for the learning experience, since the goal of most research projects is to discover what students are doing (and able to do) without a researcher sitting next to them. Interviewers can always choose to skip an interview if they have just talked to that student, considering how the triggers are distributed—both across groups of students and across time—can help to ensure that you are maximizing your opportunities in the field

Finally, **Principle 8** recommends including a *random trigger* as your last priority in every project. This failsafe strategy ensures that no matter what goes wrong with your calculations of trigger distributions across time and students or with your prioritization algorithm, you will still be able to use your time in the classroom to conduct interviewers. A *random trigger* may also help you to determine that there are things wrong with your other triggers. For example, if you know that many of the students are doing the things that you are there to interview them about (e.g., building their Minecraft bases in WHIMC), but you are only receiving r*andom triggers*, the team responsible for monitoring the backend of the QRF operation may be able to pinpoint what is going wrong with the detection algorithms.

Random triggers also allow interviews to be conducted even if the students become inactive. As long as the students are still logged in to the learning system, you should always be getting prompts for interviews. This can be especially important if none of your triggers already prompt for inactivity. For example, in the WHIMC project, we included triggers that can fire when a student is inactive, such as “student has not made an observation in the last [x] minutes,” which might fire even if the student had stepped away from their computer. However, even if some of your triggers are related to inactivity, *random triggers* can help to ensure that you are able to talk to students when they have downtime. If you do so with the QRF app, it will time stamp the interview, and you can use the textboxes to provide information about who you are interviewing and any additional context you need to collect about that interview.

### 3.4.2 Example Prioritization

Like other parts of the QRF system, the prioritization algorithm should be customized both to the learning system and the underlying research questions. Table 3.3 shows an example prioritization ranking for a pilot study conducted with the WHIMC system in April 2024. In this study, we were attempting to understand how student interest in STEM develops. Some of the details represented in this table are beyond the scope of the current manual, but readers should note a few ways in which we operationalized the principles described above. Notably, we prioritized the “No Mynoa Obs” trigger as the 25 minute window that defined it meant that, by default, it could not happen in less than 25 minutes. Note also that the triggers with higher priority (ranked 1st and 2nd in the list) were related to a lack of engagement, as (a) the goal of that research project was to find ways to scaffold low interest students, and (b) students are sometimes more willing to be interviewed. Readers may also note that some triggers are given with equal ranks in the prioritization list.

**Table 3.3:** Trigger Prioritization Ranking for the April 2024 Pilot Study

| Rank | Name | Operationalization |
|---|---|---|
| 1 | No Mynoa Obs | User is in Mynoa without doing observations for 25 minutes |
| 2 | 3rd World No Obs | User has visited 3 or more worlds without making any observations |
| | N Worlds No Obs | User has visited n worlds without making observation |
| | 3 Worlds No Tools | User has visited 3 or more worlds without using any tools |
| 3 | Close Observations | User made an observation near another observation |
| 4 | Repeated Tool (GPA) | Used gravity, pressure or atmosphere more than once in a single world |
| | All GPA | Combination of gravity, pressure and atmosphere more than once for each of those |
| 5 | Unnec. Tool Rep. | Use of a tool that is expected to be single use more than once |
| 6 | ROI Obs | Someone made an observation in a region of interest |
| | POI Obs | Someone made an observation near a point of interest |
| | Expect. Tool ROI | Someone used an expected tool in a regions of interest |
| | Expect. Tool POI | Someone used an expected tool near a point of interest |
| 7 | Obs >N by world | User has made 3 or more observations in the current world (for the first 3 worlds) |
| | Obs >5 by world | User has made more than 5 or more observations in the current world, after the user has already visited at least 3 worlds |
| | Tool>10 by world | User has used a tool 10 or more times in the current world, for one of the first 3 visited worlds (first 3 worlds) |
| | Tool>5 by world | User has used a tool 5 or more times in the current world (after first 3 worlds) |
| 8 | Exp Tool by World | User has used a tool from the list of relevant tools for an expected tool in a world |
| 10 | Random | Picks one user at random every 35 seconds if other triggers are not occurring, trigger an interview with a random student |

## 3.5 Writing Code for Triggers and the QRF Dispatcher

### 3.5.1 Overview

QRF is programmed to operate with Python scripts that define, detect, and dispatch educationally-relevant triggers in real time. These triggers are system-generated messages, extracted by code that your research team will develop, that alert the QRF server when specific learner behaviors or milestones occur during a learning session. In turn, the QRF system uses your customized prioritization algorithm to queue the trigger and transmits the highest priority triggers to connected devices via WebSocket, allowing human interviewers to conduct DDCIs.

At the heart of this system is the **QRF polling script**: a Python loop that runs every 10 seconds (configurable) within the client server (learning system), checking for specified student activity. If activity that meets prespecified conditions are met (e.g, a WHIMC student not using any in-game tools for over 20 minutes) this Python code sends a trigger to the QRF server. Each trigger is represented as a tuple and appended to a list for immediate processing and dispatch. Figure 3.1 shows a minimal version of the QRF polling script that runs continuously, checking the database every 10 seconds for user activity and applying trigger rules.

```
#src.py
from time import sleep
from datetime import datetime

while True:
    print(f"Checking triggers... {datetime.now()}")

    # 1. Pull latest student activity data
    # 2. Run logic to detect inactive users
    # 3. Append any triggers to a list
    # 4. Send triggers to database / websocket

    sleep(10)  # Wait 10 seconds before next check
```

**Figure 3.1: Sample Python Code for QRF Polling Script**

Importantly, this script is meant to be adapted to your own learning environment. You will need to replace the placeholder logic in steps 1–4 with your platform's specific data access method (e.g., pulling data from Firebase, a CSV, or a local game log), trigger definitions (e.g., inactivity thresholds), and messaging protocol (e.g., WebSocket publishing or API calls). In WHIMC, for instance, we polled student logs from a MySQL server and defined inactivity by the absence of observations or tool use over a 20-minute span.

## 3.5.2 Latency and Real-Time Availability

QRF implements a low-latency operation. From the moment a student action is detected that matches a trigger condition, the Python script processes and transmits the event to the Firebase Realtime Database and over a secure WebSocket (via PieSocket). In practice, most triggers become available to interviewers within 1-2 seconds, assuming stable network conditions. Execution time will vary depending on your own analysis, and it will also vary upon your learning system. In our research it is typically minimal (e.g., under 100ms) due to the simplicity of the logic and Python's efficient list operations. Additionally, the real-time WebSocket system allows asynchronous delivery of triggers to connected phones, decoupling server-side trigger detection from client-side availability.

## 3.5.3 Removing Triggers from the Prioritization Queue

One issue with real-time triggers is that they can accumulate quickly. In order to ensure that DDCIs are conducted *in the moment,* triggers need to expire. In the QRF system, **Trigger Expiration** is handled by the QRF dispatcher through a variable called "EXPIRATION_TIME." This variable can be modified to adjust the time between a trigger being received by the dispatcher and the trigger being removed from the priority queue.[7] This variable is customizable, and it is likely that different triggers may need different expiration times. A generic recommendation is to start with a 3 minute window for most triggers, as it may take a couple of minutes for an interviewer to get across a crowded classroom, but some rare triggers may warrant longer expiration times in order to ensure that they remain active long enough to rise to be pushed to an interviewer.

[7] For example, to set an expiration time of 3 minutes, the EXPIRATION_TIME variable should be set to 3 (minutes) * 60 (seconds) * 1000 (milliseconds) = 180,000.

### 3.5.4 Student-Level Cooldown Mechanism

One concern that arises in DDCI methodology is related to the equitable sampling of individual students. This issue is discussed in **Section 4.3.4**, where we recommend designing triggers that are likely to capture the full range of students (e.g., if you are interested in so-called careless behaviors, trigger both on carelessness and on other types of behaviors). However, equitable sampling is also ensured through a **Cooldown Mechanism** that pauses the interviewing of a particular student for a given amount of time after an interview with that student has been completed. This ensures both that interviewers are directed to a wide range of students and that any given student is not being interrupted by the interview process so frequently that it becomes disruptive to their learning process.

### 3.5.5 Avoiding Trigger Redundancy

Some triggers are complicated by the fact that they could apply for longer than the expiration time set in the QRF dispatcher. This can often be the case for triggers that are based on the learner not using a specific feature of your learning environment. For example, in WHIMC we developed a trigger indicating that a learner did not write an observation in the last 20 minutes, but a time interval like this does not (mathematically) have a clear endpoint. For example, a student who “does not do ACTION for 20 minutes” is likely to continue on for 22 or 25 or more minutes of “not doing action.” In such a case, we want to avoid repeatedly sending the same trigger to the dispatcher every couple seconds. This situation can be handled in the code that analyzes the log data from your learning environment, by avoiding resending the trigger unless a specified amount of time has elapsed. For example, Figure 3.2 shows a simplified version of a Python logic that checks whether a student has made any observations in the last 20 minutes (only sent if no similar trigger was sent in the last 20 minutes).

```
#sampletriggerlogic.py
def check_no_observations_last_20_minutes(self):
    current_time = datetime.now(pytz.timezone("America/Chicago")).timestamp()

    for user, data in self.tools_usage.items():
        last_observation_time = data.get("last_observation_time", self.start_time)

        if current_time - last_observation_time > 20 * 60:  # 20 minutes
            last_trigger_time = self.lastTriggerTimePerUser.get(user, 0)

            if current_time - last_trigger_time > 20 * 60:  # AvoidRedundantTrigger
                trigger_message = f"{user} has not made any observations in the last 20 minutes."
                self.triggers_list.append((trigger_message, user, 1))
                self.lastTriggerTimePerUser[user] = current_time
```

**Figure 3.2. Sample Python Code for the Avoid Redundant Trigger Mechanism**

This approach to writing trigger logic is both modular and extensible. Research teams using QRF are encouraged to define their own conditions based on the learning behaviors they wish to capture (e.g., detecting hesitations, sudden bursts of activity, time spent in particular activities). By using lightweight Python logic paired with WebSocket and database updates, QRF enables real-time, flexible, and scalable monitoring of student behavior across a wide variety of education technologies.

# Chapter 4: Data Driven Classroom Interviews

## 4.1 Why DDCIs?

Data Driven Classroom Interviews (DDCIs)[8] are a method that leverages a range of technological advances in education. The method is intended to capture students' in-moment thoughts based on a specific learner behavior or affect or any other meaningful construct when detected in real-time. As described in earlier chapters, DDCIs are conducted using an Android application called Quick Red Fox (QRF), so that we can pinpoint the exact time that we want to interview and/or observe a specific student.

There are several advantages to this method:

- QRF can pinpoint subtle and rare interactions that even a teacher or researcher observing a classroom might not notice (e.g., the **needle in the haystack problem** (Baker et al., 2024) or the **one shot problem** (Hutt et al., 2022)
- DDCIs can be conducted in authentic classroom environments, increasing the ecological validity over lab studies
- If students are struggling, we can talk to them before they have resolved that struggle (which may influence their memory of the interaction were we to wait for an after-the-fact interview or focus group)
- DDCIs allow the researcher to carry the burden of steering the students' talk toward important topics related to the research questions (vs. other methods, where the burden is on the student).
- QRF also allows the researcher to type in their own observations of the student in addition to capturing what the student is able to articulate.

One of the methods that most closely parallels DDCIs is the Think Aloud method (TA; Ericsson & Simon, 1993), where students verbalize their thoughts during a learning activity. TA protocols are widely regarded as reliable tools for capturing cognitive processes, as they reflect the information currently occupying the participant's attention. However, to ensure their validity, verbal reports need to be collected objectively and consistently. For some research, collecting TA data during a consistent activity or point in time may address this need, but this is less feasible for open-ended learning experiences, where a behavior of interest occurs at unique points for different learners.

To address this, DDCIs are conducted only when triggered by specific behaviors of interest as captured in students' action sequences. By anchoring questions in interviews to these observable triggers, DDCI is both more timely and directly relevant than retrospective or generic verbal protocols. These benefits are similar to those you might get from a focus group or another retrospective interview method, but DDCIs allow the researcher to observe and speak to the student before they have had a chance to process—and potentially reframe—a given event in their learning experience.

In this way, the fieldwork efforts used in DDCIs parallel methodologies like microethnography (Au and Mason 1983; Erikson, 1996; Garcez, 2008; Streek & Mehus, 2004), where the emphasis is on collecting rich, situated process data[9] from an authentic context—even if the type of data differs in terms of modality (micro-ethnographies often use video data) and goals (micro-ethnographies are interested in human-human interactions instead of human-computer interactions). Like Action Research (AR; Baskerville, 2001; Kemmis et al., 2014) and Design Based Research (DBR; Davison, 2004; Design-Based Research Collective, 2003) paradigms, DDCIs use iterative research processes to identify constructs that might be used to improve students' learning opportunities, and this is accomplished by interacting with stakeholders (in this case, students) in a way that allows researchers to build upon their growing understanding of students' learning experience (see discussion in Cole et al., 2005).

---

[8] We have developed this method for use in classrooms, but we have also referred to it as Data Driven Interviews in some research, as it is could easily be adapted to laboratory settings or other in-person contexts.

[9] Note that the terminology in this research space is deceptively overlapping with the Learning Analytics literature. Here, the "processes" are related to human-to-human "interactions" and not to the process data generated by students' interactions with a learning system.

As we will discuss in this chapter, the preparation for interviews begins with deciding what your research questions are and how they align with various student interactions within a digital learning system (Section 4.2). In this method, the researcher conducting interviews is conducting a series of short (~5 minute) interviews with students, likely over the course of several days. This means that interviewers must be prepared with a variety of strategies for rapidly building trust with the students while also eliciting the relevant contextual information needed to understand the variable(s) driving the events that triggered the interview.

## 4.2 Preparing for DDCI Fieldwork

Preparing for DDCIs is important because, while these interviews are short, conversational, and open-ended, you will likely be speaking to the same student many times over the course of the week. For example, during one of the fieldwork sessions for the WHIMC project, 2 interviewers collected data from a summer camp that had only 8 children enrolled. The kids played the game for 3 hours a day over the course of one week. By the end of the week, coauthor Barany had collected 81 interviews and coauthor Ocumpaugh had collected 83 interviews. In this case, students were almost always cooperative but also noticed when they were asked the same questions repeatedly. When preparing your own fieldwork plan, we recommend that you take the design of the learning experience into consideration (e.g., time spent, activities offered, participant/researcher ratio) and consider ways you can support participant re-engagement and avoid repetition.

You should also expect that students will change the way they interact with you over the course of the fieldwork experience. For example, some students may be shy at first—either as a reaction to you or as a reaction to being recorded. While we recommend having a variety of question strategies available, you may find that students will be more willing to share information about some topics (e.g., their struggles with the learning material) as they get to know you better and as you have proven that you are there as a helper rather than as an enforcer.

### 4.2.1 Developing Questions that Align with both QRF Triggers and Research Questions

Long before you enter a classroom (or whatever research context you are investigating), you should be thinking about the kinds of questions that align with your research question and with potential QRF triggers. As we have discussed in Section 4.2, this involves preparing a flexible bank of interview-opening questions that both center the student as a person and connect to the learning experience. Beginning with student-centered questions (e.g., "How are you feeling?" or "What's new?") helps build rapport and creates space for authentic responses, which is essential for gathering meaningful, contextual data. These can then be followed with more targeted learning-centered questions (e.g., "What strategies have you been trying?"), allowing a smooth and humane transition to discussions of the trigger event. Because students are often interviewed multiple times, it's important to have a variety of sincere, open-ended prompts that maintain engagement while aligning with your research goals. A well-rounded set of student- and learning-centered questions supports both ethical and effective interviewing.

Similar to other semi-structured interviewing approaches, having a prepared set of questions also makes it easier for researchers to follow up when a student brings up something unexpected, unique or insightful. In one case, a researcher asked a student to explain their building strategies in a learning game, which prompted a deeper follow-up discussion on the limitations of the game for mimicking real-world science and the different ways that learners can respond when they play. While not all students had this insight, the prepared questions allowed the researcher to help surface this unique perspective.

If you are using the DDCI method, you will interview a student several times over the course of your fieldwork experience—which could last for several hours across several days. Although kids will generally be happy to talk to you, there are many things that might keep you from getting the data you want. For example, self-presentation effects could make a student reluctant to:

a. Talk to someone they do not know very well (you),
b. Talk in front of their teacher,
c. Talk in front of some or all of their peers,
d. Discuss anything that they find unflattering, disrespectful, or taboo.

For this reason, you should think about the questions you use when you first meet the student, and whether those are things that would help you to establish trust.

In addition to self-presentation effects, you should also consider the degree to which learners have the metacognition to accurately report everything you would like to know. For example, asking a student "why do you think you're struggling?" is a valid question from a teacher who is trying to generate self-reflection (even if it is a potentially face-threatening question!). However, a novice learner—by definition—does not necessarily know enough about the domain to identify a misconception or to articulate what information they are missing. Even if they are willing to answer, they may not be able to do so accurately.

In fact, the likelihood of a student having low metacognition is a major justification for using DDCIs as a method. Researchers working in fields like learning analytics need contextual information about student interactions with digital learning environments—and this information is more accurately obtained before the student has had a chance to reflect about the experience. For example, if we waited to conduct a retrospective interview to ask a student about the time that they were struggling so much that they pounded their first on the table, but they had eventually gotten it correct and decided they were just being silly, they might not tell us (or even remember) how upset they were in the moment. However, as we have discussed elsewhere, Think Aloud methods (Ericsson & Simon, 1993) place a considerable metacognitive burden on the student.

Finally, we should consider cultural factors that influence the ways in which people interact with one another, particularly those that are relevant to the classroom experience. For example, many students may come from cultures where asking **known answer questions** (KAQ) is a normal conversational practice, but other students may not have that practice natively in their homes (Guitierrez, 1993; Heritage, 2005; Heritage, 2012; Higgens et al., 2003; Neissier, 2020; Rusk et al., 2017; Searle, 1969). Planning for these kinds of cultural differences is important regardless of your research question, but if part of your goal in these interviews is to find out about students' prior knowledge, you should look for ways to get students to authentically explain their knowledge.

Again, how you design such questions is highly dependent upon your underlying research question. Developing questions that ask about students' interests that are related to the current learning topic might be one strategy that would work well for a range of research questions involving both learning and motivation variables. Even if the topic is too distant from the current topic (e.g., if a student learning about astronomy has encyclopedic knowledge about chinchillas), these kinds of conversations can allow you to ask questions about their motivation for acquiring this knowledge as well as the strategies and resources they used to acquire it. You can often then pivot to the current topic (i.e., "So are dwarf planets more fun to learn about than chinchillas?" "Do you learn about them in the same way?"). Odds are that the student will incidentally report a fair amount of their knowledge by reporting on *how* they acquired that knowledge, and you'll be able to ask them about it in a more authentic way.

You should also consider strategies for getting students to continue talking without asking a new question. For example, in natural conversations, enthusiastically echoing what someone has just said to you is often an invitation for them to keep talking (Ferrara, 1994; Heritage, 1984), particularly if it sounds like you are checking to confirm your understanding (Svennevig, 2003). Likewise minimal responses (i.e., "mmmhmmm," or "oh?") or even a well-timed pause can also signal that you'd like your interlocutor to keep speaking (Bennet & Jarvis, 1991; Goodwin, 1986; Fellegy, 1995). As you will read in **Section 4.3.2**, we recommend that interviewers employ these strategies while adopting the role of **Nice Big Sister (NBS)** and not **Extra Teaching Assistant with a Microphone (ETAWAM)**. Teachers also make use of minimal responses to keep students talking (Fagan, 2012; Hosoda, 2016; Lee, 2007), and that use is sometimes very evaluative (e.g., using "oh?" to signal that the student should rethink their response).[10] It's okay to encourage students (e.g., with an occasional "awesome"), but you do not want them coming to you for answers. If your role shifts from NBS to ETAWAM, your recordings will have more information about the interviewer's knowledge of the subject than anything you might be investigating.

### 4.2.2 Technological Preparation

You should also familiarize yourself with the technology involved, including the phone, the QRF app, the QRF Dashboard, and any additional recording equipment. (Recall that we strongly recommend that you use an external, directional microphone that will maximize your ability to capture relevant data in a noisy learning environment, Section 2.3.) Testing your equipment and checking to see if the triggers are firing appropriately (ideally both weeks before and day of) is always good practice.

## 4.3 Guidelines for Conducting Fieldwork

When preparing for fieldwork, it is good to remember that you are asking these learners to trust you, which means you are asking them for a favor. Therefore, an interviewer should not be an individual already serving as any sort of teacher or authority figure to these students, as roles that invoke authority are likely to induce self-presentation effects. Likewise, as conducting this series of short interviews requires a fairly intense amount of focus and engagement with the students, you will not be able to run the back end of the QRF system at the same time. Instead, you should focus on the interviewing process itself. We have distilled the following five guidelines for DDCIs, which we explain in further detail in this section:

> **DDCI Guideline 1:** Tell the group of students you'll be interviewing about the DDCI process before you conduct any interviews.
>
> **DDCI Guideline 2:** Take a friendly and helpful but non-authoritative (Nice Big Sister) roll as an interviewer.
>
> **DDCI Guideline 3:** Confirm assent for each interview. Skip triggers if you think that an interview would be overly disruptive to the student.
>
> **DDCI Guideline 4:** Students may only provide information that you ask about. (Prepare for priming and salience effects!)
>
> **DDCI Guideline 5:** Individual questions can ask about students' struggle, but your overall protocol should take an asset-based approach.

These guidelines are also summarized in **Appendix 3** and a checklist for daily interviewer activities is included in **Appendix 4**.

### 4.3.1 Preparing Learners for DDCIs

**Guideline 1** says that you tell students about the process before you conduct DDCIs. Kids deserve to know what we are doing with these interviews, who will listen to them, and why we need this information. In fact, it is impossible for them to assent to the process if you do not provide this type of explanation to them. However, most kids will not fully understand the details of research methods, and you do not necessarily want to prime them with other specifics or provide them with information that might cause distress. For example, saying "We've built detectors that catch when you students game the system," sounds very much like you are there to get them in trouble—which you are not. Framing the same information in terms of student support, like “We want to know what you think of the game and what you and your friends do when you are struggling” is also accurate but less likely to cause fear.

---

[10] As with all conversational conventions, these strategies can feel very intuitive, and many may assume they are natural and universal. If people judge you to be well intentioned, they will be very forgiving in their responses, but be aware of possible differences if you're interviewing students from another dialect or language from your own native variety.

If you are not part of the research team that is developing the learning system, we also recommend telling students that you are part of an outside team. Making a big deal about the fact that you are there to help the software developers figure out how to improve the learning system (or to evaluate the system) is likely to make some students more likely to talk to report issues to you, which can often help you to better understand their motivations and other parts of their learning experience. It also encourages them to think of you as someone who is helpful (e.g., a safe, friendly person), without encouraging them to think that you know how to do the work or run the learning system. Remember, regardless of your position, students do not have to talk to you. They are doing you a favor if they do, and you should thank them for their help.

Finally, we recommend thinking about other ways to make students more comfortable with what you will be doing. For example, we strongly recommend using an external microphone, and this equipment can be very intimidating to some students. To help students feel more comfortable, we show them the microphone before we start the interviews, and we tell students that we use these microphones (1) so that the transcription process is easier and (2) so that we can help protect the privacy of their friends. We also show them how close the microphone should be to best capture their voice—which can sometimes be very close. We remind them that they are being asked (not required) to participate, and that if the microphone ever feels too “weird” they can tell us to stop. We find that demonstrating this ahead of time makes students feel considerably more comfortable with both participating in interviews and declining to assent if needed.

### 4.3.2 Adopting your Role as an Interviewer: The Big Sister Approach

Once you have explained to the students what the process will be like, you are ready to start conducting individual interviews. As explained previously (**Sections 4.2**), you should have already brainstormed a list of questions that you could ask a learner that could help you to answer research questions about your triggers (before you got to the classroom). However, the importance of developing rapport with the learners during the DDCI process cannot be overemphasized. Walking up to a student and immediately asking them why they are doing something risks putting them on edge. Even if the student believes that they are doing exactly what they are supposed to do, this approach may come across as “Big Brother” (in the Orwellian/authoritarian sense), as many people do not like being under high levels of scrutiny.

Instead, **Guideline 2** says that you should adopt a friendly but non-authoritative approach. To accomplish this stance, we recommend what we call the **Big Sister Approach**[11] (or **Kind Auntie/Uncle Approach**, if you prefer). Specifically, you should treat your students the way that you would a younger sibling (or niece or nephew) who you cared about. You are interested in what they are doing, and you want them to feel comfortable talking to you. You do not want to pry about deeply personal information—especially in front of their peers—but you do want them to feel comfortable telling you about relevant things in their life that connect to this learning experience.

The **Big Sister Approach** metaphor should also guide how you interact with students in terms of the learning material. For example, a cool big sister (or favorite aunt/uncle) would be enthusiastic about the things that the child is enthusiastic about. They would be supportive when the child is struggling. They might sometimes steer the conversation towards helping them towards having their own epiphany, or they might give them the occasional more obvious hint. However, **Guideline 2** says there should be clear boundaries to this sort of support. When a big sister does someone's homework for them, (a) the student doesn’t learn and (b) the teacher does not know that the student was struggling. In fact, even if the big sister just “reteaches” the material, the teacher still doesn’t know why the student is struggling. Therefore, if the student needs significant help, you should either encourage them to ask an appropriate classroom authority for it, or you should send that authority figure to the student.

---

[11] We acknowledge the cultural assumptions present in this metaphor. For our purposes, we are drawing upon the paradigm that is common among many White, middle-class families in the United States, where big sisters are often discouraged from taking on a parental role even when they are asked to do things like babysit or help with chores.

We advocate for this approach to be carried over in how you might intervene should the student do something that might elicit negative feedback from an authority figure. Unlike their parents or teachers, it is not your responsibility to make sure that they remain perfectly on-task in your presence. Instead, if you observe innocuous off-task or unproductive behaviors, you can either ignore them or use them as an opportunity to ask why they are disengaged from the learning system. In fact, a little indulgence on some off-task behaviors could potentially induce trust, and if a student is willing to be honest with you about their motivations, it could significantly help improve future learning opportunities for students.

That said, the **Big Sister Approach** does not mean that would permit a student to get away with serious infractions. Even your favorite, most silly Uncle would "rat you out"[12] if he thought you were going to hurt someone, and he might even step in himself if things were urgent (e.g., an argument between two students that is becoming increasingly aggressive). Your approach should be the same. Like a trusted older sister, your job is not to police every minor infraction, but you should call for help (or step in, if absolutely necessary), if there is a real risk of harm.

### 4.3.3. Strategies for Individual Interviews

As a trusted adult (or "Big Sister"), you should respect students' boundaries and demonstrate that they are important to you as a person. When you approach a student for a new interview, you should ask how they are doing and if this is a good time to talk (**Guideline 3**, Assent). Although some students will answer with a conventional response (e.g., "good" or "fine") to this open-ended question, many students will answer more literally. If you need to, try leading with other open-ended questions that center *the student* instead of the learning experience (e.g., Table 4.1). This strategy is important since we will need them to feel comfortable talking about themselves in order to obtain the kind of information necessary for contextualizing our data and ultimately personalizing our learning systems.

**Table 4.1. Interview-Opening Questions**

| Student-Centered Questions | Learning-Centered Questions |
|---|---|
| How are you? | What strategies have you been trying? |
| How are you feeling? | What have you been working on? |
| How do you think things are going? | What progress have you made? |
| What have you been doing? | What's been the hardest thing you've done so far? |
| How has it been going since we last talked? | What goal are you working towards? |
| How are you enjoying the activity thus far? | How would these changes help others to learn more and also have fun? |
| How is your day going? | If you designed this activity, what would you change? What would you keep the same? |
| What's new? | What strategies have you been using in the [learning experience]? |
| How do you feel about the progress have you made since we last talked? | What have you been focusing on during the [learning experience]? |
| Do you feel proud or excited when you get something right? Why? | What changes have you noticed in your understanding of [topic]? |

You can follow these student-center questions with open-ended questions that center the learning experience (also Table 4.1). This will allow you to pivot towards questions about the trigger, but your approach should always be humane. Questions like "what strategies have you been trying?" can be effective even when a child is extremely frustrated, and they are good because prompting reflection is usually an emotionally healthy form of support for the student. However, your questions need to be sincere, which means that you should be responsive both to the needs of the student and to your own interview goals.

[12] Tattle on a co-conspirator, perhaps without hesitation.

Note that like the interview questions that align to your underlying research questions, its good to have multiple ways to ask these **interview-opening questions** as you will likely talk to a child several times over the course of your study. Standbys like “How are you?” and “What strategies have you been trying” can be reused multiple times, but it’s good to have multiple ways to ask them to talk about themselves (student-centered questions) and to talk about the learning activities (learning-centered questions). Table 4.1 includes questions that are literally open-ended as well as those which are pragmatically (conventionally) invitations to share more than just a single word. Note that these are classified by the predicted type of answer (about the student or about what they are doing in the system).

Once the student has described what they are doing and/or how they are feeling in their own words, you can work to find out how that is related to the event that triggered the interview. In some cases, discussions of the trigger may emerge naturally from the opening questions, but this can be an opportunity to investigate other research questions you might have. For example, if an interview was triggered by a student building something in a learning game (e.g. placing ice blocks as a good insulator for buildings in space), starting the interview with “What have you been working on?” may prompt the student to explain what they built and why (e.g., “the class video said dry ice helps insulate buildings so I chose to cover my building in ice blocks, which is the material in the game that is closest to dry ice”). If your research goal is to understand the depth of students’ knowledge, you may then wish to ask related follow-up questions (e.g., “What are you insulating from?”). Depending on the nature of the research questions guiding your work, this line of question/answering could also facilitate a natural transition to broader questions about prior knowledge (e.g. “How does this compare to what you’ve done in your science classes?”) or student interests (e.g., “Do you usually like building things?” or “What kinds of things do you like to make?”). As we have described elsewhere (**Section 4.2.1**), your questions should be age-appropriate while also aligning to your underlying research questions.

Moreover, **Guideline 4** says that your protocol should be able to adapt to **self-presentation effects** (as discussed further in **Section 4.7**) and to **salience effects**, which refer to the tendency for some things to be more noticeable or important to students at any given time. Salience is known to be influenced by emotional intensity (Strongman & Russell, 1986), by perceptions related to how typical a given event is (Ellis, 2017), and by how recently the event occurred.

As an *in situ* method, DDCIs seek to leverage recency effects (interviewing as close to the event as possible), but in some cases, recency may not be sufficient for inducing students to talk about information that is relevant to your research question. For example, taboos, self-presentation effects, and lack of metacognition can all prevent students from discussing certain topics, particularly if they are either so typical that a student does not think you would be interested in them or so embarrassing that the student does not want to talk about them. Therefore, interviewers should be aware of what kinds of topics might be taboo across students from different backgrounds in order to ensure that questions are asked in a way that improves students’ metacognition (as necessary) about the constructs in question without asking in ways that would make students uncomfortable.

Additionally, tacit knowledge—the sort of latent information that is culturally assumed—is unlikely to be salient to many students. As a result, salience is known to be susceptible to cultural variation (Thompson & Juan, 2006), though research suggests that the perceptions that students have about the interviewer’s cultural background can sometimes influence what they might be likely to report (Ensari et al., 2022). In some cases, this may mean that having interviewers who come from more than one cultural background could be advantageous. For example, a student who does not know if the interviewer would understand normal classroom practices might be more likely to explicate them without prompting.

Likewise, we have also sought to explore the effects of having interviewers who were more or less “native” to gaming culture, employing one interviewer who had considerable Minecraft experience and another who had very little. The analysis of this effect is still forthcoming, but our hypothesis was that these researchers would get complementary information, with the expert being able to guide the interviews toward certain specific explanations of the game and the novice being able to test more authentically what was salient to these students without priming for these same specifics. For example, our Minecraft-expert interviewer strategically asked about specific block usage during building phases in the WHIMC project, following up with detailed questions about the students' decision-making strategies. In contrast, the Minecraft-novice interviewer was more likely to ask what the student was building and how they had chosen the designs. Forthcoming research will analyze these interview strategies in tandem with the content and engagement they elicited from the students. Leveraging the strengths and backgrounds of different interviewers may help to surface unique insights and minimize repetition for students.

### 4.3.4. Asset-based Interview Approaches and Shifting Your Protocol Over Time

DDCIs were developed to be used in longer stretches of fieldwork (e.g., across several days in the same classroom), allowing researchers to obtain a better sense of the learning context. In this sense, it borrows from the ethnographic tradition of adapting your interview protocols to meet the conditions on the ground, which should be done while circling back to the theoretical framework which helped to guide the design of that protocol. This timeframe is also important for **Guideline 5** (your interview protocol should seek to identify students' strengths), which are important even when your research question is related to the ways in which a student might be struggling.

As noted elsewhere in this chapter, the interviewer should work to make students as comfortable as possible before, during, and after each step of this process. In addition to the questions that align with your research questions (**Sections 4.2.1** and **4.3.2**), we recommend preparing questions about the students more generally. In some cases, it may be good to extend your efforts beyond the kinds of student-center questions shown in Table 4.1 (**Section 4.3.3**). Although we strongly recommend against overly personal topics, asking questions like "What do you like to do after school," can help you to identify students' strengths and assets so that you can better understand their perspectives on their learning experiences. For our work that explored student science interest, for example, general questions often revealed the various ways that learners chose to engage with science outside the camp setting, such as volunteering, hobbies, and in-class activities.

Specifically, we recommend an interview protocol that emphasizes students' strengths, interests, and autonomy. Although there may be good reasons to follow up with information about where they struggle, this kind of question can be very face-threatening, particularly if the students have not had a chance to see that you are not there as a teacher/authority figure. In fact, even questions that are not face threatening and possibly normal for some cultures in the United States can be very intrusive for some students. For example, "What do you want to be when you grow up?"[13] is a common question for adults to ask children in the United States, but it may not be appropriate for students from cultural backgrounds where employment is not a key part of someone's identity. Some students may be impacted by social desirability bias and may avoid negative answers or discussions of their dislikes. In these cases, reframing questions in terms of what they would change (e.g., "What would you change about the camp?") can grant greater student agency and encourage further sharing. In general, it is very common for younger students to be told not to share information with strangers, and some students may have complicated relationships with institutional authority (e.g., schools and school personnel).

Therefore, in addition to thinking about how your questions align to the learning experience and your research questions, you should also think about the times that it would be most appropriate to ask them. We strongly suggest asking students about their strengths earlier in the process for two reasons: (1) identifying learner strengths is often important for research purposes and (2) it is essential to ensure that a student never feels like an interviewer is accusing them of failure if they struggle during a learning experience. For this reason, we also recommend proceeding cautiously with questions about help seeking, difficulty, and any question that implies that a student has trouble with authority. Although these are all important topics, confiding this information to a stranger can be very face threatening, and this effect is likely to vary across different groups of students.

[13] Note that Table 4.2 shows "Who do you want to be when you grow up?" as a potential question. This twist on the standard question pivots you out of asking about careers, which—while a common question—can be very difficult for some children and culturally inappropriate for others. Some students will be surprised that you ask the question this way. If so, just reassure them that you're trying to make the question more open-ended than their career goals. Other students respond well to this option without prompting, as they are more comfortable talking about life goals outside of specific careers.

**Table 4.2** walks through a sorting process of the kinds of questions you might ask earlier rather than later in the interviewing process. You'll note that there are some questions that specifically ask about struggle, obstacles, or conflict. For most kids, these questions are fine as long as they are asked in a light-hearted way by an interviewer who has made it clear that the goal is to be supportive. However, some students are going to be more sensitive to these questions than others, and students should **never** be pushed to talk about things that they find uncomfortable.

**Table 4.2: Adapting Your Question Strategy Over Time**

| Timeframe | Example Question |
|---|---|
| **Earlier** | How are you doing? |
| | What have you been working on? |
| | What strategies have you been using? |
| | What has been working well? |
| | What do you want to do next? |
| | Is this (learning software) like anything (else) you do at school? |
| **Middle** | Does this remind you of anything you've read about/studied/seen on TV/done at home? |
| | What kinds of games do you like to play? What makes them fun? |
| | Do you like working with your friends when you're learning? What makes them good to work with? What do you do to help them? |
| | What kinds of books do you like to read [shows do you like to watch]? What makes them fun? How does that compare to what you're doing now? |
| | How do you usually learn about this topic at school? What do you like about that? ***What do you dislike about that? |
| | What do you want to do next? |
| | What's the hardest thing you've done so far. |
| | What did [your teacher] say about what you should be working on? ***Are you finding that helpful?/***Are you doing what they suggested? |
| | ***Who do you want to be when you grow up? |
| | ***Do you feel like you need help with anything? |
| | ***Do you like this software better than [however you usually learn in school]? |
| | ***Does the stuff you're learning about here feel relevant for anything you (like to) do outside of school? |
| | ***What kinds of things do you like to do outside of school? What makes that fun/important? |
| **Later** | What have you learned this week? What has been most surprising? |
| | How have your strategies changed over the week? |
| | What's the coolest thing you've done this week? |
| | What do you wish you could learn more about? |
| | ***What's the worst thing you've had to deal with in this class/camp/experience? |
| | ***How hard have things been this week? |
| **Cautiously** | ****Do you usually like school? What do you like best about it? |
| | ***What do you usually do when you feel like you need help? |
| | ***Have you ever really struggled at something at first and then got really good at it? What happened? |

*****Requires cultural adaption**

## 4.4 Choosing When to Skip a Student or Override a Trigger (e.g., Managing Multiple Triggers and/or Resampling the Same Student)

One challenge of this method is that it can be difficult to know how frequently a given trigger will occur in a live educational experience. Even if you have access to historical data, different groups of students often do not behave as expected. There is also always the possibility that calculations were wrong when developing the QRF triggers or when setting the prioritization algorithm. These issues (which we covered in **Section 3.4**) can sometimes be corrected by other members of your research team, but in the meantime, you should be prepared to move forward.

Even if all triggers are being perfectly operationalized, there could be times where you feel like this might not be the right time for an interview on that exact trigger with that exact student. For example:

(a) you feel like you have enough information about that trigger
(b) you have just talked to that student—either about that trigger or for some other reason
(c) the learner seems distressed
(d) the learner is highly engaged, and you don't want to interrupt

(e) the learner is already being interviewed by another researcher
(f) the learner is talking with a peer or teacher in a way that precludes an interview
(h) the trigger is not relevant to the activity right now
(i) there is not enough time to meaningfully engage with the student (e.g., end of class)

In particular, reason (c) deserves further discussion. Although there are times when a distressed learner (especially a young learner) could benefit from the self-reflection that an interview might provide, you should avoid creating additional stress for the student. If you have any reason to suspect that asking them for an interview would make them feel worse, you should skip the interview.

There is no hard and fast rule for what to do in these situations, and you have a variety of choices:

(i) You could use the skip button in the QRF app to get the next trigger/student pair
(ii) You could override who you are choosing to interview using the appropriate text boxes in the app to note the student and the reason that you have selected them
(iii) You could try to interview anyway, focusing more on building rapport with the student than on investigating the specific trigger.

If the student seems particularly distressed, you should alert the teacher/facilitator.

Your decision will need to be guided by the conditions in the classroom at a given time, but you should err towards making students comfortable. As we described above (**Section 4.2.1**), we strongly suggest preparing a range of interview questions. Not only will these allow you to build trust over time with the students, they will also allow you to pivot should either technological or contextual circumstances leave you in a situation where you are not fully satisfied with the triggering events.

## 4.5 Post Interview

The conclusion of the interview is an opportunity to build further rapport with students and also to leave the door open for future follow-up interviews and check-ins. Depending on the number of days that you are conducting interviews, you may interview this child many times, and it is important that they know that you appreciate their time and expertise. Expressing gratitude for their time and attention during interviews is valuable.

If you can do this while expressing interest in the things that are important to them, you'll have a "hook" for beginning your next interview. For example, when students discussed exploration activities during the WHIMC study, the interviewers sometimes concluded an interview by saying things like "Thanks for sharing, I'll let you get back to exploring. I'll stop by later to see what other things you've found." However, in studies where students were more likely to be experiencing frustration or struggle, the interviewers sometimes used a slightly different tactic, validating how difficult it can be to learn and promising to see if the student was doing better the next time they talked.

After you step away from each interview, you can leave notes in the app before advancing to the next trigger. Important information to document includes, but is not limited to: (1) whether another student (or students) joined the interview, (2) if the student that you interviewed was not the student you received a trigger for, and (3) anything else of note that you observed during the interview. Each of these text boxes are optional, in that you can advance to the next interview without filling anything in. However, the more complete you can be in these notes, the better off your research team will be later during data cleaning and processing.

## 4.6 Post Fieldwork Event

The QRF app will take care of aggregating many of the notes and data that you need from fieldwork events, but it can still be a good idea for the research team to regroup at the end of each day and the end of each fieldwork event. Table 4.3 provides an overview of the data that will be collected in this process, and **Appendix 4** provides a checklist to help streamline daily efforts to manage that data.

**Table 4.3 QRF Details that are/not automatically logged**

| Data Tracking | Research Information |
|---|---|
| **Automatically Stored on the QRF App/Phone** | All audio recordings |
| | |
| **Automatically Logged within the QRF server** | Each time a trigger fired, regardless of whether or not an interview was conducted, or notes were taken |
| | All associated metadata (time stamps, trigger info, interviewer, unique trigger ID, username, username overrides, additional student input, etc.) |
| | Information about each time a trigger was skipped, including (a) the student was just interviewed, (b) another student was just interviewed on this trigger, (c) the student is unavailable, etc. It is possible to customize this list. |
| **Requires Manual Effort** | Seating maps |
| | Relevant information about relationships between students (team members, siblings, conflicts, etc.) |
| | Linking between usernames and student names used in consent/assent forms. |
| | Disruptions to individual or group activities (e.g., students who do not complete a task because they have to leave early that day, fire drills, etc.) including student absences |
| | Concerns that the team has about over-representation or under-representation of the triggers that are firing that day |
| | Other observational data made by the research team |
| | Any other data that is not automatically logged in the online learning system (surveys, hands-on activities, etc.) |

## 4.6.1 Daily Efforts

As with any research involving fieldwork, it is a good idea to check in with the research team at the end of each day and to document things like attendance, activity completion, etc. For QRF, this may be a good time to double check how many recordings you've obtained and for what triggers. It is also a good time to work on aligning data from sources that are external to QRF and the online learning system.

In particular, it is a good time for interviewers to check in with the members of the research team who are writing triggers and controlling the prioritization algorithm. This is because it can be difficult to anticipate how frequently a given student might execute an interaction related to a given QRF trigger. For example, if DDCIs are being used to study student interest, your team may have selected behaviors that you thought would be associated with high, low, and mid-range levels of interest so that you were sure to sample from both ends of that spectrum. However, even if you have substantial historical data to support this data selection, you may not know how frequently a new group of students will engage in those same behaviors.

Moreover, even if the prioritization algorithms have worked exactly as you hoped, you may find that you have discovered slightly new questions that you would like to try with students. In this case, you may have new reasons to increase the priority of triggers that you had initially given a lower priority. For example, if you discovered that students were consistently describing a behavior you had not anticipated when you triggered based on their frustration (e.g., as we found in Andres et al., 2022), you might want to increase the number of interviews related to frustration. These changes to the prioritization algorithm should be documented.

The end of each day is also a good time for interviewers to check in with one another to see who seems to be getting the best interviews from each student. For example, it may be that a given student has bonded with a particular interviewer over a shared interest or a favorite book. Conversely, if an interviewer has had to intervene in a student conflict (which should be avoided, whenever possible), they may need another interviewer to prioritize taking those students' triggers the next day. Even if the day has been conflict-free, some interview strategies just work better with some kids, and the best way to find that out is to talk with your fellow interviewers.

Finally, it's a good idea to take notes that you'll be able to find later about important classroom events. For example, if you observe something important outside of an interview (e.g., one student working to help another, a student telling their teacher about a meaningful event that is relevant to the learning context, important patterns of participation with the class that are not being logged by the online learning system, a fire alarm, etc.). You may also wish to note details about students (e.g. specific needs, relevant patterns of behavior) and social relationships (e.g., friendships, conflicts, family relationships) that could impact or lend context to interviews and outcomes.

### 4.6.2 Fieldwork Final Efforts

At the end of a week of fieldwork (or however long you are working with a particular group of students), you should consider ways to fully document and aggregate all of your data sources (seating charts, prioritization algorithms, field notes, student attendance/activity completion, etc.). How you go about doing so should be catered to your analysis plans and to the commitments you have made during your consent/assent processes, but having a plan to do so before you go into the field will help.

In addition, it may also be a good idea to have your interviewers reflect upon and report back on their experiences in a more holistic fashion. For example, did they notice that students who talked about a particular topic seemed to also behave in specific ways? In general, it is good practice to have researchers who were not interviewers to conduct the analysis of interview transcripts (see **Chapter 5**). In some cases, interviewers may provide important insights and context to the analysis process.

## 4.7 Additional Resources for Interview Techniques

Although DDCIs differ in key ways from other kinds of interviewing practices, it can help to consult other sources for best practices when it comes to interviewing. In the course of developing what we are referring to as the **Big Sister Approach**, we were strongly influenced by the field of sociolinguistics, which studies how various groups of speakers differ in their linguistic practices. Although sociolinguistic research often does not require the interviewer to steer the conversation towards a particular topic (e.g., to you acquire specific kinds of information), it is nonetheless a relevant area of research to draw upon because the interviewer must make people comfortable enough to speak in their native dialect. Moreover, when specific topics are important, they can sometimes include very sensitive subjects (e.g., which groups of people do you usually talk to in your daily lives), which necessitates the establishment of trust (Bowren, 2005; Hazen, 2000; Samarin, 1967; Zadrożna, 2016). Therefore, the sociolinguistics research community is highly concerned with the **observer's paradox** (Labov, 1966; Cukor-Avila, 2000; Tovares & Kulbayeva, 2022; St. Pierre, 2024), which articulates concerns about the change that is induced into a "naturalistic" research setting when an interviewer is present. Your goal with the Big Sister Approach is to keep the setting that you are working in as naturalistic as possible—which is why we stress that you should attempt to minimize your role as an authority figure. However, no interview can be done without some interference in the learning environment. Therefore, you should work to be more aware of the kinds of effects you might have (Hazen, 2000), particularly those that are related to students' **self-presentation goals** (Lewis & Neighbors, 2005).

Even in research areas that depend very heavily upon ethnographic interviews, learning how to navigate the nuances of fieldwork (including the observer's paradox) is challenging, and there is often a "sink or swim" approach to teaching it (Macaulay, 2004). In part, this is because variation across both individuals and cultures can be wide-ranging and unpredictable. Moreover, your needs may change depending on your research question. For example, as we discussed above, many linguists are more interested in how people talk than in documenting information on a specific subject, so many publications on fieldwork training discuss how techniques for eliciting phonological (sound) patterns differ from those related to lexical (word), syntactic (sentence) or pragmatic (discourse-level) differences (e.g., Crowley, 2007; Shilling, 2013; Hyman, 2001; Guo, 2022; Chellah & Deruse, 2010).

These issues may seem far removed from the goals of QRF research, and indeed it is difficult to argue that we care about pronunciation differences among our students—as long as the researchers are able to understand them. However, these kinds of publications often also document differences in the kind of data you might get depending on how you ask your questions. Skimming references like these can help you to think about these issues, since even well-intentioned efforts may have unintended consequences. For example, in Anderson's (2008) sociolinguistic study of Appalachian migrants, she positioned herself as an altruistic volunteer who was working on a family history of that community. This role provided everyone with a strong purpose when sitting down for an interview, and indeed she was able to use this effort to fulfill the goals of the principle of linguistic gratuity—a guideline among sociolinguistic researchers that we should work to give back to the community who is volunteering their time (Wolfram, 1995). However, this technique also biases the data towards events that happened in the past, which can be an important consideration even for linguists who are "topic neutral" (e.g., studying grammatical features rather than content).

For that reason, we have given considerable thought to the role that we recommend adopting in QRF fieldwork. We suggest a **Big Sister Approach** so that the interviewer (a) does not become the go-to for students' questions about the learning domain or the learning software (as discussed in **Section 4.3.2**), and (b) so that the interviewer remains approachable for talking about things that are seemingly unrelated to the learning efforts. For example, if you are studying any kind of motivational construct, asking students what they like to read, watch, or do in their free time can help you to better understand how they are approaching the learning domain. These are all topics that a child might discuss with their friends and their siblings, but they may adjust how they report them to someone who was in a position of power over them. Likewise, some students might be reluctant to provide information about struggle to someone who controls their grade, but they would potentially be more open to someone who was supportive and non-judgmental. Teasing out these issues is an area of serious research in and of itself (e.g., Bucholtz et al., 2023; Feagin, 2013; St. Pierre et al., 2024; Tovares & Kulbayeva, 2022), but the goal of the Big Sister Approach (or **Fun Auntie/Uncle Approach**) is to make sure that you are positioned—truthfully—as someone who cares about what the student thinks, even if you see that they are struggling, misbehaving, griefing other students, off-task, or otherwise just having a bad day.

Publications that cover general fieldwork practices could help you better prepare for this approach (e.g., Becker, 2013; Bowern, 2015; Cameron et al., 2018; Cukor-Avila & Bailey, 2001; De Reuse, 2010; Duranti, 2012), particularly if they emphasize the theoretical foundations of these approaches (e.g., McGregor & Fernández, 2019), but so could research on **speech acts**[14] (i.e., how we do things with words) and other research on pragmatics (Fuller, 2000; Gumperz & Cook-Gumperz, 2008; Hymes, 2013; Johnstone, 2005; Searle, 1969) or on discourse analysis more generally (Fairclough, 2003; Kress, 1990; Holliday, 2021; Gee, 2014; Hodge, 2017; Johnstone, 2005; Johnstone & Andrus, 2024; Pillar 2017; Van Djik, 1995; Van Djik, 2011).

If you will be working with younger students, we recommend looking at publications that discuss the needs of children who are being interviewed. These can sometimes be found in research related to sociolinguistics and anthropology (e.g., Johnstone, 1999), but may also be found in research on speech pathology (Augustine, 1995; Barone, 2006; Hoepner, 2024; McFarlane, 2012), medicine more broadly (Aldiss et al., 2009), in legal contexts ( Bull, 2008; Lamb, 1995; Pool & Lamb 1998) and in other areas of research that are interested in fieldwork methods (Danby et al., 2011; Einarsdóttir, 2007; Gibson, 2012; Greig et al., 2007; Hatch, 1995; Jorgensen et al., 2005; Schiller & Einarsdóttir, 2009;) or childhood development (Cameron, 2005; Fargas-Malet et al., 2010; Koivula et al., 2020; Morrison, 2013). Again, your goal is not to have every interview technique and question mapped out, but to be aware of critical issues related to interviewing kids so that you are not caught off-guard by the unique challenges that younger interviewees may present.

[14] Speech acts describe things that you do with words, e.g., verbs related to language: ask, declare, pronounce, imply, explain, etc., but see canonical work in this area.

# Chapter 5: Analyses of Interview Data

## 5.1 Overview

Once you have completed your fieldwork, you can download the interviews from the phone (**Section 5.2**), transcribe the interviews (**Section 5.3**), and begin your analyses. As we have discussed in previous chapters, you should already have some plans for these steps in place, but as we will discuss in this chapter, even the most deductive approaches will likely need iterative rounds of coding in order to understand how your theoretical framework will emerge in your novel research context. This chapter seeks to provide an overview of that process.

## 5.2 Exporting Data from the Phones

QRF data storage may vary by device, but it can be accessed in most Android phones by through the "My Files" application, where you can navigate to "Internal Storage." If the phone's internal system allows it, the data should be stored in its Downloads folder. If not, it should be found in the Android/Data file (e.g., Android/Data/com.Upenn.qrf). It can be difficult to anticipate what new development changes to the Android platform may further impact these storage locations, so we recommend searching for resources on the web that are specific to your phones should an issue emerge that keeps you from immediately locating your data.[15] Once you have located the data on the phone, you can export it to your laptop. Note that Android phones may also require you to download a special app to your laptop in order to export those files.

## 5.3 Transcription

Before analysis can begin, all data collected from the interview sessions needs to be accurately transcribed and proofread. Transcription can be done in a fully automated way (e.g., the assistance of apps like *Whisper AI)* or simply by using a normal word processing tool (e.g., *Word* or *Google Docs*). Many professional transcriptionists also prefer to use a semi-automated transcription tool (e.g., FTW Transcriber), which allows you to automate common functions, including the inclusion of time stamps and to install macros for common speech patterns (e.g., pauses, redactions, etc.). In this section we will walk through choosing a transcription process (**Section 5.3.1**), as well as considerations related to deidentification (**Section 5.3.2**), the careful documentation of speech patterns (**Section 5.3.**3), and handling inaudible data (**Section 5.3.4**).

### 5.3.1 Choosing a Transcription Process

Choosing how to transcribe is an important step in the process. Currently, automatic systems like Whisper AI are not highly accurate for the kinds of conditions that DDCIs are likely to be recorded under. Even with directional microphones, there are many physiological differences in children—including their naturally higher pitch—that increase the variability of their speech patterns and make them notoriously difficult to analyze acoustically even with modern tools (Lee et al., 199; Huber et al., 1999; Sobti et al., 2024). There are also still problems with the recognition of dialect differences that should be considered, depending on your population (Wassnick et al., 2022; Ocumpaugh et al., 2025). When you add to that fundamental difficulty the recording conditions of DDCIs (e.g., multiple voices in noisy classrooms), it is likely that any automatic transcriptions will need significant corrections or revisions. In fact, our understanding of the current state of the art is that disambiguating adult voices (which are far less variable than children's) is only possible with multimodal data and multiple directional microphones, since then vision sensors can help the system to know which microphone to use (Lieberman, 2024).

---

[15] At the time of publication for this manual, this particular tutorial is potentially useful for locating the data on your phone: How To Access App Data On Android! (2025)

If you choose to use human transcription, you should note that this is an important and potentially time-consuming skill. Certification for court transcriptionists, for example, can require weeks or months of training. Even with practice, you should budget considerably more time for transcription than you might imagine. Our professionally-trained transcriptionist still works at a 5:1 ratio (e.g., 25 minutes to transcribe a 5-minute recording), while using a preferred transcriptionist app (e.g., *FTW Transcription*), and this ratio increases for noisy recording conditions.

## 5.3.2 Deidentification

When analyzing educational data collected from in-class activities, particularly when working with children, it is critical to ensure that all personally identifiable information (PII) is thoroughly de-identified prior to any data sharing or further analysis. This step is essential not only for protecting the privacy and confidentiality of student participants (or any participants) but also for adhering to ethical research standards, including institutional review board (IRB) guidelines. When minors are involved, these concerns can become even more important.

There are typically multiple stages of the de-identification process. First, during the interview stage and throughout the research process, care should be taken to protect the PII of the students. If possible, students should not be referred to by their legal name, and interviewers should work to avoid questions that are likely to elicit highly personal information. However, even when these precautions are well-executed, it could be very likely that a child might mention something relatively benign that nonetheless contains PII. For example, they could mention a sibling, incidentally, providing you with names of their family members. Or, if they are reporting the speech of someone else, they might mention their own name.

In other words, interview strategies are an important step for ensuring deidentification, but they are not sufficient in and of themselves. Instead, your transcription process should include the removal of any PII. Student's real names should be replaced by a pseudonym, and other references should be transcribed as [REDACTED] so that readers know that PII has been protected and removed. There are different ways to approach the de-identification process (e.g., through manual review, or by using automated de-identification tools (Zambrano et al., 2025) followed by human checking). In each of these methods used, the goal of de-identification is to effectively remove or anonymize any direct or indirect identifiers (e.g., names, location data, or combinations of attributes that could lead to re-identification), thereby protecting participants while enabling researchers to conduct responsible analysis. However, redaction efforts should still be recorded in the transcript, as we show in Figures 5.1 and 5.2.

| | |
|---|---|
| Interviewer | Hello Octopus! How is everything going? |
| Octopus | Hi |
| Interviewer | "How's your build going?" |
| Octopus | "Fine, but I wish that [REDACTED] would let me build where I want to" |
| Wolf | "Find somewhere else to build!" |
| Interviewer | "Octopus, it looks like Wolf is building there, is there something special about this spot?" |
| Octopus | "Yes, I found it first, it's closest to the sun" |
| Wolf | "You weren't here first!" |

**Figure 5.1: Example Transcription, Showing Deidentification During the Interview (i.e., "Octopus" and "Wolf") and During Transcription (i.e., "[REDACTED]")**

| | |
|---|---|
| Interviewer | You are back. |
| Eagle | Hello Back, I'm [REDACTED]. |
| Interviewer | You are? I thought you were Eagle. |

**Figure 5.2: Example Transcription, Deidentified During Transcription**

### 5.3.3 Documentation of other important language features

Documentation of speech features requires a number of considerations that are different from normal writing. People rarely speak in complete sentences, and we advise you to follow a protocol such as the one developed by the Penn Linguistic Data Consortium (PLDC) to help you make consistent judgements related to these issues. Transcripts should document features that occur in natural speech, including:

I. ***Pauses*** (Figure 5.3): can help you to better understand how the student is responding during the interview Long pauses during interviews can provide a lot of information and it is important to code them appropriately. Pauses can illustrate thinking work, confusion, frustration, shyness, and more. Although not shown here, timestamping your transcripts can also help to account for pause length.

II. ***Hesitation Markers* or *Fillers*** (e.g., “um” or “uh,” Figure 5.4): although many transcription conventions remove hesitation markers like “Um” or “hmm” or false starts, we recommend recording them as they can be indicative of both cognitive and motivational constructs that are important to understanding learning.

III. ***Connected Speech Patterns*** (e.g., “gonna” vs. “going to”, as in Figure 5,5): Many things that people might label as slang are actually just examples of connected or informal speech (e.g., “gunna” for “going to,” “‘cause” for “because,” etc.) Since (in)formality can sometimes be used to convey attitude, and because syllable counts could help you to better estimate length/time-related constructs, we recommend transcribing these as accurately as possible.

IV. ***Speech Overlap*** (Figure 5.6): Showing speech overlap can help coders to better understand the degree to which the student is cooperatively participating in the interview as well as their level of enthusiasm

V. ***Slang*** (Figure 5.7): Slang is a technical term that refers to the innovative vocabulary/language use of a particular group, often young people, that is typically perceived as informal. Transcribing it accurately can help you to better interpret their meaning later. This could be especially important if the interviewer did not realize what the student was saying at the moment, as in this hypothetical example.

VI. ***Jargon*** (Figure 5.8): Although the distinction may or may not be relevant to your underlying research questions, jargon and slang are different constructs. Jargon refers to specialized terms that are used within certain groups. It often refers to technical vocabulary used by specific groups of professionals,[16] but could also refer to the language of video games (e.g., *griefers, glitches, grind, easter egg, AFK, creepers, mod, nerf, noob, redstone*, etc.) or to language specific to the learning domain (e.g., *dwarf planets, apulsar, centaur, airlock, isochrone,* etc.). Capturing jargon accurately helps you to better interpret students’ meaning and to assess the degree to which they have the skills (including the vocabulary) to engage deeply with the learning system.

VII. ***Words Being Spelled Out*** (Figure 5.9: if a student spells out a word, it could indicate a number of things, including their inability to pronounce the word or (as shown in this example) an affective experience

| | |
|---|---|
| Interviewer | “Hi Dolphin, why aren’t you working with your group?” |
| Dolphin | [long pause] |
| Interviewer | Is your internet working? |

**Figure 5.3: Example Transcription, Long Pause**

[16] For example, the distinction between consent and assent, discussed in another footnote of this document, is an instance of legal jargon. Many adults in the US would probably refer to the process of asking students to agree to be interviewed as asking for their “consent.” Typically, only those who are legal professionals or who have training in the ethics of human-subjects research would use the word “assent” to describe this process.

| | |
|---|---|
| Interviewer | Hi Marlin, what are you up to? |
| Marlin | Um…. |

**Figure 5.4: Example Transcription, Hesitation Markers**

| | |
|---|---|
| Interviewer | So what do you think- |
| Kangaroo | It just needs one more block. |
| Interviewer | … you're gonna do next? |

**Figure 5.5: Example Transcription, Speech Overlap**

| | |
|---|---|
| Interviewer | What is your plan, what will you build next? |
| Dolphin | I'm gonna build a tower, I shoulda probably built it first. Cause it'd be cool to be able to use it to find the base. |

**Figure 5.6: Example Transcription, Connected Speech**

| | |
|---|---|
| Interviewer | What's going on? |
| Bear | Dolphin and Fox are fighting. |
| Interviewer | Oh no? What happened? |
| Bear | Fox was helping me build a landing pad and Dolphin took down his airlock. |
| Interviewer | Awww! That doesn't sound very nice! So are you helping him? |
| Bear | Base. [long pause] He's helping me now. |
| Interviewer | He's helping with your base? |
| Bear | No. Well, yeah. He's helping with my landing pad. |

**Figure 5.7: Example Transcription, Slang**

| | |
|---|---|
| Interviewer | What happened here? |
| Fox | I was building my base but Bear's a griefer! |
| Interviewer | What do you mean? |
| Fox | I was doing what I was supposed to do! And he griefed me! And now I have to put more redstone in. |
| Interviewer | Oh no! What's the redstone going to do? |
| Fox | It's going to do what it was doing before! |
| Interviewer | Ah, so you have to start over!  What is your goal? |
| Fox | I'm trying to mod the base so it can have an airlock. |

**Figure 5.8: Example Transcription, Jargon**

| | |
|---|---|
| Interviewer | How are you doing? |
| Tardigrade | I am M-A-D! |
| Interviewer | M-A-D? What do you mean? |
| Tardigrade | I hate my user name! |
| Interviewer | Oh, you're mad cause we're calling you tardigrade? |
| Tardigrade | Yeah |

**Figure 5.9: Example Transcription, Spelling (Hypothetical)**

## 5.3.4 Handing Inaudible Data

Given that the interviews were sometimes conducted in multi-participant settings or within real classroom environments, there is a likelihood of background noise or overlapping conversations. These factors can occasionally make portions of the audio/video difficult to hear or interpret. In such cases, rather than attempting to guess what was said, that section of the transcription should be labeled as inaudible (Figures 5.10 and 5.11).

| | |
|---|---|
| Interviewer | “3D printer” |
| Butterfly | So you’re [inaudible] around the 3D printer |
| Interviewer | That’s what this is, cool! |

**Figure 5.10: Example Transcription, Documenting Inaudible Data**

| | |
|---|---|
| Interviewer | Hello, Mister Octopus. |
| Octopus | Hello. |
| Interviewer | How's your [inaudible] working out? |
| Octopus | All of these! |
| Interviewer | All of those? Wow, you filled up your whole slip there. |
| Octopus | Yes! And now I have the freedom to explore this really ... |
| Interviewer | Since you've done your homework? |
| Octopus | Yeah! Even though I hate homework. |

**Figure 5.11: Example Transcription, Inaudible**

## 5.4 Coding Interviews

As with other parts of this process, the codes that you will use to analyze your interviews will depend on your specific project, the learning opportunities available in your learning system, and your exact research questions. Although researchers sometimes make distinctions between inductive and deductive coding practices, we typically recommend an iterative process, which allows you to both examine the data more inductively while aligning with the theoretical frameworks that have guided your research question and your interview strategies.

Whether you are using a bottom-up or top-down approach to code the data, we recommend following key steps: first, take time to familiarize yourself with the data, including its context, target audiences, data type, and the overall scope or length of the dataset. Second, we encourage you to take notes while reading and reviewing the data. It is perfectly fine to start with more codes than you may ultimately use, as codes can be refined or re-consolidated throughout the coding process. Third, as mentioned above, an iterative coding process is suggested for coder(s) to make any necessary adjustments to the codebook and improve coding consistency over the course of analysis.

As discussed in other parts of this manual, the QRF method has been influenced by a wide range of different areas of research, including those related to discourse analysis. Whether you are taking a more qualitative or quantitative approach to analyzing your data, it is worth thinking about how those priorities might influence your analysis. Certainly, it is possible to have important data that is infrequent, and there is important work in the areas of discourse analysis and intercultural communication that could help you to consider these concerns. For examples in the broader literature, see Van Dijk, 2013; Chiu, 2008; Scollon & Scollon, 2011; Dunn & Neumann, 2016, but also see our discussion in **Section 4.7**.

**Table 5.1: Example Coding: Explanation, Enthusiastic, Brief, Neutral, and Uncertainty**[17]

| Speaker | Text | Explanation | Enthusiastic Response | Brief Response | Neutral Response | Uncertainty |
|---|---|---|---|---|---|---|
| Interviewer | So, what, what's the difference here that you guys are exploring? | NA | NA | NA | NA | NA |
| Lion | Uh ... Uh, what if Earth had, was orbiting a star half the size of ours? | 1 | 0 | 0 | 0 | 1 |
| Interviewer | Did you ever think about this before? | NA | NA | NA | NA | NA |
| Lion | Nah | 0 | 0 | 1 | 1 | 0 |
| Lion | [long pause] | NA | NA | NA | NA | NA |
| Interviewer | But you're thinking about it now? What do you think about it? | NA | NA | NA | NA | NA |
| Lion | Well, it's super cool cause I get to run around and figure out what happens to the tides and also all the frogs are dead. | 0 | 1 | 0 | 0 | 0 |
| Interviewer | Oh? You like dead frogs? | NA | NA | NA | NA | NA |
| Lion | I love dead frogs! I'm going to feed them to Bear! | 0 | 1 | 0 | 0 | 1 |

Table 5.1 shows an example of a coded interview segment, which was transcribed using a method adapted from CITE and described in **Section 5,3.** Here, we have coded for constructs related to how receptive a student was to being interviewed at that particular moment. Although the codes you apply in your research will typically be more specific to your research questions than these, we chose to apply these codes in order to to have a way to control for whether or not the student was interested in talking to us during our investigation of science interest development among students using the WHIMC learning system. The coding scheme includes six constructs each reflecting a different communication style: *Disconnected Responses (*students introduce ideas unrelated to the interviewer's question), *Explanation (*students provide added detail or reasoning of their actions), *Enthusiastic Response* (verbal or emotional expressions of strong interest or excitement), *Brief Response* (minimal replies), *Neutral Response* (factual answers delivered in a flat or emotionally neutral tone), and *Uncertainty / Hesitation* (pauses, hedging, or language that indicates the student is unsure). Together, these categories help us analyze the degree to which students are willing to be interviewed, but they also provide some information about how enthusiastic they are about talking about the WHIMC (Minecraft) game, which we hoped could further contextualize other codes in our data.

## 5.5 Priming Effects and Analyzing Null Instances of Codes

During the process of coding interview transcripts, researchers will produce what we refer to as *"null instances of codes," or* times when a given code does not apply. This should not be surprising, as the fact that we are coding for the occurrence of something means that we expect variability. There may be many reasons that we find these gaps in the data, and it's worth considering the causes of those as we are picking our methods of analyses

In some cases, our methodological approach may influence what gaps we uncover. Notably, the interview protocol can influence the sort of topics that students find salient. **Salience** can be generated by conversational coherence expectations. For example, a student who has just been asked if their teacher usually gives homework is more likely to give an opinion about homework than a student who has just been asked if they have any pets. However, salience can also be generated by other kinds of recency effects (e.g., did their teacher just assign homework), including those that might induce strong emotions (e.g., did the teacher just cancel the class's pizza party because the students hadn't turned in their homework). Salience can also be related to novelty (e.g., did the teacher just cancel the students' daily homework assignment), with events that are considered typical (e.g., the teacher wrote the students daily homework assignment on the board—as expected every day), less likely to be noticed and/or discussed unless the student thinks that interviewer is not likely to understand what the teacher is doing. Salience can also be generated by activating other parts of the perceptual system (e.g., MacLeod, 2015; Wang et al., 2018), including information that is related to the self (as opposed to information related to others) and other factors that help to guide the cognitive/attentional systems (see discussion in Liu and Sui, 2016).

[17] These codes were developed for use in Wei et al., (2025) and Ocumpaugh et al., (2025), which investigate how interviewer strategies are related to student responses.

Because salience can be hard to predict and document, interpreting null codes is challenging. In fact, their interpretation may depend on several factors, including the relevance of the code to the theoretical constructs you're investigating and whether the student was primed to provide an utterance that would elicit that code. Being aware of these issues may be particularly important when you are employing a top-down (deductive) approach to the codebook development process, as the theoretical framework being used may not always fully reflect the nuances present in the analyzed data.

## 5.5.1 Analyzing Null Instances of Relatively Common/Expected Codes

Even if you are conducting an inductive thematic analysis of your data, there will likely be codes that you anticipated when you were designing your interview questions. For example, consider the "explanation" code from the previous section. Obviously, we do not expect every utterance that a student produces to be an explanation. In fact, the occurrence of explanations is highly conditioned by the kind of question the interviewer asks. However, interpreting what it means when we have a null instance of this code is complicated, as it could mean:

> (a) the interviewer didn't ask a question that would typically elicit an explanation, and the student was following conventional conversation patterns (e.g., a close-ended yes/no question that is more likely to prompt a binary answer from a student),
>
> (b) the interviewer asked a question that would elicit an explanation, and the student was unable to provide one (e.g., a student may not able to explain why water boils at a lower temperature on Mars than on Earth at that given moment),
>
> (c) the interviewer asked a question that would elicit an explanation, and the student was unwilling to provide one (e.g., a student focused on the gameplay and therefore unwilling to be distracted).

For codes that are closely tied to theoretical frameworks, like explanation (which is associated with knowledge measures), one might be able to anticipate the interpretations and code accordingly.

In these cases, how you code will depend upon the level at which your statistical analyses will be structured (see further discussion of this in **Sections 5.1 and 5.3**). If your interview questions were designed in a way that you expect students to be asked about their prior knowledge in every interview, you may only want to apply a binary code (presence or absence) to each transcript. However, if you want to analyze how many times it appeared in each interview, you'll likely want to control for both the length of the interview and whether or not the student was primed to give an explanation in the previous interview question. Table 5.2 shows a table coding an interviewee for the "I hate homework" code and coding the interviewer for whether or not they had primed the student to talk about homework.

**Table 5.2 Table Coding Transcript for Student Utterance and Priming Effect**

| Speaker | Text | Priming-Homework |
|---|---|---:|
| Interviewer | Hey Bear! How is the Minecraft going? | Not Primed |
| Bear | I love it! I usually have to do homework after school. | 1 |
| Interviewer | Is this better than homework? | Primed |
| Bear | I love Minecraft | 0 |
| Interviewer | Aww! Yay! Do you love homework, too? | Primed |
| Bear | I hate homework. This is so much better! | 1 |
| Interviewer | Oh yeah? What do you like about Minecraft? | Not Primed |
| Bear | I really hate homework. | 1 |

## 5.5.2 Analyzing Null Instances of Relatively Uncommon and/or Unexpected Codes

When analyzing codes that were not common or expected, you must think about whether or not the codes occur frequently enough to warrant your attention. Again, these decisions may be related to the type of analyses that you want to run, but they may also require additional nuanced decisions about the (a) your beliefs about why this code emerged so infrequently, (b) it's relationship to other constructs that are important to your research question, and (c) relevance of the code to your constructs. This kind of information can help you to decide whether or not to keep the code independently, and if so, how to treat null instances of the code.

In many cases, if a code was highly infrequent, it may be related to the design of your interviews. Take, for example, a student expressing, *"I hate homework!"* in an interview. If your interview protocols were not designed to ask students about homework, this sentiment may only emerge in a handful of interviews. Rare codes should not automatically be excluded, but it does complicate its interpretation and how you choose to treat the code during your analyses.

The examples in Tables 5.3-5.6 illustrate these concerns. Both of Tables 5.3 and 5.4 have null instances of the "I hate homework" code, and they are differentiated by whether or not the interviewer **primed** the student for talking about homework. Priming effects are important to interpreting null instances of the code. A spontaneous eruption of "I HATE HOMEWORK" in an interview where the child was not primed to talk about that activity (Table 5.3) might be qualitatively different than an instance like that in Table 5.4 where the interviewer mentioned the word homework in reference to what the students had been told to look at while interacting with the learning game.

**Table 5.3: Applying the "I hate homework code," Unprimed Condition (Hypothetical)**

| Speaker | Utterance | Hate Homework |
|---|---|---|
| Interviewer | Hey, okay if we talk? | Not primed |
| Quipper | Sure... | 0 |
| Interviewer | Do you like the Mars world? How is your base-build going? | Not primed |
| Quipper | I HAAAAAAATE HOMEWORK!!!!!!!!!! This is soooo much better. | 1 |

**Table 5.4: Applying "I hate homework code," Primed Condition (from WHIMC project)**

| Speaker | Utterance | Hate Homework |
|---|---|---|
| Interviewer | All of those? Wow, you filled up your whole slip there. | Not primed |
| Elephant | Yes! And now I have the freedom to explore this really ... | 0 |
| Interviewer | Since you've done your homework? | Primed |
| Elephant | Yeah! Even though I hate homework. | 1 |

Now contrast these transcripts (Table 5.3 and Table 5.4) with Table 5.5, where the interviewer primed the student to talk about homework, but the student did not directly specify their feelings about homework.

**Table 5.5: The "I hate homework code," Primed Condition (Hypothetical of Table 5.4)**

| Speaker | Utterance | Hate Homework |
|---|---|---|
| Interviewer | All of those? Wow, you filled up your whole slip there. | Not primed |
| Panda | Yes! And now I have the freedom to explore this really ... | 0 |
| Interviewer | Since you've done your homework? | Primed |
| Panda | Yeah! I did my homework and now I get to go check on other things. | 0 |

Next, consider the hypothetical example in Table 5.6, which has also been coded for both the "I hate homework" code and whether or not the interview had primed the student to talk about homework. In this longer example, all of the student's utterances represent a null instance of the "I hate homework" code, but that is not surprising because the student was being asked about their learning-game experience rather than about their opinions of homework.

**Table 5.6: Hypothetical null instance of the "I hate homework code," Unprimed Condition**

| Speaker | Utterance | Hate Homework |
|---|---|---|
| Interviewer | Hey, okay if we talk? | Not primed |
| Caterpillar | Sure... | 0 |
| Interviewer | How are you doing? | Not primed |
| Caterpillar | I'm good. I've been building the gym for our Mars base. | 0 |
| Interviewer | Do you enjoy that? | Not primed |
| Caterpillar | It's great! I'm working on making it pretty. And I'm putting up signs to tell the other teams to stay out. | 0 |
| Interviewer | Cool! What do you want to work on next? | Not primed |
| Caterpillar | I don't know yet. I like being able to just decide what to do next whenever I'm done. | 0 |
| Interviewer | What's been the best thing that you have gotten to do in WHIMC so far? | Not primed |
| Caterpillar | I liked exploring, and I liked playing with my friends while I was learning things. | 0 |
| Interviewer | So are there any things that you get to do in school that are like this? | Not primed |
| Caterpillar | No…well yes. On the playground we chase people. And sometimes I get to read whatever I want to when we go to the library. | 0 |
| Interviewer | What do you like to read? | Not primed |
| Caterpillar | I like Geek Girls. And Percy Jackson. Stuff like that. | 0 |
| Interviewer | Those sound fun! Do any of those talk about science stuff? | Not primed |
| Caterpillar | Yeah, but I like it anyway. | 0 |
| Interviewer | You like it anyway? Do you not usually like science? | Not primed |
| Caterpillar | I do, but it's messy and I don't like frogs and slimy things and my brother says we have to do that in high school. | 0 |
| Interviewer | Gotcha. So, would you say you like the kind of science you're learning about here better? | Not primed |
| Caterpillar | Yes because it isn't slimy. And I can help people. | 0 |
| Interviewer | So what parts of the base that you're working on can help people? | Not primed |
| Caterpillar | Well, the gym is definitely super important because we need people to be healthy and gravity and Mars. And also the airlock, cause we don't want to lose our oxygen. And the green place. I mean, the greenhouse. | 0 |

As Table 5.7 summarizes, these scenarios represent four possible conditions related to the "I hate homework" code (not primed and expressed, not primed and not expressed, primed and expressed, prime and not expressed). If most of your interviews do not show priming (the first row), it may be tempting to throw out a code like this, particularly if it is very rare. And, indeed, **since you have not explicitly asked most of the students about their feelings towards homework, you cannot accurately interpret the null instances of this code**. It's possible that students who did not express their feelings about homework (e.g., the student in Table 5.6), also hates homework, but they did not express it in the interview. This could be justification for dropping the "I hate homework" code from your analyses.

**Table 5.7: Four Conditions Related to Priming Effects**

| | Student's Utterance: I hate Homework Code | |
|---|---|---|
| **Interviewer's Utterance** | **Expressed** | **Not Expressed** |
| Not Primed | H1 | H4 |
| Primed | H2 | H3 |

However, dropping the code because it is rare and unprompted is not the only option. Another option is to fold that code into another that is related to your research questions. For example, if "(lack of) affinity for school" had emerged in your data, you could choose to recode the "I hate homework" code under that broader category, even if this instantiation seems unique.

A third option is to keep the "I hate homework" code as its own stand-alone variable, despite the concerns that you have about interpreting null instances of it. This may sometimes be an appropriate choice—especially for an utterance that demonstrates strong and negative arousal. For example, you may find later that the student in Table 5.6, who provided us with an emphatic and spontaneous opinion about homework, is an outlier in some of your analyses, even among students who do not have a strong affinity for school. If that is the case, the student's spontaneous announcement about their feelings towards homework may be pointing towards evidence that this student is qualitatively different from their peers, opening up new avenues for investigations of their data.

## 5.6 Construct Validity

In addition to making sure that your codes align with your research questions, you also need to ensure that they are valid and reliable. **Construct Validity**[18] refers to the degree to which you have identified a single construct (Chronbach & Meehl, 1955). How this is proven varies by field, with many qualitative studies (particularly in research domains where data is hard to obtain and analyzed by a single researcher), a strong, theoretically aligned definition may be sufficient (Lavery et al., 2018).

There are good reasons, however, to consider methods for demonstrating that the definitions of your qualitative codes have achieved clarity. Two methods associated with the testing/development of construct validity are **social moderation** and **interrater reliability (IRR)**. These two distinct methods, described in more detail below, are distinct, but related. With social moderation, researchers work together to discuss codes until they reach consensus, whereas with IRR, researchers code independently. Neither method proves construct validity, but they are sometimes used to improve the definitions of the qualitative labels being used in research and to demonstrate the degree to which they can be reliably applied by others.

### 5.6.1 Social Moderation

In **social moderation**, two or more researchers go through the data either to generate new codes or to discuss existing codes. In this dialogic process, researchers may challenge each other's interpretations, clarify the application of codes, and collectively refine their understanding of the data (Shaffer et al., 2016). This collaborative negotiation helps surface assumptions, sharpen conceptual definitions, and ensure greater consistency across coders. Rather than seeking immediate consensus, social moderation emphasizes mutual sensemaking and iterative development of the coding scheme, treating disagreement as a productive opportunity to deepen analytical rigor.

### 5.6.2 Interrater Reliability (IRR)

In contrast with social moderation, interrater reliability (IRR) tests whether or not two coders can agree on the same code without talking to one another about it. IRR does not prove construct validity, but it does demonstrate the clarity of a given code since it signals that your efforts could be reproduced by others without needing you there to explain it to them.

This approach is particularly useful when the goal is to establish replicability or when codes will be applied across a large dataset by multiple analysts (human or automated). While high IRR scores suggest that a set of codes (codebook) is well-defined and internally consistent, they do not necessarily capture the nuance or depth of the meaning-making process. As such, IRR is often used in combination with social moderation: researchers might first co-develop a shared understanding of the codes through discussion, and then test the consistency of their application independently using IRR procedures.

Interrater reliability is the most common assessment of the reproducibility of coding results, but you may want to consider other metrics such as test-retest reliability (assessing whether the same coders apply codes in the same way at different points in time). As Stemler (2004) argues, you may also want to assess interrater reliability in more than one manner.

[18] **Construct Validity** is a notion introduced to the social science community in the 1950s alongside several other types of validity (Becholdt, 1959; Chronbach & Meehl, 1955), and was initially used to refer to attempts to show that a test (i.e., a survey related to personality distinctions) was measuring a single, latent construct that researchers were struggling to define. Over the years, its meaning has evolved (Messick, 1995), and it has been borrowed into several different disciplines, including those related to qualitative coding.

There are several ways to calculate interrater reliability, some of which depends on what kind of data you are working with. As Table 5.8 shows, these include Cohen's (1960) Kappa (κ), Fliess' (1971) Kappa (κ), Kendall's (1938) τ, Krippendorff's (1970) Alpha (α), and Cronbach's (1951) Alpha (α), but Pearson's (1900) Rho (ρ) and Spearman's (1904) Rho (ρ) can also be used. All are preferred over accuracy measures, as accuracy is more susceptible to agreement occurring by chance. If this does not make sense, consider a context where you are applying binary labels to data. For example, if you had created an "I hate homework" code for a research context where most students were not talking about homework. If 98% of the data does not mention homework, your accuracy is likely to be extremely high even if both of your coders just code every utterance as "not hating homework."

**Table 5.8: Select Measures used to Check Inter Rater Reliability (IRR)**

| Description | Evaluation and Criticisms |
|---|---|
| **Cohen's κ:** (-1 ≥ κ ≤ 1) Measures agreement between two raters on categorical items, adjusting for chance agreement. Use when you have two raters and nominal (categorical) data. | Banerjee et al. (1999); Brenner & Prediger (1981); Carey & Gottesman (1978); Flak et al. (1988); Feuerman & Miller (2008); Guggenmoos-Holzmann (1996); Hsu & Field (2003); Powers (2012); Vach (2005); Von Eye (2005); Warren (2015) |
| **Fleiss' κ:** (-1 ≥ κ ≤ 1) Extension of Cohen's κ for more than two raters. Adjusts for chance agreement on categorical items. Use with three or more raters and nominal data. | Banerjee et al. (1999); Flalotico & Quatto (2015); Hsu & Field (2003); Powers (2012); Warrens (2010); Quarfoot & Levine (2016) |
| **Kendall's τ:** (-1 ≥ τ ≤ 1) Rank-based correlation that compares the number of concordant and discordant pairs. Use with ordinal data or when examining rank agreement between two raters. | Arndt et al. (1999); Bonnet & Wright (2000); Chen et al. (2022); Essam et al. (2022); Hamed (2011); Xu et al., (2013) |
| **Krippendorff's α:** (-1 ≥ α ≤ 1) A flexible reliability coefficient for any type of measurement (nominal, ordinal, interval, ratio), and any number of raters. Handles missing data. Use when multiple raters, varied data types, or missing data are involved. | Antoine et al. (2014); Madugu & Ibrahim (2024); Quarfoot & Levine (2016); Zapf et al. (2016) |
| **Cronbach's α:** (-1 ≥ α ≤ 1) Measures internal consistency across items to indicate how well a set of items measure the same concept. Use for assessing consistency of scale items, not interrater agreement *per se*. | Becker (2000); Dunn et al., (2014); Kalkbrenner (2023); Malkewitz et al., (2023); McNEish (2018); Sjitsma (2009) |
| **Pearson's ρ:** (-1 ≥ ρ ≤ 1) Linear correlation coefficient for continuous variables. Indicates strength and direction of a linear relationship. Use when raters give continuous (interval/ratio) ratings and assumptions of linearity are met. | Bonnet & Wright (2000); Chok et al. (2010); Essam et al. (2022); Rupinski & Dunlap (1996) |
| **Spearman's ρ:** (-1 ≥ ρ ≤ 1)Non-parametric rank correlation for ordinal data or non-linear relationships. Use for ordinal data or when assumptions of linearity for Pearson's r are violated. | Bonnet & Wright (2000); Arndt et al. (1999); Cliff & Charlin (1991); Essam et al. (2022); Myers & Sirois (2014); Xu et al., (2013) |

In contrast, metrics like Cohen's (1960) Kappa (**κ**) adjust for these base rates so that we do not overrepresent how often two coders agree with one another. (See for example, Dewey, 1983; Fleiss, 1981; Perreault & Leigh, 1989; Tooth & Ottenbacher, 2004; Sadatsafavi, Najafzadeh, Lynd, & Marra, 2008; Wirtz & Kutschmann, 2007.) Kappa scales from -1 to +1, where **κ =** -1 means that the coders have shown perfect *dis*agreement, while **κ** = 0 indicates chance levels of agreement and a **κ =** 1 indicates perfect agreement. A "good **κ** score," however, is still fairly arbitrary, as **κ**'s meaning (how difficult it is to achieve) varies across different data sets.

Researchers have recommended IRR cut-offs ranging from a Kappa of 0.4 to 0.8 (Di Eugenio & Glass 2004; Fleiss, 1971; Landis & Koch, 1977; etc.), but more conservative recommendations (e.g., **κ** = 0.8) are probably more appropriate in domains where coders are working using definitions that are not open to judgment and where the consequences of being wrong are quite high (e.g., research where the implications of being wrong could cause substantial harm). Meanwhile, lower thresholds (e.g., **κ** = 0.4) are more likely when the underlying truth is uncertain or the coding schemes are holistic, and—even when the consequences of being wrong is higher—they may be more acceptable in larger data sets since where aggregation can help to wash out imperfections (e.g., labeling errors in 2,000,000 photos of cancerous tumors may still result in a machine-learned model performing better than one being trained on a small amount of excellent data).

### 5.6.3 Choosing when to apply social moderation and/or IRR techniques

Qualitative researchers have strong opinions about the use of techniques like social moderation or IRR, and you will want to make sure that the methods you use are appropriate for your targeted publication venue. However, it is not unusual to iterate across these methods, particularly since IRR metrics like Kappa do not reveal whether coding differences are random or caused by systematic differences between coders; this determination requires additional qualitative examinations of the data (Sim & Wright, 2005), which social moderation can facilitate. In some cases, this kind of iteration can be important even if you do reach an acceptable IRR.

The process for iteration may differ depending on whether or not you are using an inductive or deductive approach to codebook development. For example, researchers who are developing inductive codes may use some form of social moderation while developing their codebook, and then stop that process to code independently for an IRR check. Likewise, even researchers who are using a more deductive approach may find that their initial codes are under-specified, making it difficult to achieve IRR. In such cases, it is not unusual to (1) pause your efforts to achieve IRR, (2) use a social moderation process to refine those codes, and then (3) resume independently coding the data to re-check IRR.

In many quantitative research communities, being able to achieve IRR is preferred. However, in many qualitative fields it is neither standard practice nor particularly preferred. For example, in many qualitative fields, where it is more unusual to have large teams collaborating, IRR is simply not feasible. Likewise, if your data set is quite small and (especially) if a particular code is exceptionally infrequent, achieving IRR is going to be impossible. Similarly, IRR is not a reasonable standard in research spaces where you would need years of training to be able to apply a particular code (e.g., a researcher studying a dying language might not have someone who is fluent enough in that language to apply that code).

For researchers hoping to conduct thematic analyses, there are many resources that cover the processes more broadly (Bryant & Charmaz, 2007; Corbin & Strauss, 2015; Locke et al., 2022; Miles et al., 2014; Saldana, 2016), and several that propose specific methods or strategies (Eisenhardt, 1989; Gioia et al., 2013; Yin, 2015). Although there are strong arguments for not coding your data (Packer, 2017; St. Pierre & Jackson, 2014), the practice of coding DDCIs is likely to help you to better contextualize the events you have chosen for triggering interviews, which is likely to be a major goal of your work. **Appendix 5** provides an abbreviated overview of the process of working with your data from the transcription to the coding process.

## 5.7 Aligning Codes and Metadata

Each interview is associated with a unique student ID (which does not contain personally identifiable information) and a timestamp marking the moment the interview began. Using these student IDs, it is possible to link interviews to external measures collected through other instruments (e.g., knowledge assessments or motivational measures), as well as to specific in-game actions made by each student. While the in-game interaction logs are inherently synchronized with the interview data—since they serve as the trigger for the interviews—external measures may not always align with the user IDs, especially when collected outside the learning platform. Therefore, it is essential to encourage students to use their designated IDs when completing these external instruments and to ensure they are entering them correctly.

To facilitate the likelihood that students enter these IDs correctly, we recommend assigning child-friendly IDs, such as simple animal names. Ideally, when filling in these questionnaires, students should be able to select their ID from a predefined list that matches those used in the logs and interview triggers.

However, if technical constraints require the creation of new IDs in real-time, providing a text box or a blank space for students to manually enter their IDs can be a practical alternative. In such cases, manual review and mapping between the IDs used in external instruments and those from logs and triggers will be necessary for accurate analysis.

Timestamps also enable researchers to identify when interviews occurred and to align these moments with students' in-game actions. This alignment is crucial, as gameplay activity may be reduced or paused while students are engaged in an interview, potentially having some impact on the analysis of logs. Moreover, each utterance within the interview transcripts should be time stamped at a fine granularity—ideally down to the second or millisecond if feasible. Doing so facilitates synchronization with in-game logs, making it possible to analyze what students were doing in the game at the exact moment they responded to interview prompts, mapping logs and coded interviews.

## 5.7 Possible Analyses

Once the interviews are coded, there are many possible ways to analyze the data, depending on whether or not you believe that the phenomena in your data require the most simple of analysis (e.g., the presence or absence of a topic emerging in the interview) or whether the data warrants more complicated analysis that examines the frequency or the structural patterns that emerge for a given topic in the DDCIs.

One method of analysis that is still relatively simple is correlation. Correlating aggregated student response patterns (e.g., the number of times that a particular code emerged in each student's interviews) with independent measures such as interest levels from the survey or in-game behavior can provide information about whether students with higher interest levels are more likely to discuss certain topics in their interviews. For example, students who show more elaboration or enthusiasm during interviews may also display stronger astronomy interest or more interaction with specific points of interest or tools within particular worlds of an educational game. Such relationships could offer insight into how student responses and behaviors reflect underlying motivation or engagement with the learning experience.

Another useful angle is to examine temporal patterns across the interview. We can ask whether certain types of responses—such as hesitation, elaboration, or silences—tend to cluster at the beginning, middle, or end of the interview, which may reflect how students warm up to the conversation or respond to different phases of questioning. If interviewer lines are also coded, we can investigate which kinds of prompts or strategies tend to trigger specific student responses, such as elaboration following open-ended encouragement or hesitation after evaluative questions.

It is also valuable to consider structural patterns using techniques like Ordered Network Analysis (ONA, Tan et al., 2023) or Sequential Pattern Mining (SPM, Srikant & Agrawal, 1996). Analyzing transitions between student response types or coded utterances allows us to visualize how students move from one type of response behavior to another, such as from brief to elaborative responses, from hesitation to enthusiasm, or from one topic to another across the interview (Ocumpaugh et al., 2025).

**A similar analysis can be done also considering in-game actions to determine specific behavioral patterns that precede or follow certain responses beyond the specific triggers that prompted the interview.** These transitions might correlate with other outcomes, such as interest development or learning gains. Response patterns can be used to cluster students into profiles—for instance, students who primarily give brief replies versus those who elaborate frequently, or students who give brief responses while performing multiple actions within the game versus those who give brief replies but without doing multiple actions beyond wandering within the game. These profiles can then be compared across game performance, survey data, or demographic variables. Similarly, clustering students by in-game behaviors and then comparing response types across multiple behavioral groups might also provide valuable insights into the association between behavioral patterns and student responses.

Finally, even without additional coding, we can revisit segments that contain elaboration or emotional intensity to identify moments where students express identity, motivation, or metacognition. These can be qualitatively examined for how students frame their relationship to the subject matter, their learning process, or the interview context itself. Altogether, combining coded interview data with other sources opens up rich possibilities for understanding not just what students say, but how they participate in a learning conversation.

# Chapter 6: Ethical Considerations

## 6.1 Intro

We have discussed several ethical issues related to interview strategies and the handling of data in earlier chapters. This chapter summarizes those issues, covering information related to informed consent/assent and a discussion of potential immediate benefits for study participants. It also discusses issues related to confidentiality and data storage.

## 6.2 Informed Consent and Assent

Legally, this research cannot be conducted in the United States unless (a) your project has been approved by an independent Institutional Review Board (IRB). If you are planning to interview children, the IRB will likely require you to obtain both (b) parental consent and (c) student assent[19] for this research. Most IRBs will likely require parents to sign paperwork showing that they have consented, but—particularly with very young students—the IRB may be less likely to ask you to obtain children's assent with a signature on a piece of paper. In part, this is because it is unusual (in the US) to ask children to sign legal documents, but in part this is because assent should be an ongoing process.

For legal and ethical reasons, it is important that students understand that they are allowed to decline an interview at any time for any reason. As we described in **Section 4.3.1** that information should be explained to them at the beginning of your fieldwork effort, when the interviewer explains to students what assent means (as many will not have heard the term before). It should be very clear to students that they will not be in trouble with either the interviewer or their teachers if they decide they no longer want to be interviewed.

In fact, it is good practice to re-confirm assent throughout the interview process. This can be done in a very child-friendly way by asking a question like "Hey, is it cool if I talk to you now?" as opposed to a more legalistic "Do I have your assent to conduct this interview?". Confirming assent is important for ethical reasons, and it can also demonstrate to the students that you are not there to patrol them.

In some cases, a student might decline assent for a particular interview because they are—in that moment—very upset. We have recommended elsewhere (**Section 4.4**) that you may want to skip an interview with a child who looks particularly upset, but some kids might not be overtly expressing distress. If a child declines assent, it is good practice to (a) confirm that they are allowed to do so at any time and (b) let them know that you will still be happy to talk to them again if they ever change their mind, but it may also be appropriate to (c) gently ask if they are doing okay—without recording their audio.

Likewise, an ongoing consent process requires that you stop the interview at any time upon the student's request. If a student appears uncomfortable at any point, it is essential to reiterate that participation is voluntary and that they are not obligated to answer any questions. As described in **Section 4.3.2**, the interviewer's role and tone should always be that of a supportive adult, but if *for any reason* a student interprets a question in a way that suggests they feel uncomfortable or ashamed, you should remind them that they do not have to answer any questions. As a trusted adult, you should reassure them that the goal of the study is to figure out how to improve the learning system.

Likewise, an ongoing consent process requires that the interview be paused or terminated immediately upon the student's request. If a student appears uncomfortable at any point, the interviewer should reiterate that participation is voluntary and that the student is not obligated to respond to any questions. Interviewers should maintain a neutral and supportive tone throughout and must take care to avoid any language or behavior that could be perceived as shaming. In the event that a student interprets a question as critical or confrontational, the interviewer should promptly offer a clear and respectful apology, clarifying the intent behind the question. A supportive and truthful way to do so is to tell the student you are sorry and then to remind them that (a) everyone struggles at some point during learning and (b) your goal is to make sure that the learning system works for all students.

---

[19] The word ***consent*** has legal implications that do not apply to minors in the United States. In the US, minors can only ***assent*** because ***consent*** must be granted by a guardian. That said, even many adults in the US do not understand this distinction, so explaining this in a child-friendly manner is important.

## 6.3 Immediate Benefits to the Students

Your institutional review board may also require you to state that participating in your study will have no immediate benefits to the students. Although most researchers would not be conducting research unless they believed that the learning software would improve the students educational experience, your consent paperwork—a legal document—will likely be required to say that it does not (especially if you are investigating a learning software where the learning outcomes have not yet been well-researched).

That said, researchers in anthropology and related fields often look for ways that they might give back to the population who is volunteering for their study (Cameron et al., 2018; Wolfram 1993) It is worth thinking about the ways in which DDCI research might be of benefit to the students in your study—as opposed to the students who you hope will benefit from any future improvements to the learning system. Our early research showed that DDCIs may help students to productively resolve confusion (Bosch et al., 2021), which is in line with more general research that shows that talking through problems can improve metacognition (Chan 2012; Smith & Mancy, 2018; Wall & Higgins, 2006).

Making sure that interviewers are supportive of students is an important aspect of ethically conducting DDCI research. Although we strongly recommend against interviewers taking on the role of teacher, providing light motivational support is an ethical thing to do. Indeed, we encourage you to validate students' experiences of struggle and to respond in ways that acknowledge those students who persist through difficult material, in line with research on growth mindset (Burnette et al., 2023), which we know to can be influenced by the adults that students interact with in their learning environments (Bardach et al., 2024; Mesler et al., 2021; Yu et al., 2022).

## 6.4 Privacy and Confidentiality

There are multiple points of data collection and storage in a QRF study. Students engage on client machines that may collect identifiable information. It is advisable to add layers of obfuscation to each layer of data capture to help ensure privacy. This could include loaning students computers (rather than having them use their own), providing pre-created accounts in the learning environment and deleting automatically collected data when it is no longer required.

The QRF app does not use student names, and—when possible—we encourage interviewers to do the same. For example, in the WHIMC project, each student was assigned an animal for their username. This allowed interviewers to greet the students by their username (e.g., "Hi Octopus! How is it going?"), further preserving anonymity. The same animal name was used to label subsequent data analyses and provided an easy shorthand to refer to students without exposing identities. Participants often have a lot of fun using their animal moniker, though picking labels that students are familiar with is probably advisable. (Humorously, many WHIMC players had never heard of a "tardigrade" or a quipper.")

## 6.5 Data Storage

As described in **Chapter 2**, QRF maintains a separate server from that of the client server (i.e., the learning environment being investigated). Most learning systems collect real-time data about student behaviors within this system, some of which are used to adapt the learning environment to the students' needs and others of which are maintained as records (i.e., log files) that can be analyzed later.

The database(s) for your learning system may vary, but best practices for securing this data involve limiting access and control of that data as well as limiting the amount of personally identifying information (PII) collected. In the WHIMC learning system, for instance, student accounts are assigned unique user identifications (UUIDs), and students data types are separated into many *documented* linked tables and. The MySQL version is regularly updated and backed up both on and off of the server machine. Researchers are only able to access this data using secure systems, including multi factor authentication, and they are given “read-only” access to data so as to not accidentally alter it.

The QRF server for each project is secured through a multi factor authentication password protection system. Trigger information used by the QRF server’s dispatcher is stored in QRF’s FireBase, which is also password protected. Interview data is recorded to the phones used for conducting interviews, which are also password protected. That data can be uploaded to your university's preferred data storage system for transcription (e.g., DropBox). Once transcription is complete, audio recordings can be destroyed to help further ensure that reidentification is not possible. Although the transcription process should also include deidentification, best practice includes ensuring that they are also stored in secure, password-protected systems that meet the legal requirements of your university for storing educational data (e.g., FERPA laws in the United States).

Any scripts or plugins pulling database entries for QRF triggers should be backed up and documented as well, typically in a repository like Github. Such repositories might also be used for documenting changes to trigger codes, although we recommend keeping a running list of such changes in your daily notes (see discussion in **Appendix 4**) as well.

# Chapter 7: Past, Present, and Future Research

## 7.1 Overview

DDCI research has been used to investigate a variety of different student interactions within online learning systems, and it shows the flexibility to adapt to many others. For example, it could also be useful in non-educational contexts, helping user design experts to pinpoint specific, infrequent interactions with other digital systems that warrant further investigation.

In fact, first DDCI study resulted in an unusually high number of studies (Andres et al., 2019; Baker et al., 2021; Baker et al., 2024; Biswas et al., 2017; Bosch & Paquette, 2017; Cloude et al., 2024; Jaeger et al., 2019; Jiang et al., 2018a; Jiang et al., 2018b; Hutt et al., 2021a; Hutt et al., 2021b; Hutt et al, 2022; Munshi & Biswas, 2019; Munshi et al., 2020; Munshi et al., 2018a; Munshi et al., 2018b; Munshi et al., 2018c; Munshi et al., 2020; Munshi et al., 2022; Munshi et al., 2023; Ocumpaugh et al., 2021; Paquette et al., 2021; Rajendran et al., 2018; van Stee et al., 2023; Zhang et al., 2020; Zhang et al., 2021; Zhang et al., 2022; Zhang et al., 2023; Zhang et al., 2025; Zhang & Paquette, 2020). Although not all of these studies were a direct result of DDCI data, we believe that the amalgamation of interdisciplinary specialists required to facilitate such a study lead to strong collaborations and insights that might not have been reached were we working in isolation.[20]

## 7.2 Modifications on the Fly

### 7.2.1 Changes to the Online Learning System

In addition to being used to collect evidence for longer-term research projects, DDCI data has also been used to make rapid design changes to a learning system. For example, in Ocumpaugh et al., (2021), DDCI interviews revealed that students were particularly unimpressed by interactions with a non-player character (NPC) named Mr. Davis. Specifically, they disliked the fact that Mr. Davis would sometimes say "Hmmf" to them. This concern arose across multiple interviewees in multiple classes, allowing us to ask students what they would rather have Mr. Davis say instead. Those changes were very quickly made through minor modifications to Mr. Davis' script, and when we returned to the fieldwork site a few months later, not only did the students report liking Mr. Davis better, but this affinity was correlated with stronger learning performance measures in that fieldwork session.

### 7.2.2 Changes to the Research Protocol

In addition to making changes to the research system, the advantage of conducting interviews *in situ* is that you can adapt your research protocol as new information emerges as well. For example, in the Betty's Brain study, as it became clear that some students' self-regulation strategies were related to intrinsic interest, the lead interviewer (Ocumpaugh) began asking DDCI questions that went beyond students' immediate gameplay activities to include what kinds of science they were most interested in. Likewise, as evidence about students' trait-level anxiety emerged, the research team was able to add a survey scale that ensured that we were able to capture the anxiety of all students—and not just those whose anxiety was salient during the individual interviews.

## 7.3 Analysis of Transcripts and Student Interactions with the Learning System (To Date)

Not surprisingly, much of the research that has been conducted so far has involved the examination of interview content. In addition to research that centers directly on the analysis of student interviews, DDCIs have also led to deeper investigations of students' interactions with the learning system, and some research has also investigated the relationship between these two data sources (interview data and logged interactions) with external measures of learning and motivation (e.g., surveys and external learning measures).

[20] And we are grateful for those partnerships.

To date, much of the published research has been within the Betty's Brain online learning system. This research has shown that students who verbally expressed goals—be they mastery goals or achievement goals—were / more likely to experience confusion and more likely to engage in certain productive behaviors related to self-regulation strategies like reflection (Andres et al., 2022). We also found that students whose DDCI transcripts included evidence of metacognition while experiencing confusion were more likely to show learning (Bosch et al., 2021).

In contrast, DDCIs uncovered evidence that some students using Betty's Brain likely had state-level science anxiety, and our analysis of that data (Andres et al., 2022) allowed us to show that anxious students described engaging in less productive behaviors (e.g., deleting responses out of frustration or retaking past quizzes). Specifically, when it became clear that trait-level science anxiety (as opposed to in-the-moment instances of being nervous or anxious) was common among students in the Betty's Brain study, we added a validated survey of this trait and compared it to students' interactions within the learning system. Instead of using quizzes as tools to identify potential areas for improvement, anxious students described using quizzes as a means for confirm information they were confident about, as a method for quelling the distress over unfamiliar content as opposed to a method for learning whether they had mastered new information. In fact, we found that anxious students were less likely to take quizzes, and that they had distinct patterns of frustration, as measured by the automated detectors that were triggering some of these interviews (Andres et al., 2022). Additionally, our analysis showed that these students had higher rates of deletion behaviors (e.g., removing constructs from their causal maps) and posited that these behaviors were induced by anxiety related to uncertainty and difficulty self-assessing their responses or identifying solutions.

Emerging research using DDCIs to investigate the WHIMC learning system is currently in its early phases. Interest development codes are being applied to the interviews that have been transcribed so far, but that effort is still underway. Instead, our focus has been on how the interview behaviors (e.g., speech acts) of students and interviewers differ in the earliest fieldwork conducted in this project. For example, Wei et al., (under review) explores the relationship between student interest and these patterns, and Ocumpaugh et al., (2025) describes how these patterns emerge more synchronically during the interview process. These papers rely on the same transcription data, but the latter seeks to examine how students respond to different strategies employed by the interviewer in an effort to advise the use of those interview strategies in future DDCI research.

## 7.4 Potential for Future Research

The flexibility of this method suggests that it could be useful within a variety of research paradigms, both within the learning analytics community and beyond. QRF can help to address the **needle in a haystack problem** (Baker et al., 2024) or **one shot problem** (Hutt et al., 2022), helping researchers to locate specific interactions with a digital learning system that may be infrequent and easy to miss by an observer in the room. Therefore, it is likely that the QRF system could be useful with other kinds of Human Computer Interaction (HCI) research questions (e.g., those that are related to online systems that meet consumer or business needs, rather than educational ones).

Currently, the Penn Center for Learning Analytics (PCLA) has partnerships with two other learning systems where DDCIs have been deployed. As discussed in several parts of this manual, this includes the WHIMC system, where DDCIs are being used to investigate student interest development. It also includes CMU's Decimal Point learning system, where DDCIs were conducted in order to better understand why students may be gaming the system (e.g., Baker et al., 2024). As we have described, developing triggers for a more amorphous set of research questions (e.g., why are some students (not) increasing their interest in STEM during gameplay?) may be more complicated than developing triggers to investigate a specific behavior (e.g., gaming the system). However, in both cases, the goal is to better understand why students are (or are not) participating in certain behaviors within the learning system so that we can better support their learning.

The potential for this methodology to generalize to new research questions is as strong as many other interview-based methodologies. It is likely to adapt well to a variety of phenomena as long as researchers are respectful of students and their need for autonomy while they are working through online learning systems. In this manual, we have laid out guidelines for the approach that interviewers should take while approaching students (see **Sections 4.3** and **Appendix 3**). We have also outlined advice for designing interview triggers and for developing the QRF prioritization algorithm (see **Chapter 3).** We are excited to see this method expand to new research groups and welcome questions from those who are interested in adapting QRF and DDCis to their own research questions.

# Glossary

**Big Sister Approach -** this is the role that we advise you to adopt when interviewing students. As described extensively in **Chapter 4**, taking on the role of big sister (or Kind Auntie/Uncle) gives you minimal obligations for enforcing classroom activities or learning goals, making it more likely that students will provide you with authentic information about their learning experience.

**Client Server -** the server for the digital learning system you are investigating (e.g., Betty's Brain, WHIMC, or Decimal Point).

**Cooldown Mechanism -** coding within the QRF system that keeps triggers from firing too often for the same student (see **Section 3.5.4**).

**Constructs -** the concepts under investigation in your study; a term relevant both for defining your research question, identifying and developing triggers, coding transcripts, etc.

**Construct Validity -** refers to whether or not you have identified a single, well-defined construct (see discussion in **Section 5.6**), which may be important when determining how to shape your analyses of interview data.

**Data Driven Classroom Interviews (DDCIs)/Data Driven Interviews (DDI):** This manual was written for conducting in-the moment interviews with students, triggered by their actions or experiences within a digital learning system. To date, this work has taken place in classrooms, but we have sometimes referred to it as Data Driven Interviews (DDIs), as the method could easily be adapted for laboratory and other in person, non-classroom settings.

**ETAWAM (Extra Teaching Assistant With A Microphone) -** this acronym describes an approach where the interviewer conducting DDCIs takes on the role of a teacher or teaching assistant (TA). This approach, while seemingly benevolent, can prevent the interviewer from gathering the kind of data necessary to improve the learning system—both because being an ETAWAM changes your role with the student to that of an authority figure and because the interviewer is actively working to scaffold the students' learning process. As tempting and altruistic as this role appears, we urge you instead take on the role of Big Sister (see Big Sister Approach).

**Firebase Database -** The central storage database for each QRF project, where all triggers and dispatcher metadata are saved. Dispatcher metadata is a temporary copy of the trigger data, used to guide how each trigger is processed, prioritized, and cleaned up.

**Interrater Reliability (IRR) -** one method for ensuring that the constructs you are coding for in your interview transcripts can generalize to new circumstances is to check for IRR. As **Section 5.6.2** describes, there are multiple algorithms for checking IRR, and it is often considered good practice.

**Known Answer Questions (KAQ) -** answers that are asked even though the asker knows the answer. KAQs can be common in many US learning contexts, where they are often used to check whether students were understanding material that was just covered or whether the retained information that was covered before. However, KAQs are considered disingenuous and potentially even inappropriate in some cultures. Being aware of these kinds of cultural expectations can help you to better modulate your interviewing strategies as needed.

**Needle in the Haystack Problem (NHP) -** As described by Baker et al., (2024), it can be challenging to conduct qualitative interviews on phenomena that occur infrequently in the classroom—especially if a student is afraid they might get in trouble or be perceived of poorly if they are caught doing actions associated with that phenomenon. DDCIs make it possible to study any such phenomena as long as it can be detected by the learning system because the QRF app is able to immediately alert an interviewer when such a phenomenon occurs. See also: **one-shot problem**.

**Observer's Paradox -** the idea that having a researcher/outsider in the learning experience fundamentally changes the learning experience. In this manual, we provide suggestions for minimizing observer effects (see **Section 4.7**).

**One-shot Problem (OSP) -** As described by Hutt et al., (2022), this refers to the idea that some phenomena of interest to researchers are so rare that if they are missed, they may not occur again for a significant amount of time. Akin to Baker et al.'s (2024) Needle in the Haystack Problem, QRF helps to address the OSP.

**Presentation Effects -** the idea that people change their behaviors (and the ways they describe them) in order to be perceived in a more positive light. In this manual, we provide suggestions for minimizing these effects

**Priming Effects -** the effect of introducing certain topics or reactions based on the way an interview is structured and/or the individual questions in an interview are asked. Some priming effects are closely related to salience. See **Section 5.5.**

**Prioritization Algorithm -** Located within the QRF server, the prioritization algorithm is applied to the triggers that have been harvested from the client server in the order in which you have previously specified. This ensures that the most important triggers (as you have defined before entering the field) reach the interviewer in a timely fashion.

**QRF Android App:** The mobile application that interviewers use while collecting data. This application receives and sends signals to the QRF Dispatcher, facilitating user interactions.

**QRF Dashboard:** A web-based interface that retrieves data from the Firebase Database, providing a real-time overview of stored triggers and system activity.

**QRF Dispatcher:** The core routing system of QRF, responsible for receiving, organizing, and transmitting triggers between users and the database.

**QRF Polling Script –** the Python script that is installed inside the client server (learning software) that pushes potential interview triggers to the QRF server.

**QRF WebSocket:** The communication service that connects the QRF Android app, Dispatcher, Firebase Database, and Dashboard in real-time.

**Salience Effects -** the tendency of certain topics to be more relevant to a given student at a given time. In interview research, salience is likely to be increased by emotion, recency, event typicality, and other factors that increase the likelihood that an event will be vividly remembered. See the discussion of **Sections 4.3.3, 4.7,** and **5.5,** including the discussion of **Guideline 4**.

**Social Moderation** - a method used when coding qualitative data where two or more coders discuss the coding strategies and definitions they are using. As described in Section 5.x, social moderation is sometimes used as an alternative to interrater reliability (IRR), but it is sometimes a secondary step in achieving IRR (e.g., used when IRR calculations suggest that the definition being used is ambiguous). In the latter case, coders may use social moderation to improve the construct definition before re-checking IRR. See **Section 5.6.1.**

**Speech Acts -** What people do with words (declare, ask, invite, imply, etc.). See **Section 4.7**.

**Triggers -** in the DDCI method, these are the events (e.g., specific behaviors) within your learning system that you want to investigate with your interviews. They are referred to as triggers because the QRF server tracks these and sends them to an interviewer in order to trigger an individual DDCI.

**Trigger Expiration -** As the goal is to conduct interviews *in the moment,* it is important that triggers that are not immediately selected are able to expire. This is handled by the QRF dispatcher, which can be adjusted so that some triggers expire more quickly than others.

# Appendix 1: Technical Troubleshooting FAQ

- The triggers aren't firing or random triggers are firing to often
  - Check firewall permissions at your field site. For example, are all websites associated with the client server and the QRF server permitted by the local school's firewall system?
  - Check to ensure that all variables have been finalized in your triggers. For example, if the trigger includes something like "interview a student when they are within [x] blocks of an NPC," did the [x] get filled in to the code?
  - Check to make sure that the prioritization algorithm in the QRF server matches the prioritization algorithm in the Client server. For example, does one rank things 1-10 where 10 is most important and the other ranks them 1-10 where 1 is most important?
  - Otherwise check the prioritization algorithm?
- Certain triggers are NOT firing
  - Checked to ensure that variables have been filled in. For example, your random trigger likely states that the trigger should fire "every [x] seconds" if no other trigger is available
  - Adjust the trigger to fire more frequently
- Certain triggers are firing too often
  - Adjust the trigger to fire less frequently in the trigger code
  - Adjust the prioritization algorithm to reduce that trigger's priority in the queue
- The recordings are too quiet to be heard
  - Check connections with microphones and associated equipment (e.g., a USB pre used to connect microphones to the phones)
  - Check and/or change batteries in the recording equipment.
- What's my password?
  - Your research team will put together a password for you
- The screen keeps shifting sideways and I cannot advance to the next trigger
  - Go into your Android settings and change the lock screen function
- I'm having trouble finding the recordings on the phone
  - see **Section 5.2**.

# Appendix 2: Installing the QRF Server

## Overview of Steps

1. **Create a Firebase Realtime Database**
   Set up a Firebase database to store and manage QRF triggers.

2. **Create a Personal Channel within the QRF Secure WebSocket Link**
   Generate a unique WebSocket channel for secure communication.

3. **Remix the QRF Dispatcher Code on Replit and Configure the Dispatcher to use your Personal WebSocket Channel**
   Clone the QRF Dispatcher repository to create your own instance. Modify the code to connect with your WebSocket.

4. **Configure the QRF Android App to use your Personal WebSocket Channel and Server Storage Path**
   Modify the QRF app settings to connect with your WebSocket and database.

5. **Set up the QRF Dashboard on your Server**
   Deploy the dashboard code and link it to your Firebase Database for real-time monitoring.

# 1 Create a Firebase Realtime Database

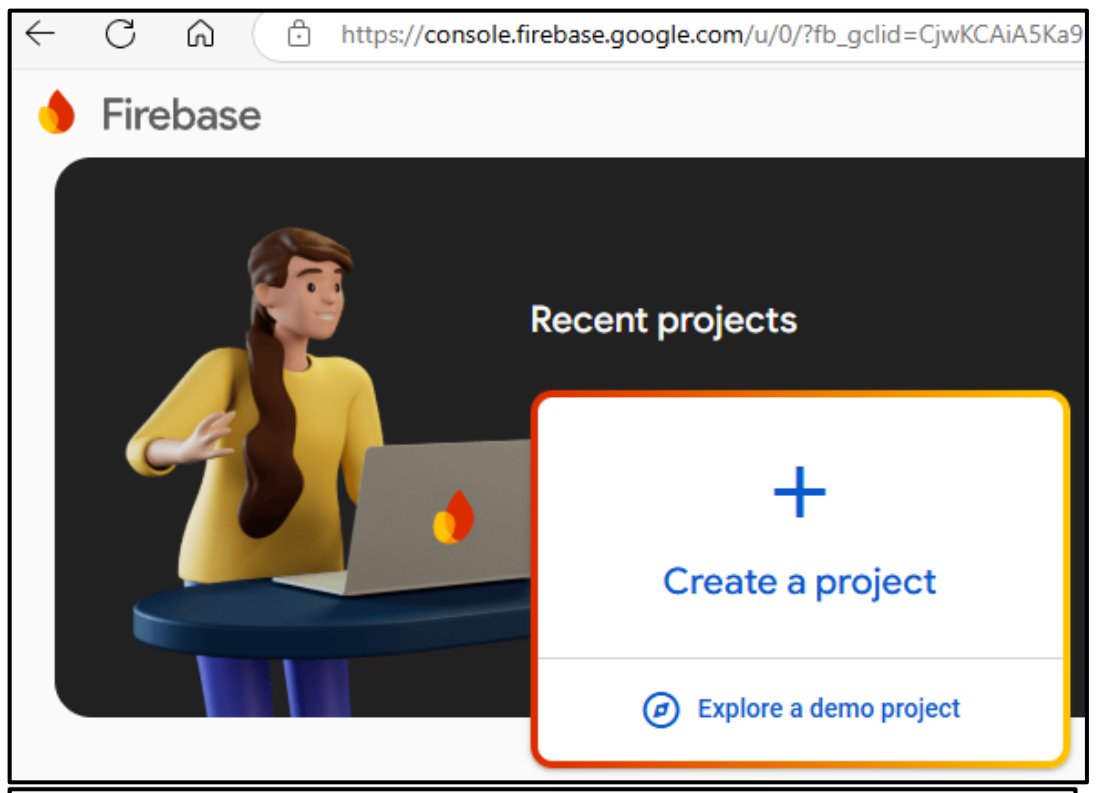

Go to [console.firebase.google.com](console.firebase.google.com) and create a project

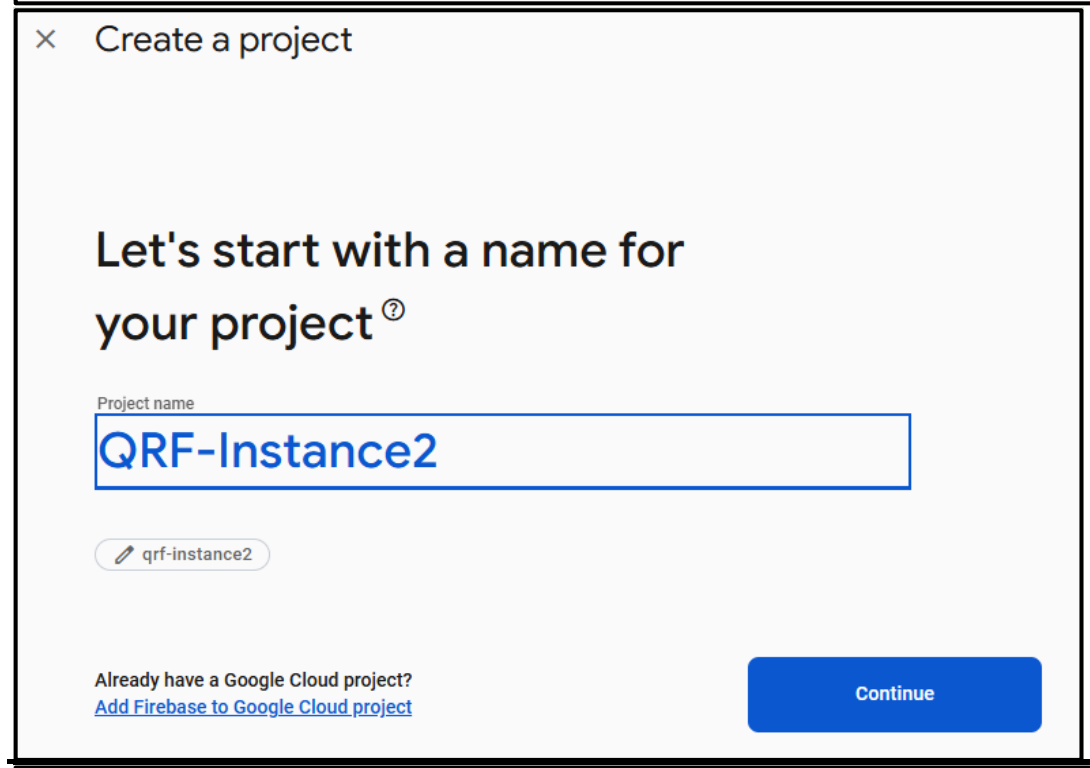

Use a unique name for your project. You will be asked if you want to enable Gemini integration or Analytics in the next step, you may just opt out of these additional features.

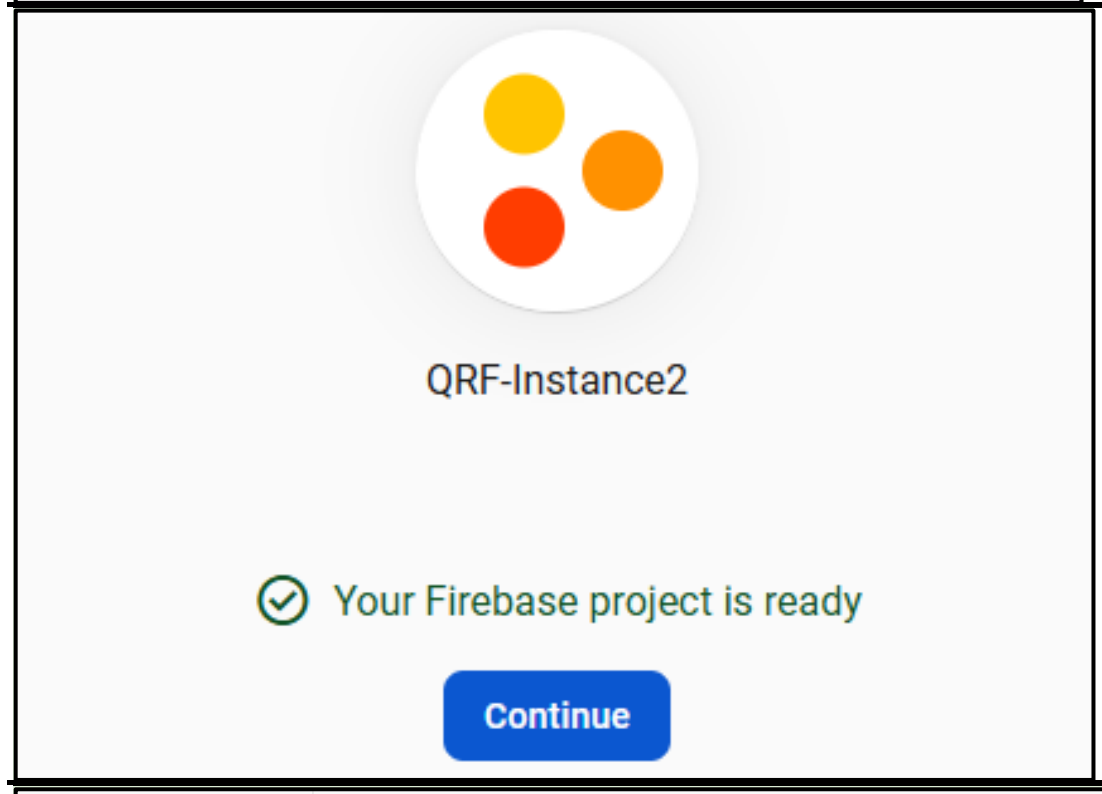

When your project is ready click Continue

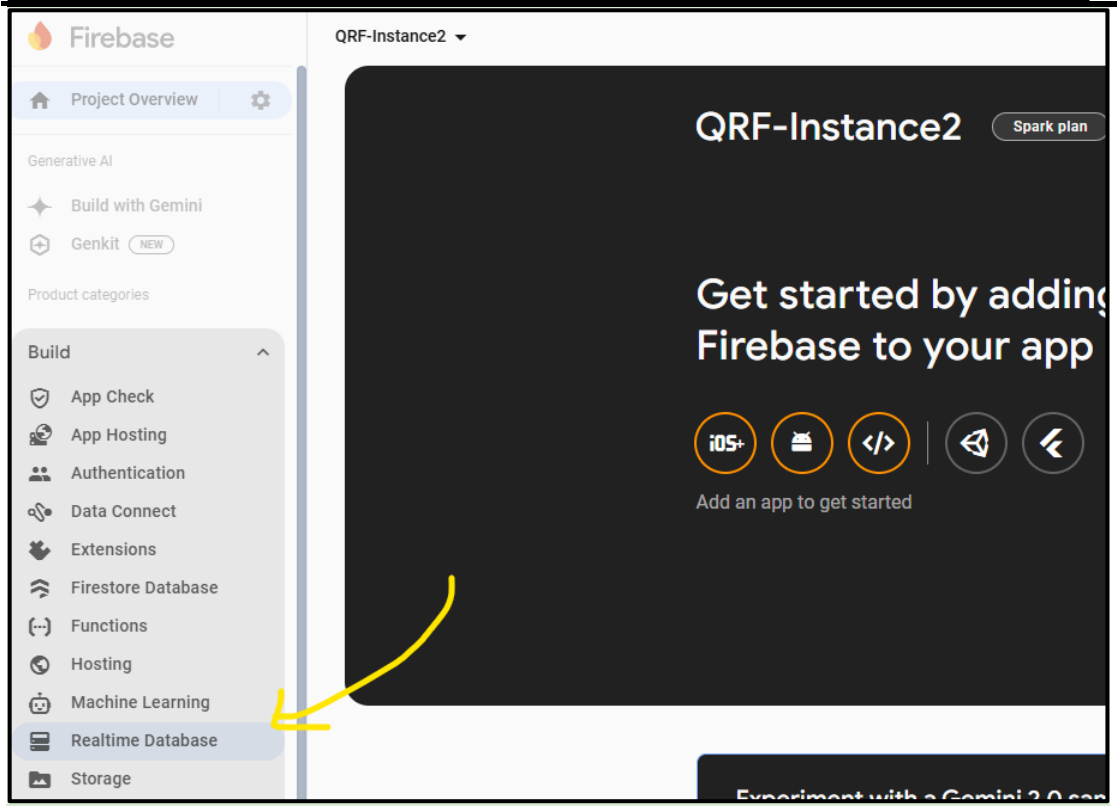

Create a Realtime Database by clicking on Build->Realtime Database in the left sidebar of your new project

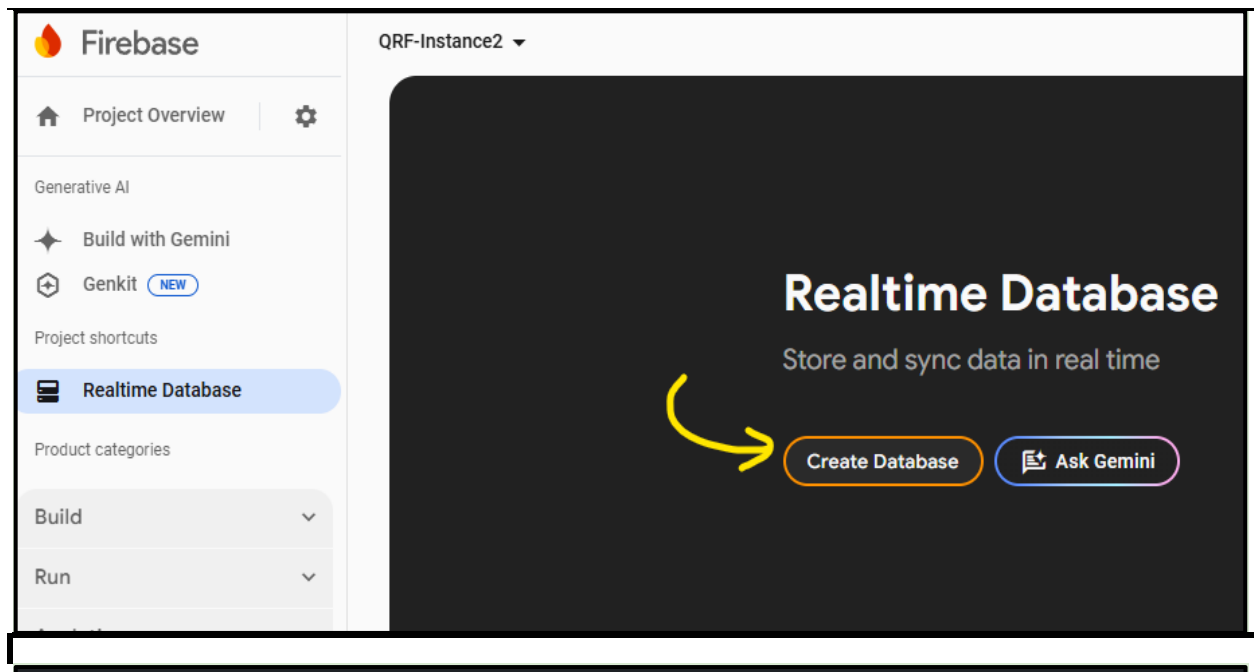

Choose Create Database

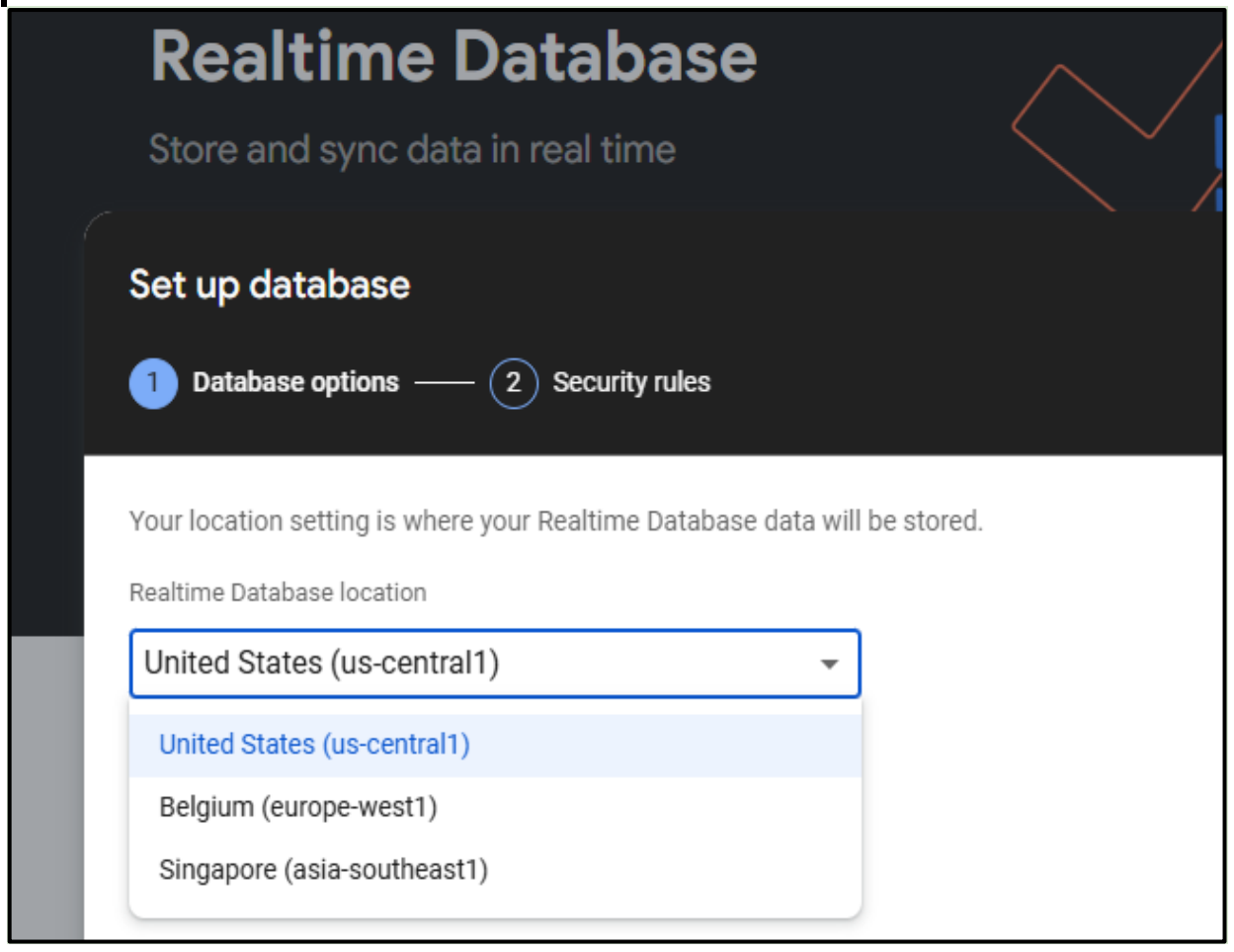

Choose where you would like the server of the database to be set up geographically. We recommend choosing the server closest to you.

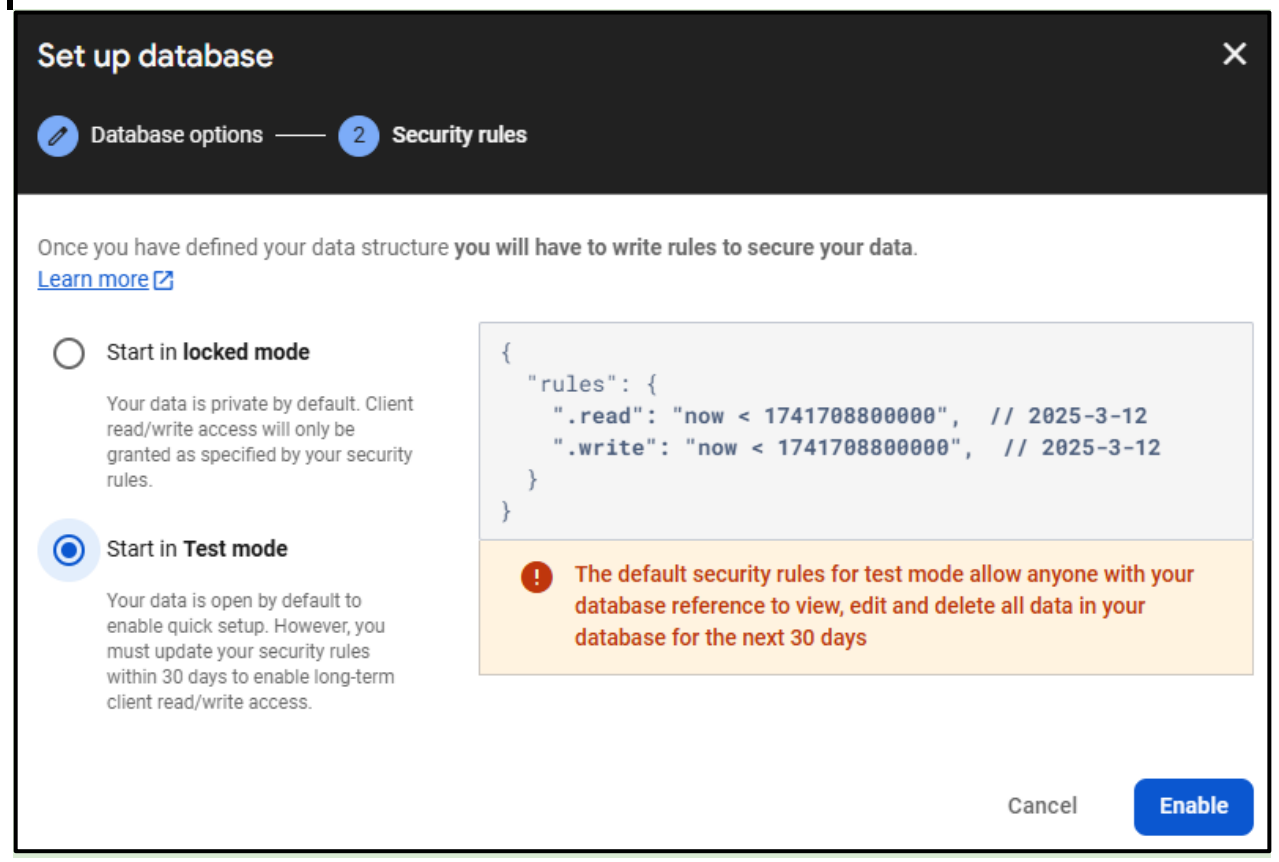

Choose Start in Test Mode

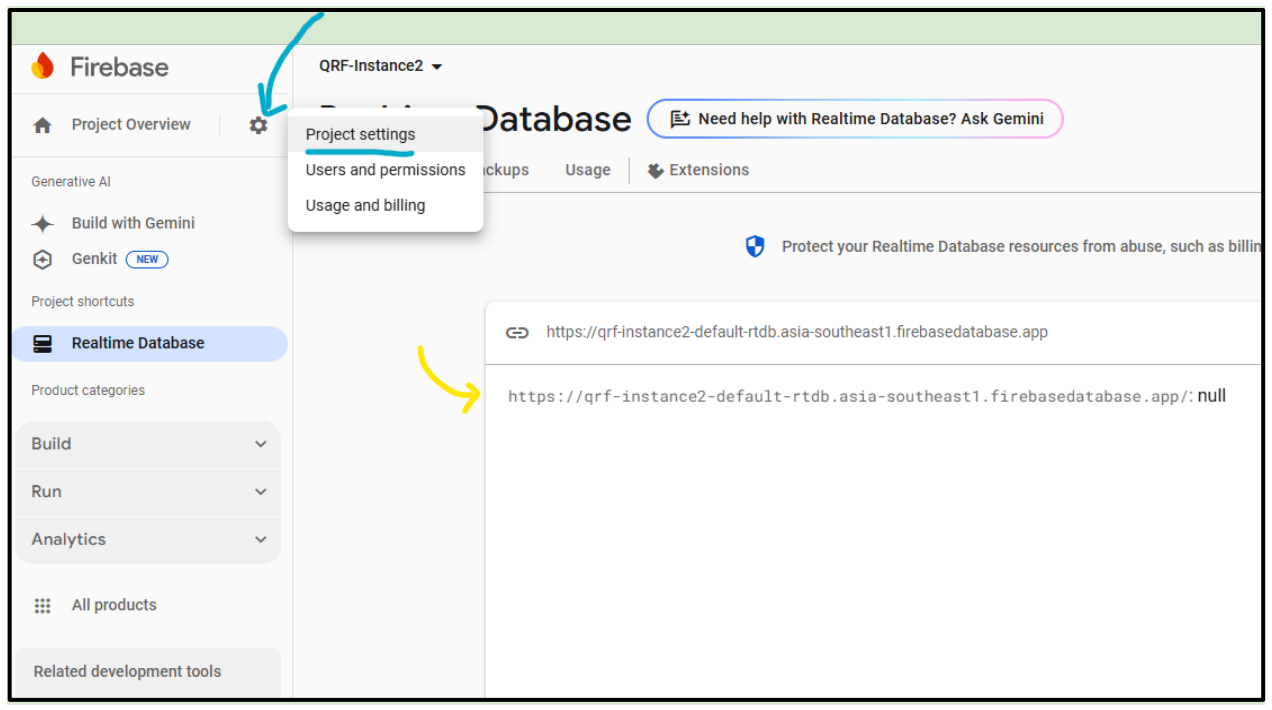

Once you can see a link like this: [https://qrf-instance2-default-rtdb.asia-southeast1.firebasedatabase.app/](https://qrf-instance2-default-rtdb.asia-southeast1.firebasedatabase.app/)
You have successfully created a Firebase Realtime Database (See yellow Arrow)

The next step is to find the API key for your Realtime database. Click the gear icon beside the Project Overview Menu and choose Project Settings. (See blue arrow.)

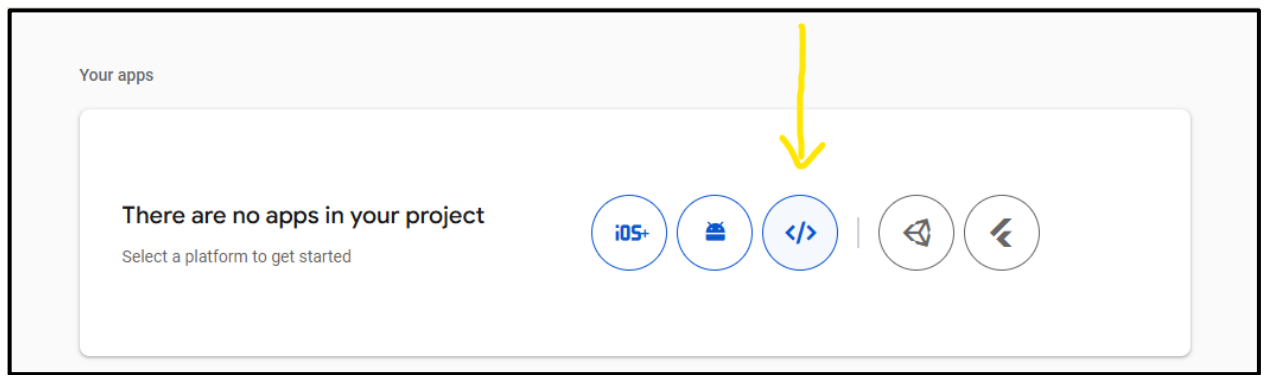

Under the “Your Apps” section click on the Web API’s button.

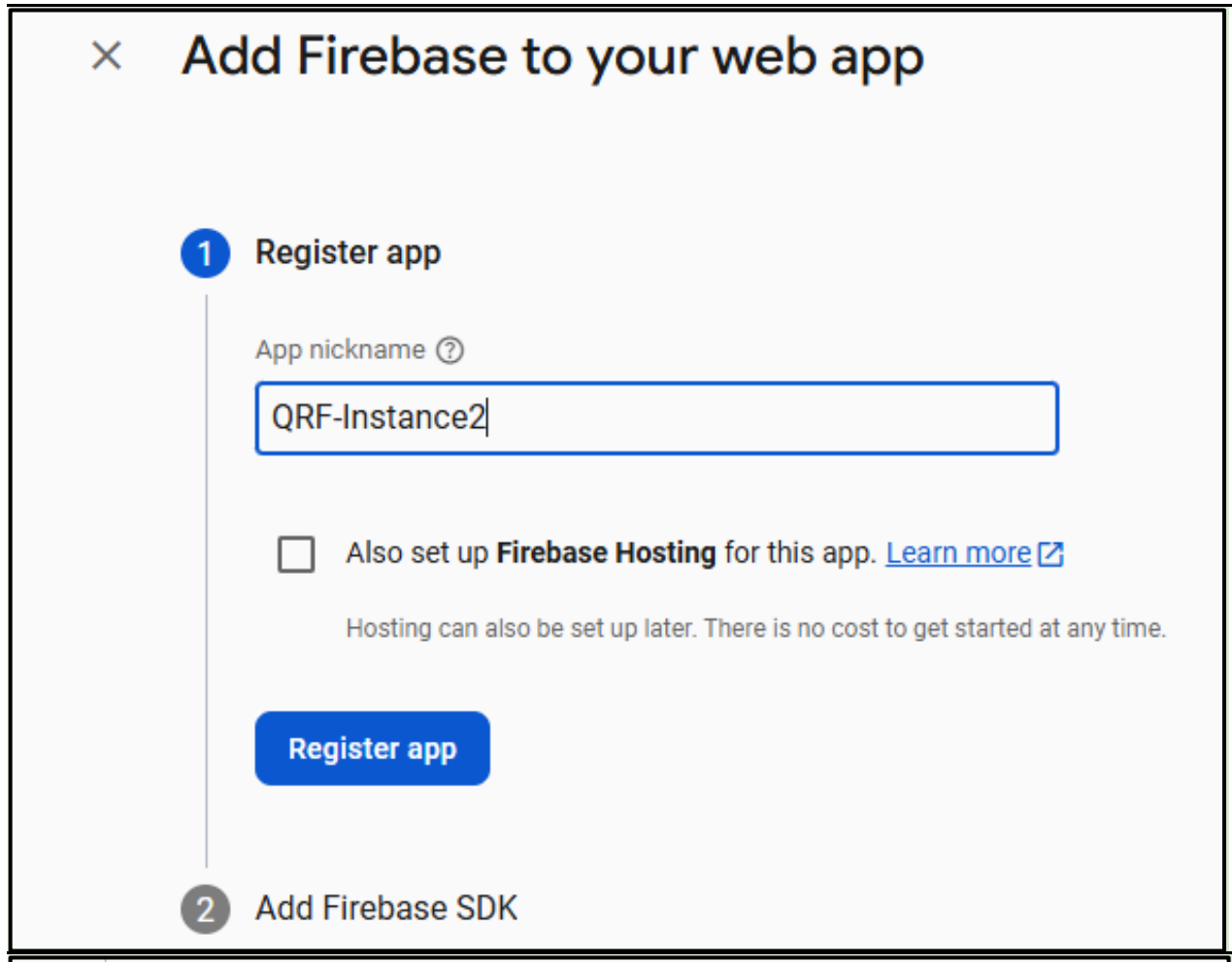

Register your app with a unique identifier. Firebase hosting is not required. Click on Register app.

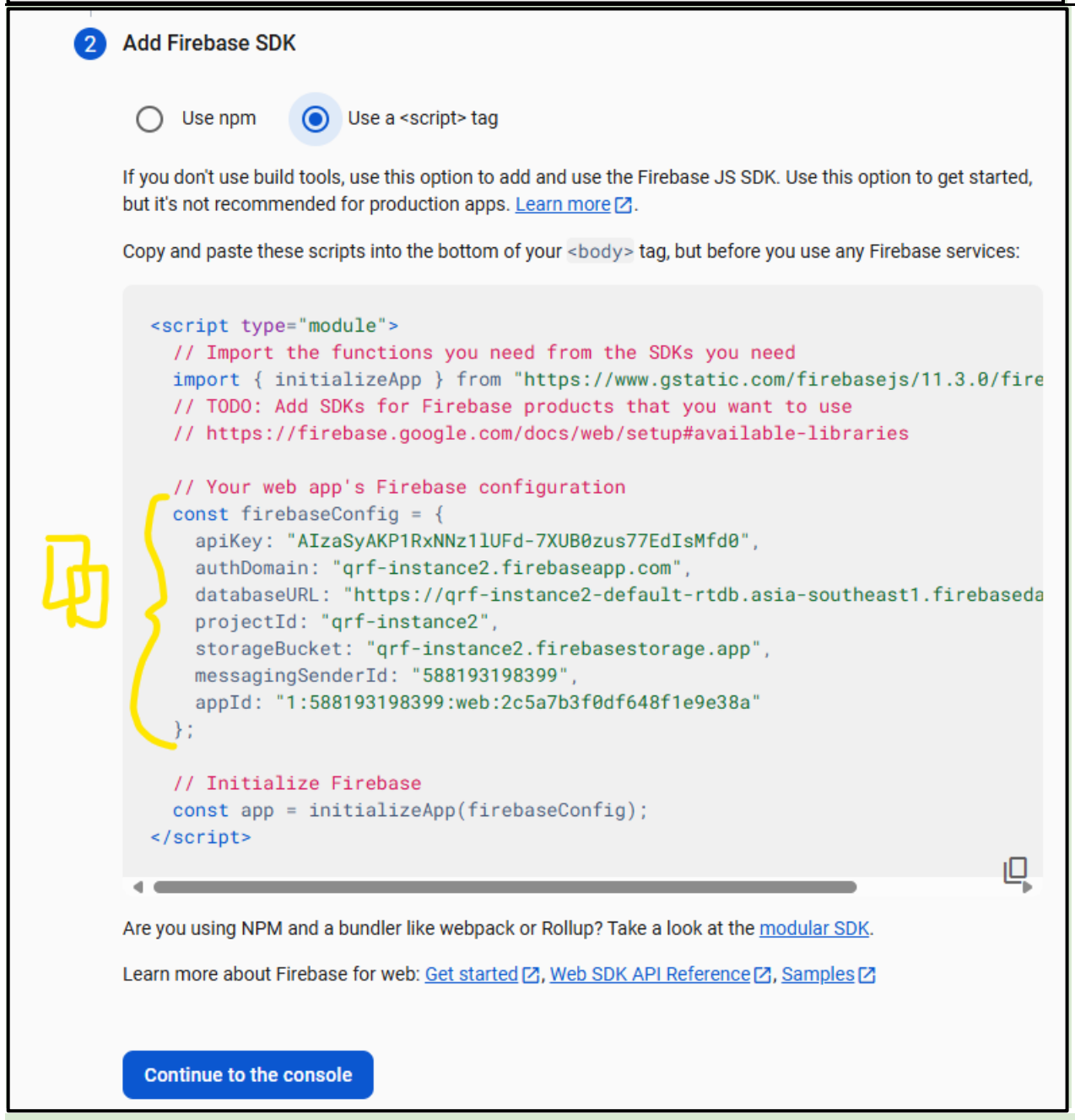

Choose “Use a <script> tag” and copy the const Firebase config codeblock. Keep this code block in a secure place and make sure you have a copy. Click continue to the console.

**YOU ARE NOW DONE WITH STEP 1**

# 2 Create a Personal Channel within the QRF Secure WebSocket Link

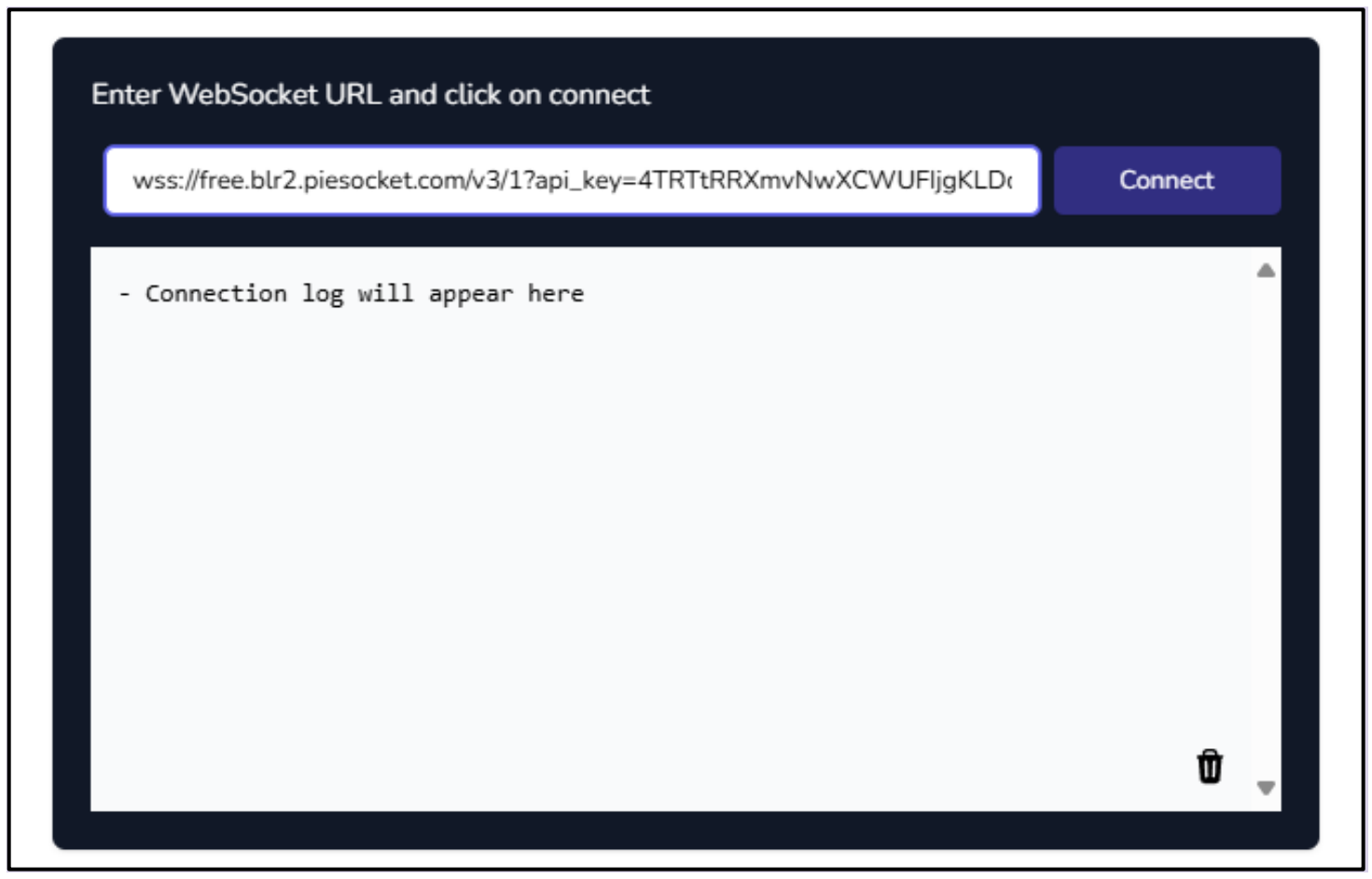

Go to this link official Websocket Tester link for the main QRF Websocket:

[https://piehost.com/websocket-tester?url=wss%3A%2F%2Ffree.blr2.piesocket.com%2Fv3%2F1%3Fapi_key%3D4TRTtRRXmvNwXCWUFIjgKLDdZJ0zwoKpzn5ydd7Y%26notify_self%3D1](https://piehost.com/websocket-tester?url=wss%3A%2F%2Ffree.blr2.piesocket.com%2Fv3%2F1%3Fapi_key%3D4TRTtRRXmvNwXCWUFIjgKLDdZJ0zwoKpzn5ydd7Y%26notify_self%3D1)

You should be able to see an interface similar to what's shown in the screenshot

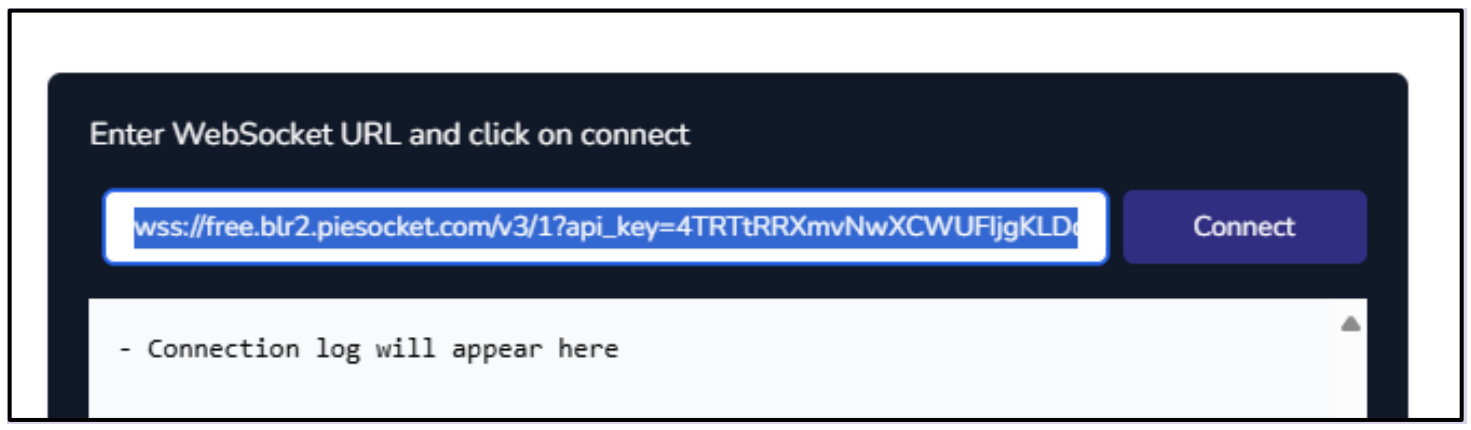

Copy the following WebSocket link into the input text box:

wss://free.blr2.piesocket.com/v3/1?api_key=4TRTtRRXmvNwXCWUFIjgKLDdZJ0zwoKpzn5ydd7Y&notify_self=1

Replace the number before the question mark (1) with a unique WebSocket identifier of your choice. This identifier will be used for all your QRF data collection activities.

**Example:**

wss://free.blr2.piesocket.com/v3/StarScience2025?api_key=4TRTtRRXmvNwXCWUFIjgKLDdZJ0zwoKpzn5ydd7Y&notify_self=1

Click Connect. If no error appears, your chosen WebSocket identifier is available.

**Important:** Take note of this WebSocket link—this will be your official instance identifier and will be required in later configuration steps.

YOU ARE NOW DONE WITH STEP 2

## Step 3: Remix the QRF Dispatcher Code on Replit and Configure the Dispatcher to use your Personal WebSocket

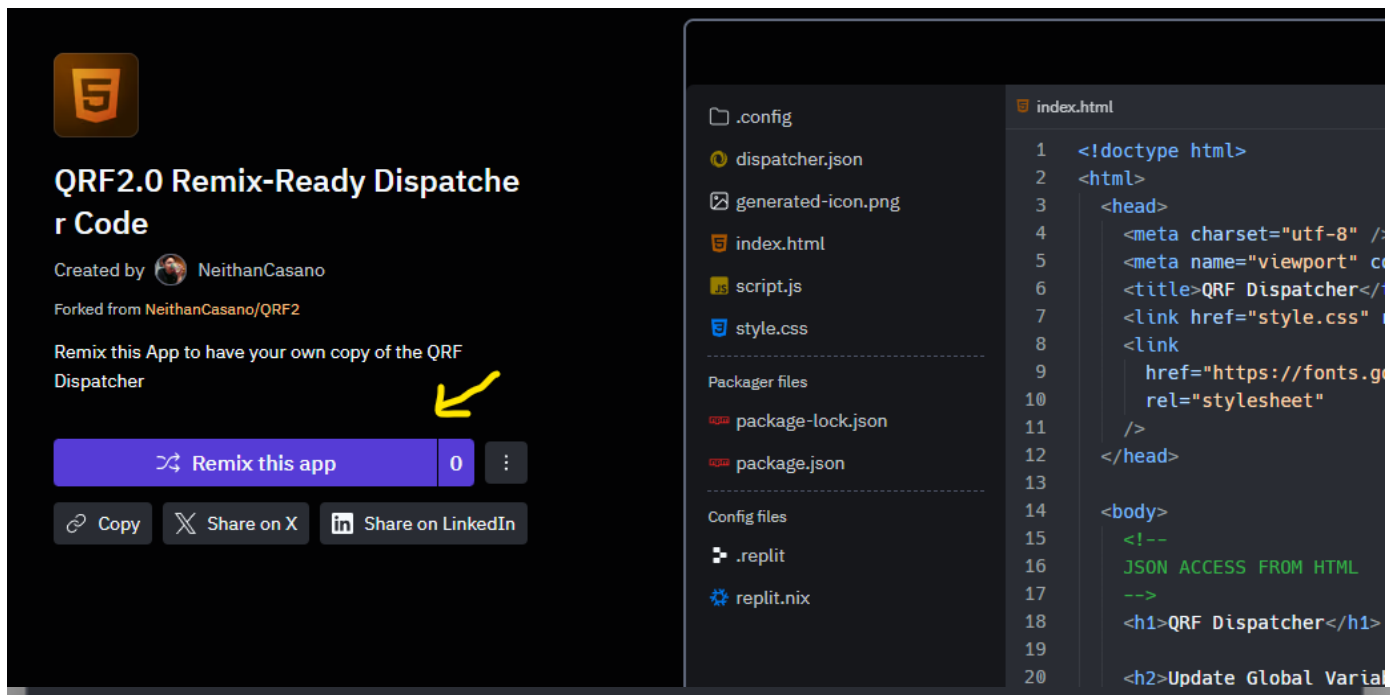

Create an account on Replit and Remix this project using the Remix this app button:

https://replit.com/@NeithanCasano/QRF20-Remix-Ready-Dispatcher-Code?v=1#index.html

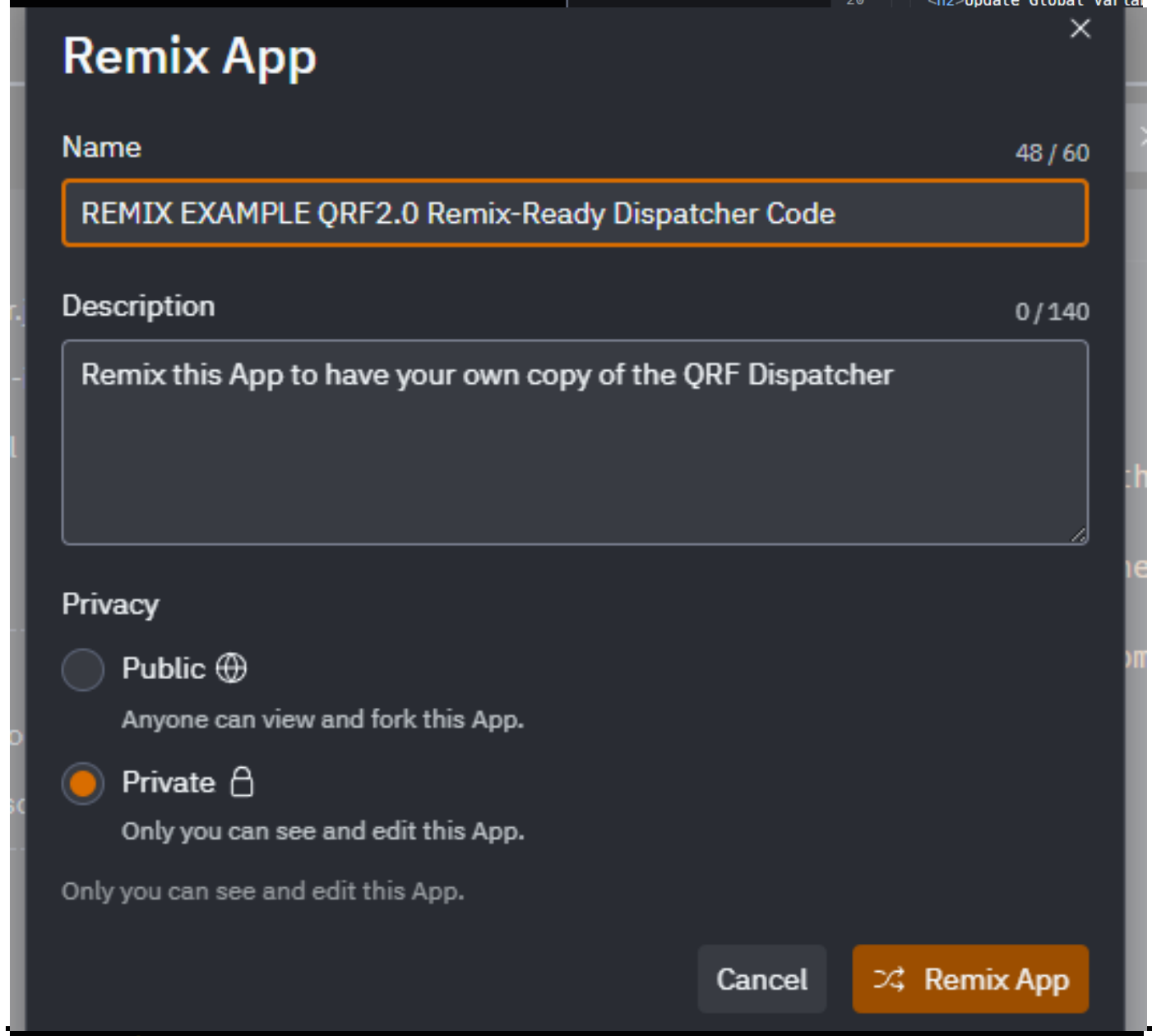

Give your Remix a unique name and set the privacy to Private so you are the only one who can access your copy and make sure the API keys included in it are secure. Choose Remix App.

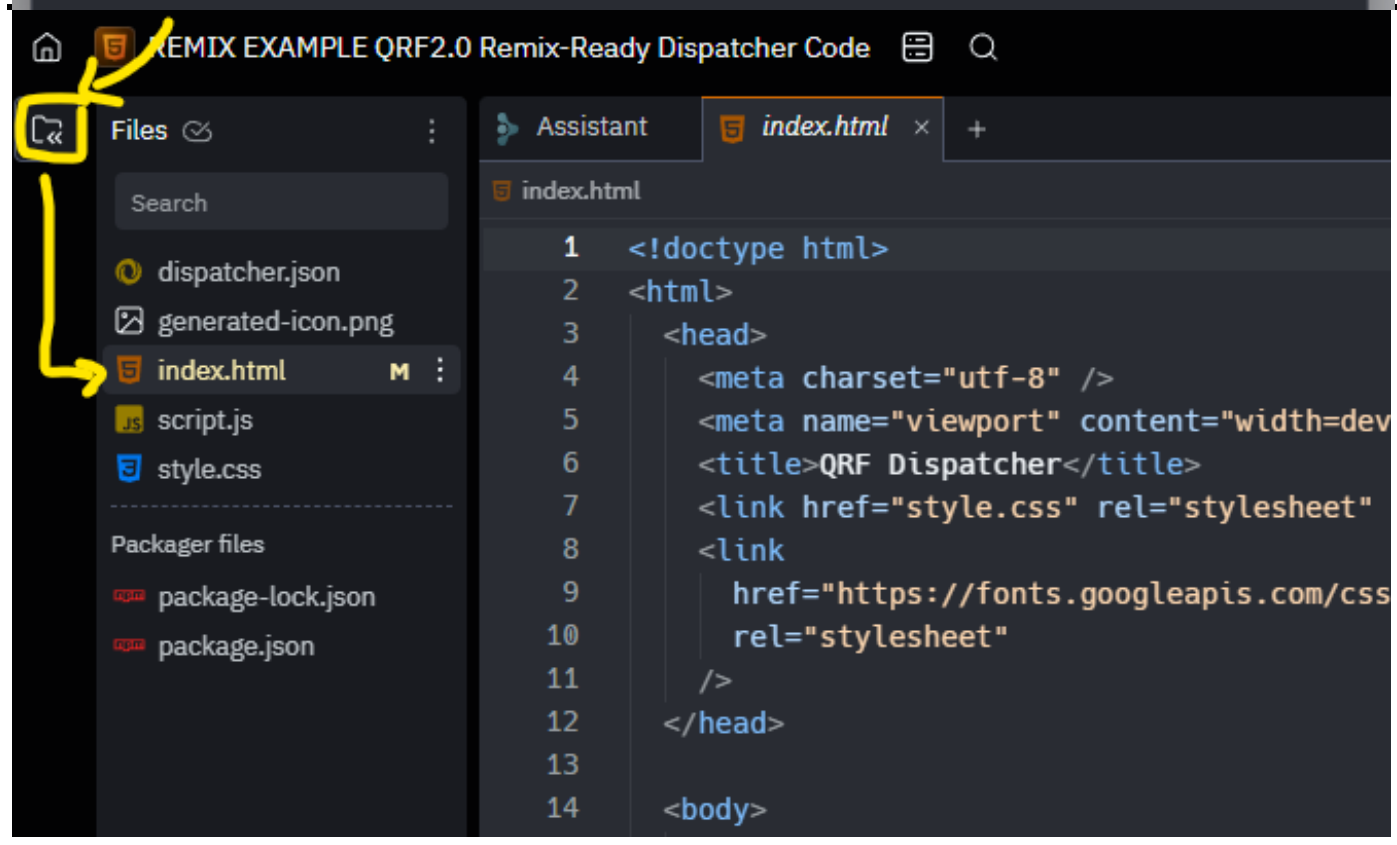

If not yet shown by default, click the folder icon to display the files in the remixed Dispatcher code. Navigate to index.html

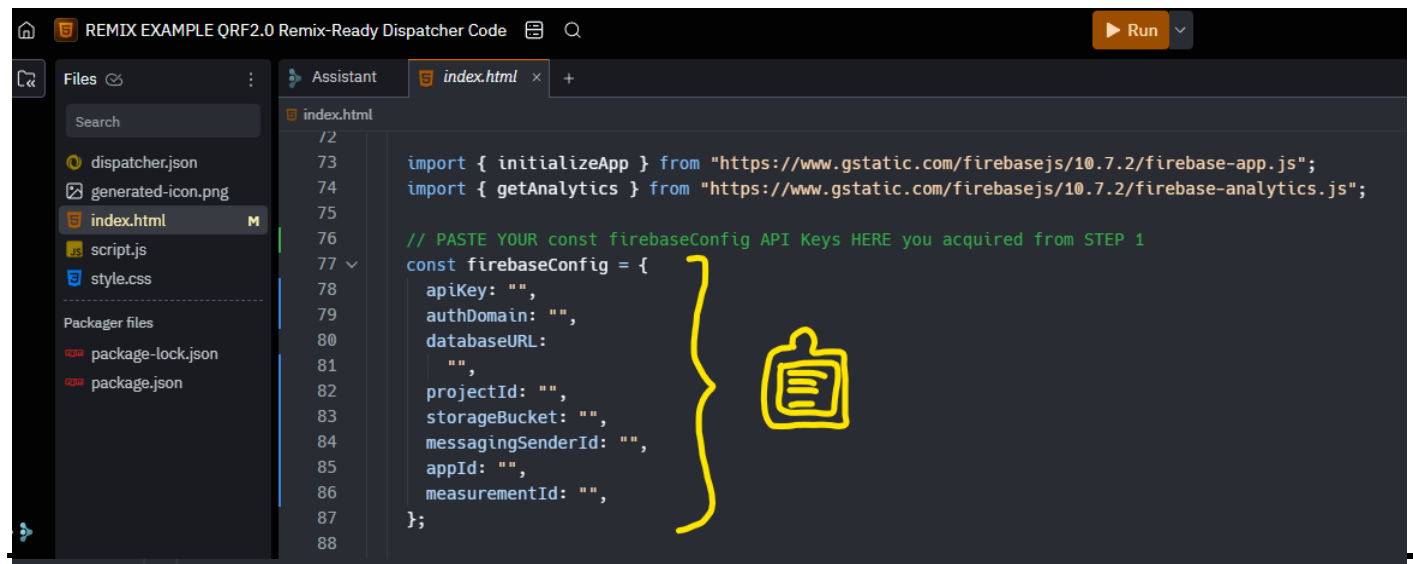

In line 77 of the Dispatcher's index.html file paste the const firebaseConfig code block you acquired from STEP 1.

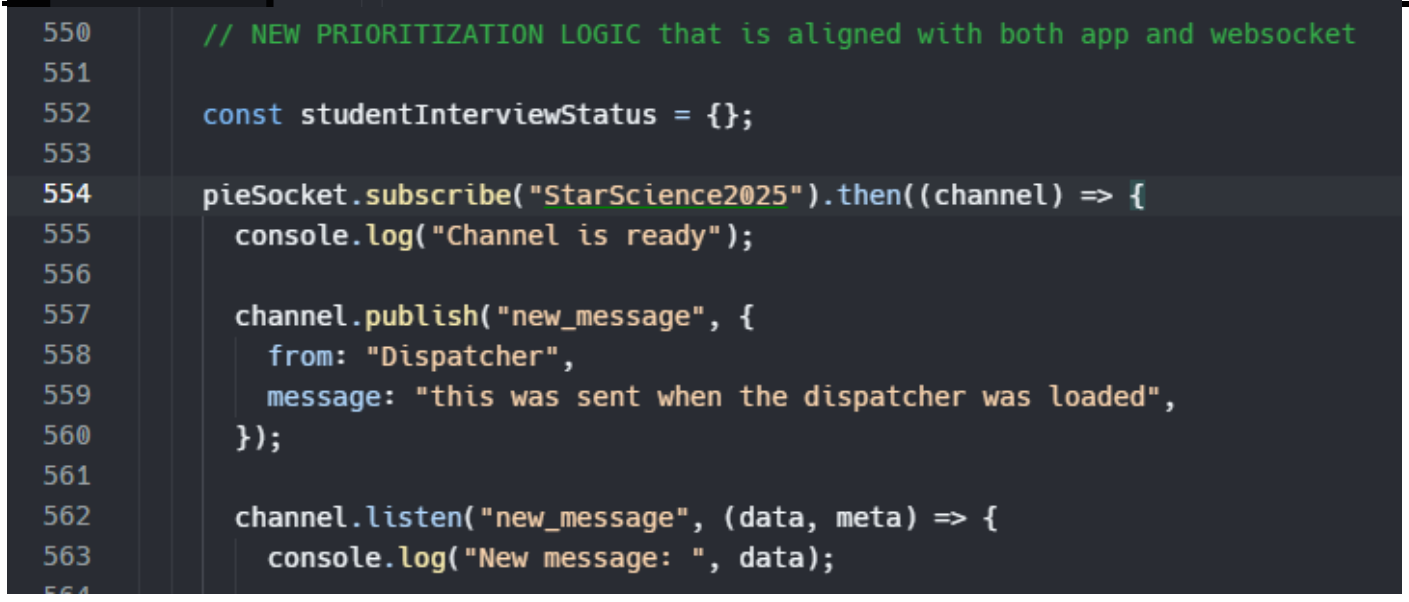

In line 554 of the Dispatcher's index.html file, indicate your websocket unique channel identifier in the pieSocket.subscribe function. In this example we used the StarScience2025 unique name we have also secured in STEP 2

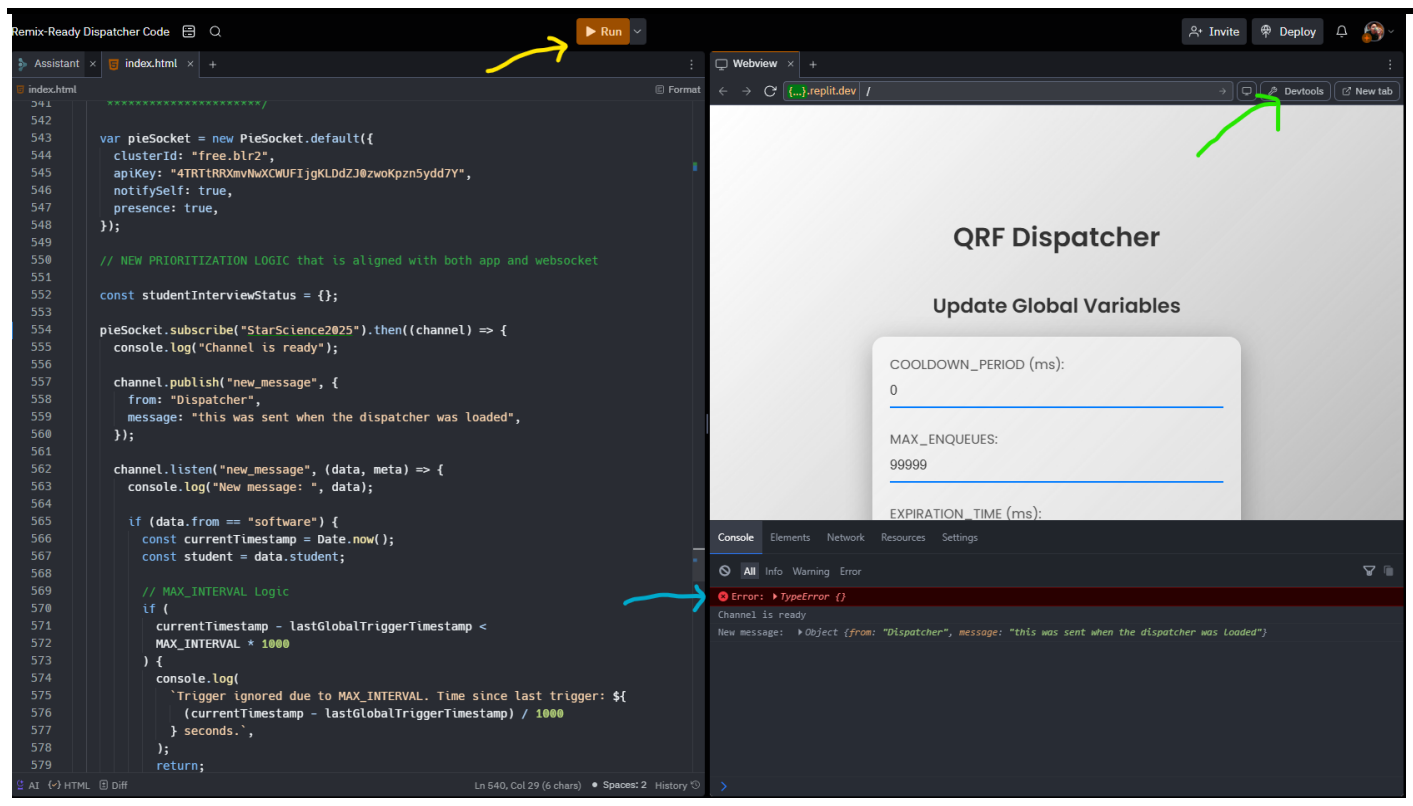

Click the Run button at the top to commit the changes (yellow arrow). You should see an interface similar to the one shown at the right.

To check if the websocket connection went through, click on the Dev Tools button (green arrow).

You should be able to see an error Error: TypeError {} followed by the message:

Channel is ready
New message:
Object {from: "Dispatcher", message: "this was sent when the dispatcher was loaded"}

This means the Dispatcher was loaded correctly and working as expected.

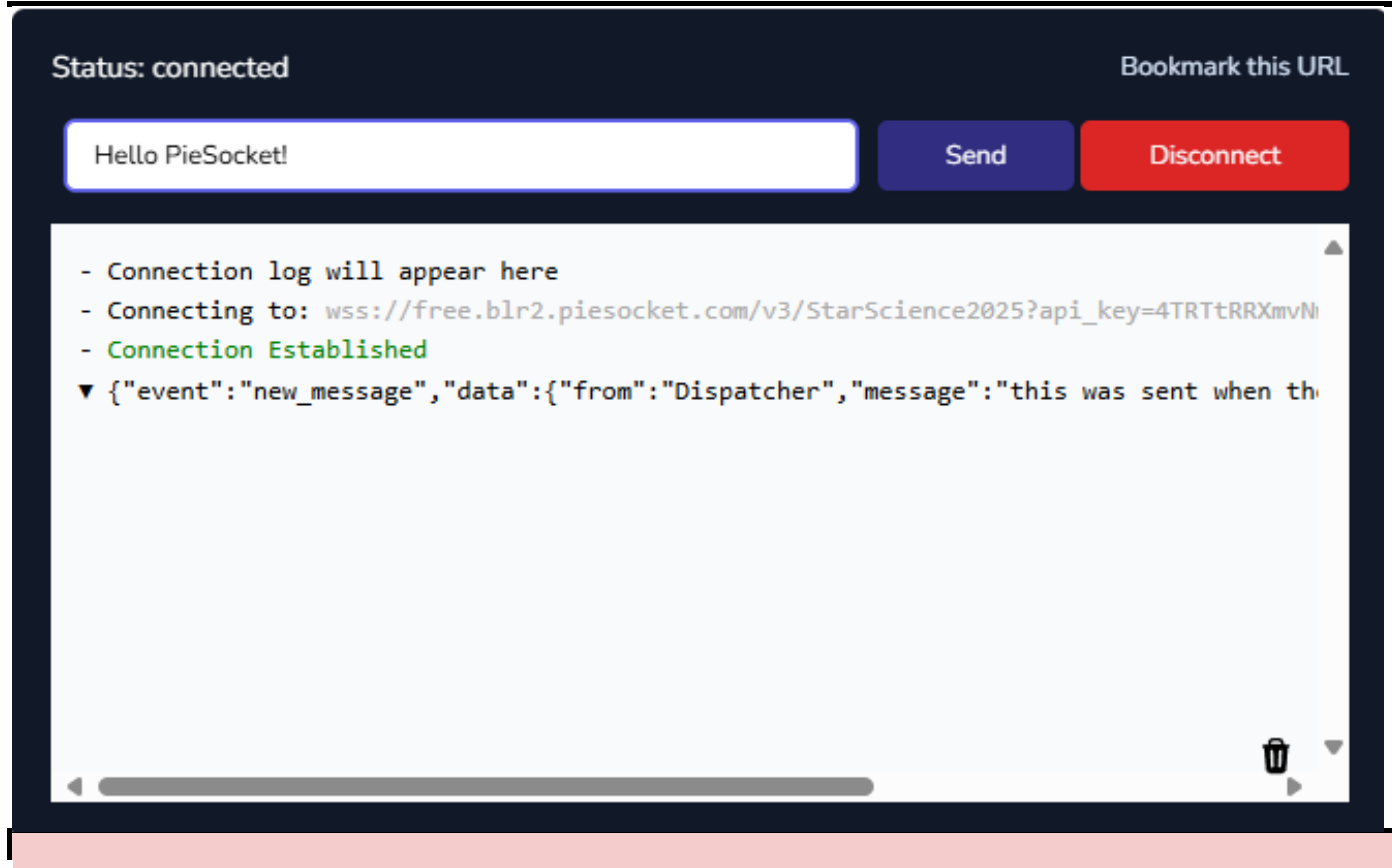

To double check if the two-way connection between websocket and dispatcher is working go back to the websocket tester link and check if a corresponding confirmation message was received (It is assumed you have already connected to your channel identifier at step 2).

If you see the message, this should mean the connections were established correctly.

YOU ARE NOW DONE WITH STEP 3

# 4 Configure the QRF Android App to use your Personal Websocket Channel and Server Storage Path

## USE YOUR GITHUB TO BEGIN THE CUSTOMIZATION PROCESS

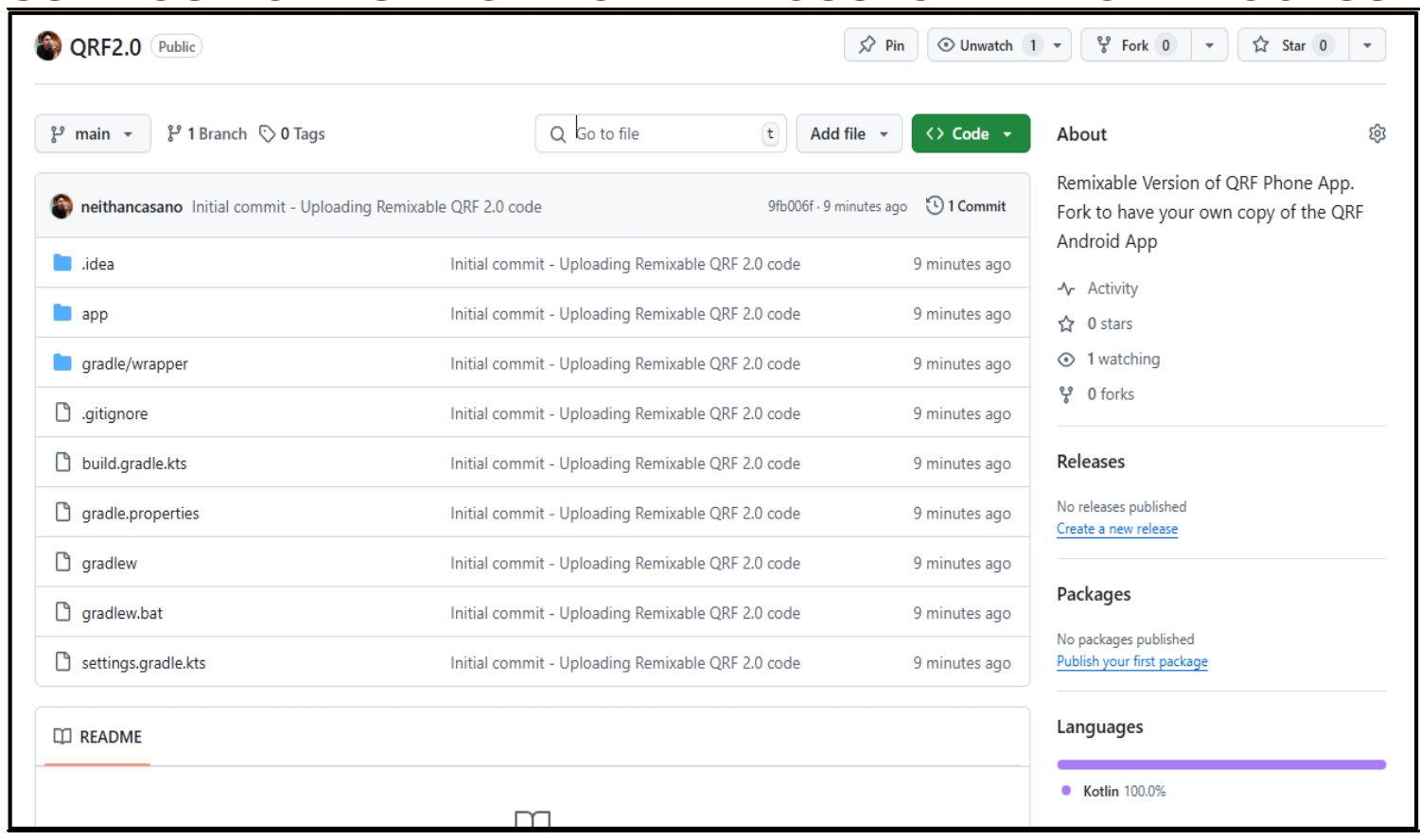

For this next step you need to have a Github account. Login to your Github and fork the QRF2.0 Project:

https://github.com/neithancasano/QRF2.0

## CONVERT THE EXISTING QRF CODE TO ADD YOUR OWN WEBSOCKET CHANNEL

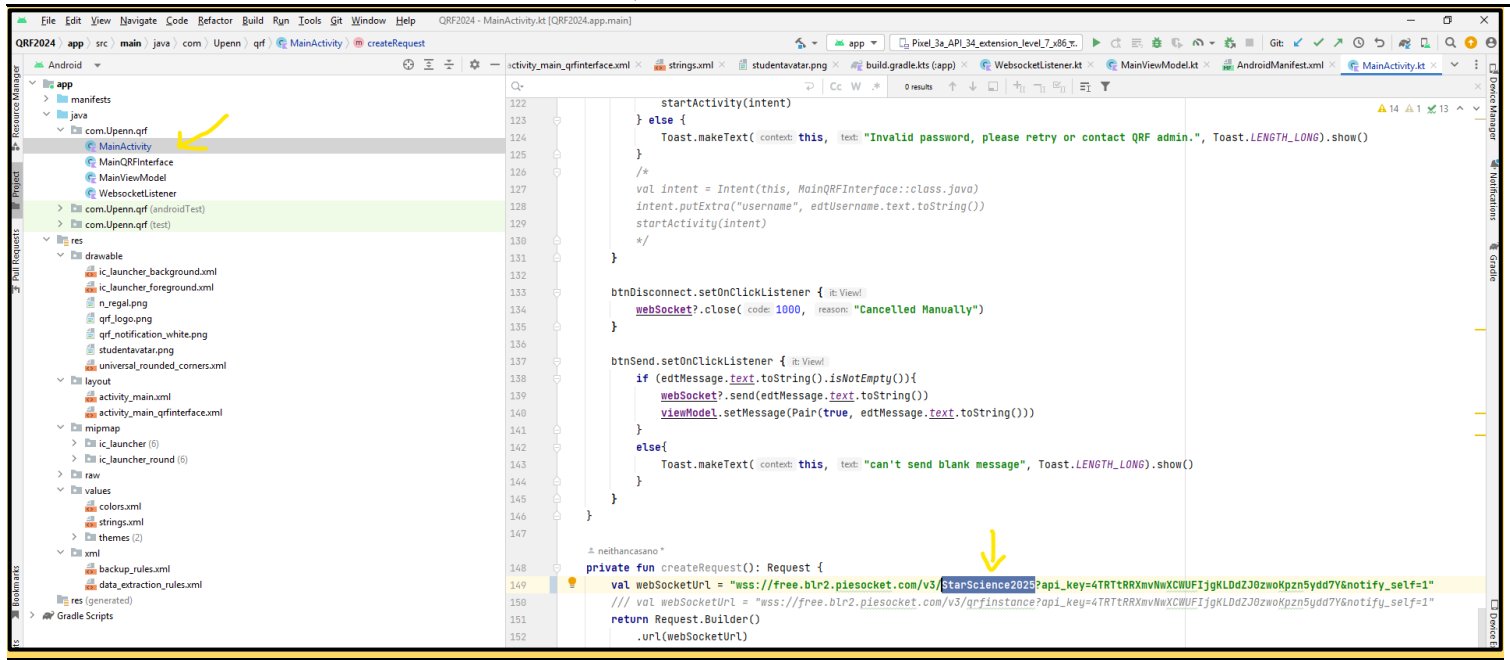

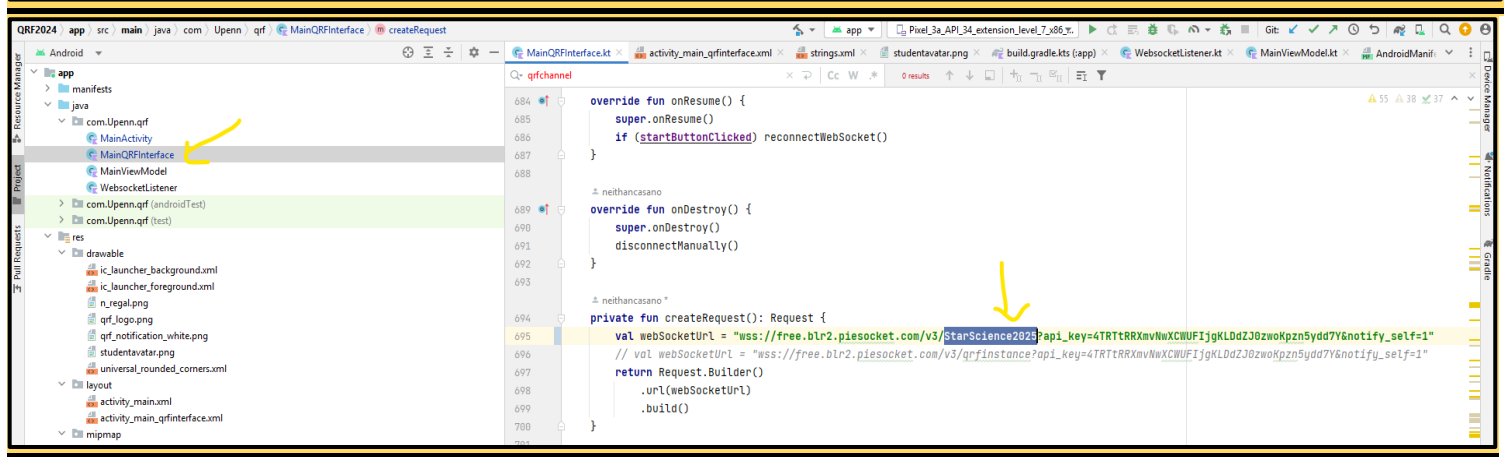

Once downloaded open the project in Android Studio and edit the following files to use your Personal Websocket Channel Identifier.

1. MainActivity line 149
2. MainQRFInterface line 695

## IF YOU WISH TO SAVE AUDIO FILES TO A SERVER

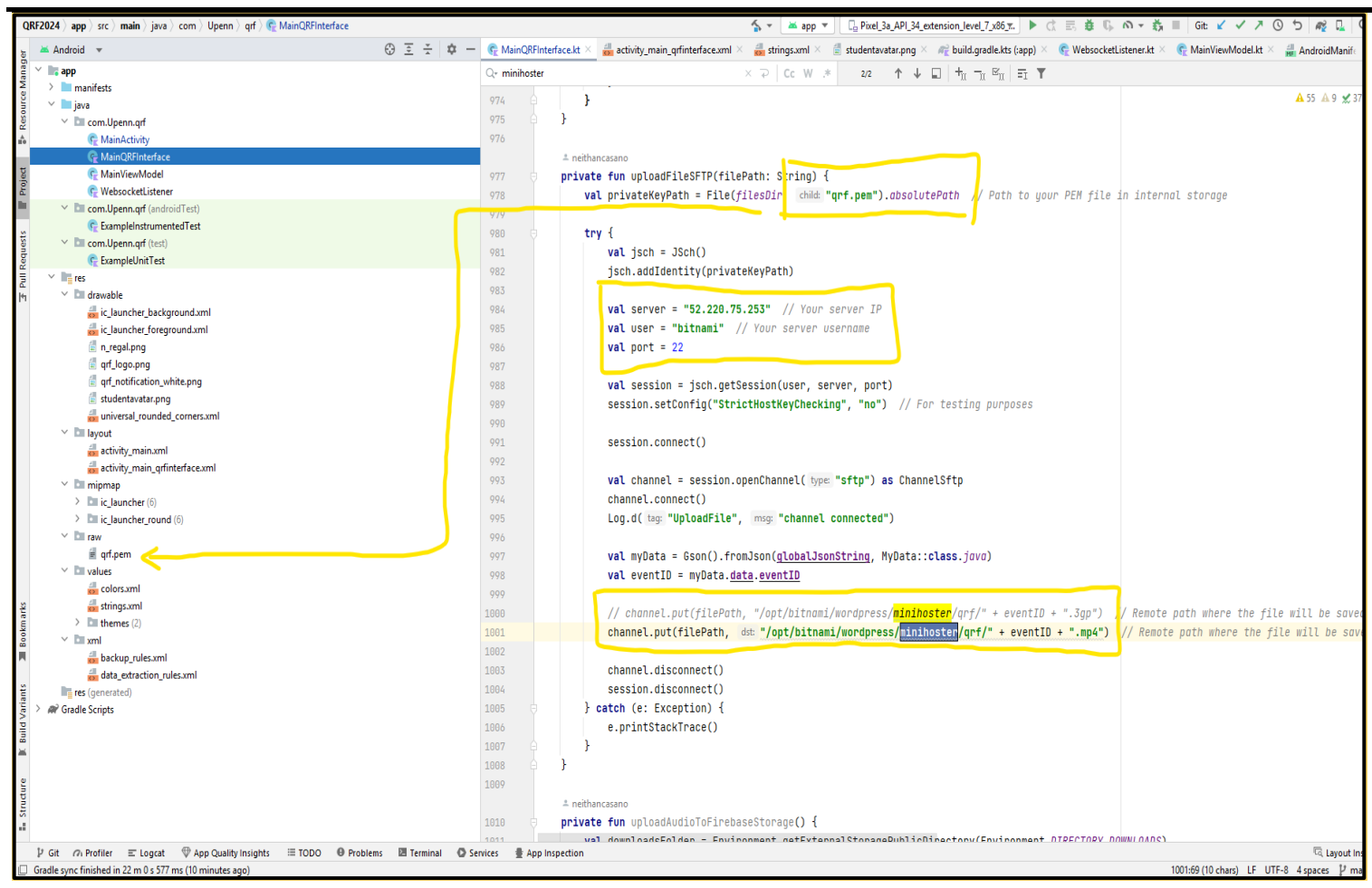

In MainQRFInterface, also edit the following lines to contain the relevant server information where the audio files will be kept (This is the same server storage the Dashboard will work with in Step 5)

(1) MainQRFInterface line 978 (indicate your server's .pem authenticator file, contact your server manager, all servers have this)

(2) MainQRFInterface lines 984 to 986, to contain your server IP (DNS will not work, it must be the raw IP), master user (or user with admin privileges chown 755),

(3) MainQRFInterface lines 1001, to contain the path where the QRF audio files will be saved.

## ALTERNATIVELY, IF YOU WISH TO JUST SAVE AUDIO FILES TO THE PHONE

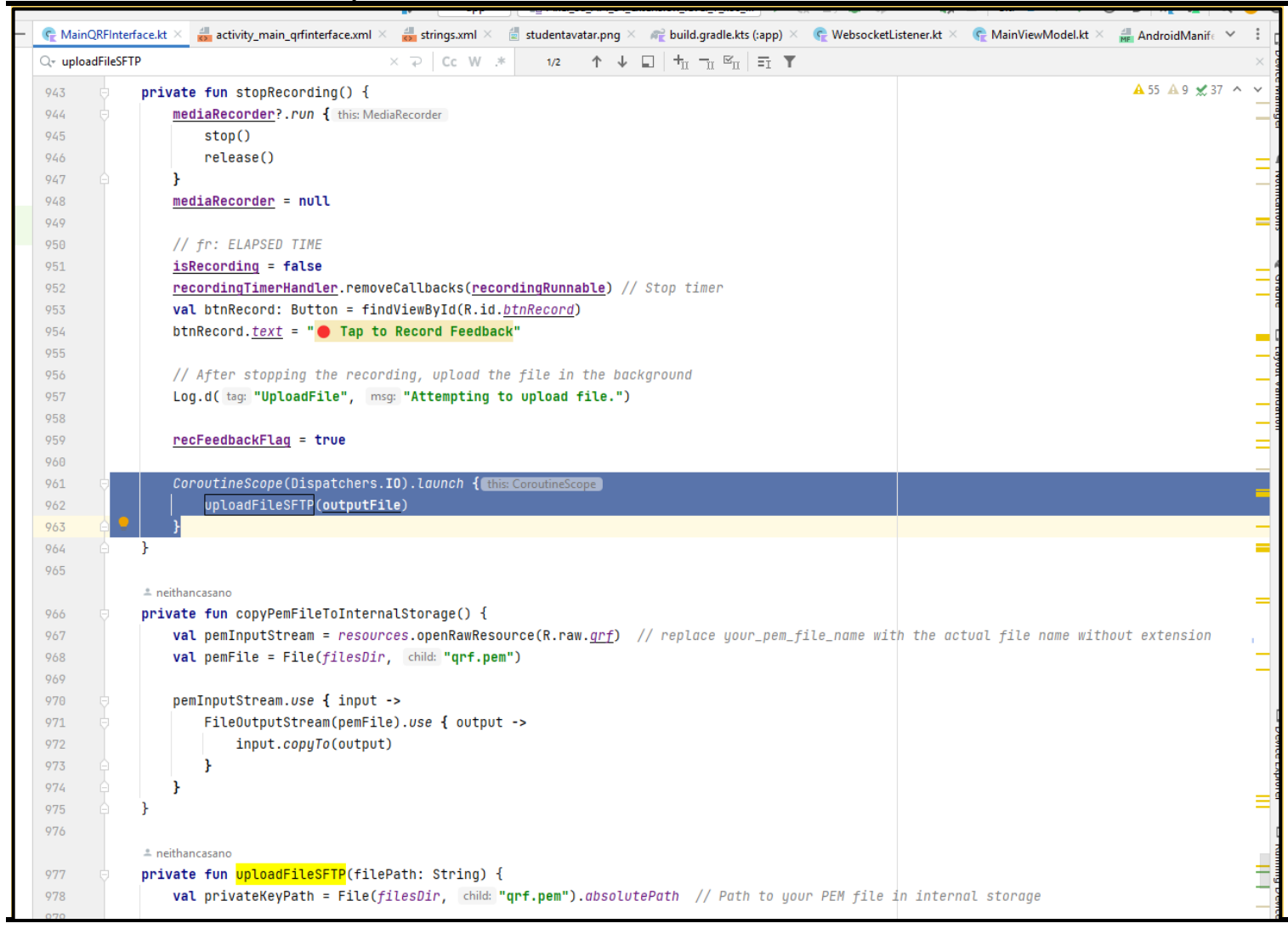

Comment out lines 961 to 963 in MainQRFInterface.kt to disable the feature that saves audio data to a specified server.

*Note that when saving to phone only, you don't need to configure a server for storage described in the previous step (you may ignore that step).

## DEPLOYING THE APP

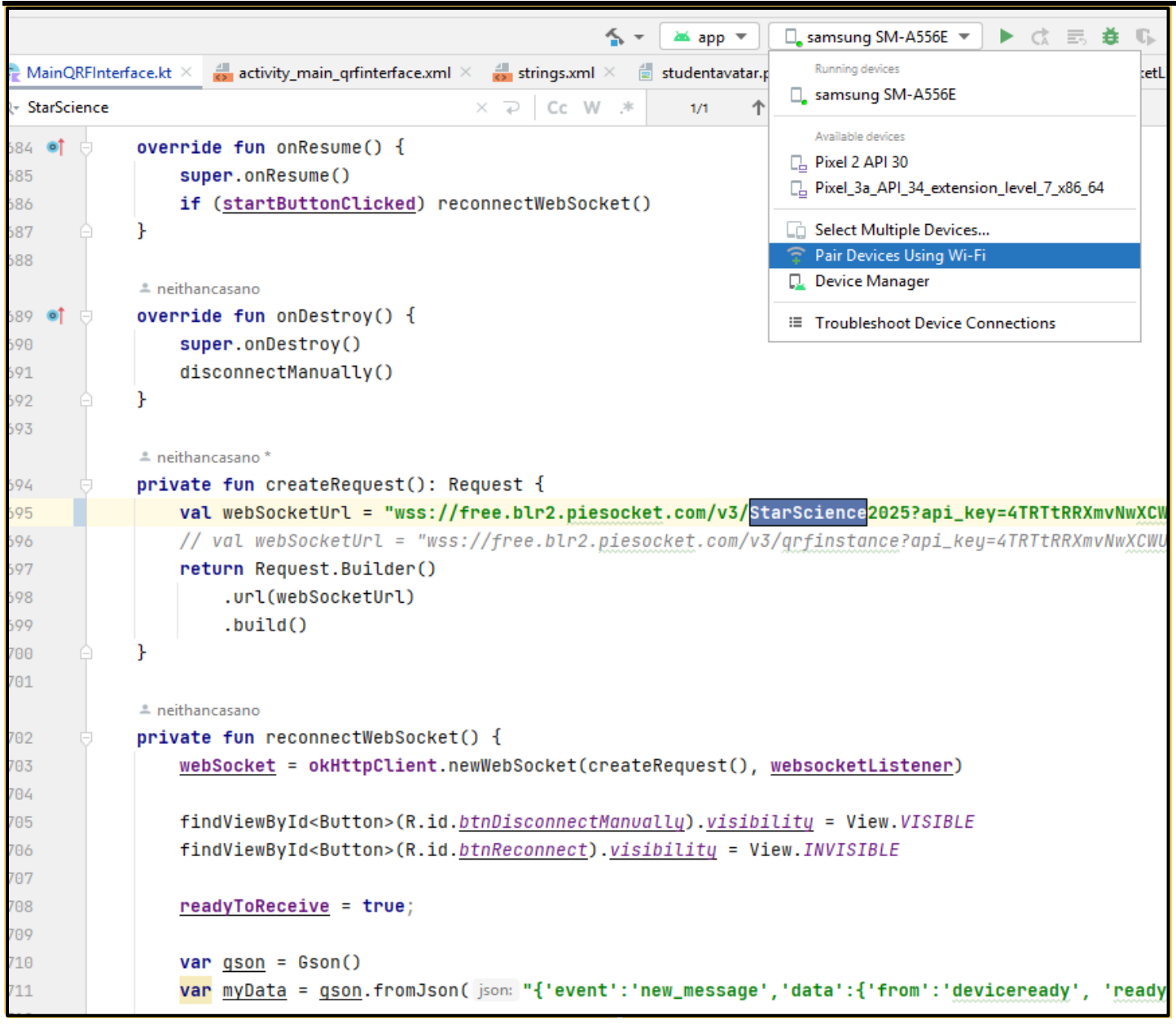

You can now deploy the app to your phone. Find the pair devices using Wifi Option to connect your phone to your computer.

Enable Developer mode in your Android Phone by tapping the Build number of your phone 7 times under Settings -> About Phone -> Software Information.

(1) In Developer Options, choose Wireless Debugging and tap Pair device with QR code.

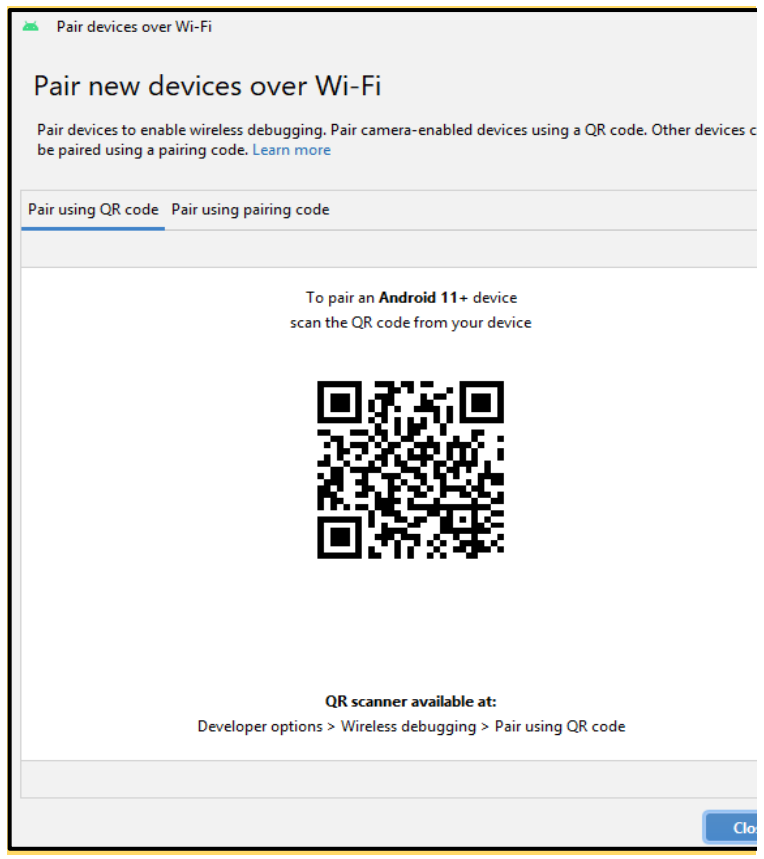

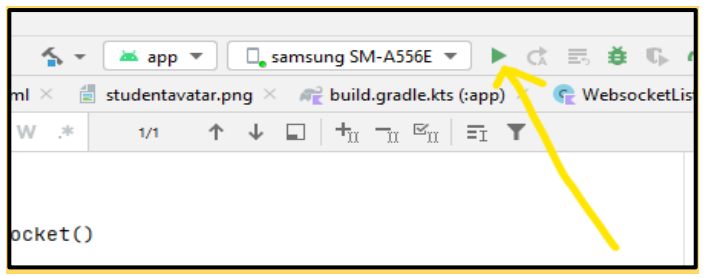

(2) Scan the QR Code shown on the computer screen served by Android Studio.

(3) Once connected, hit the play button beside the phone model to transfer the new QRF code to your phone.

**At this point the Dispatcher, the Android App, and the WebSocket are now connected.**

## TO TEST IF EVERYTHING IS WORKING, DO THE FOLLOWING.

(1)

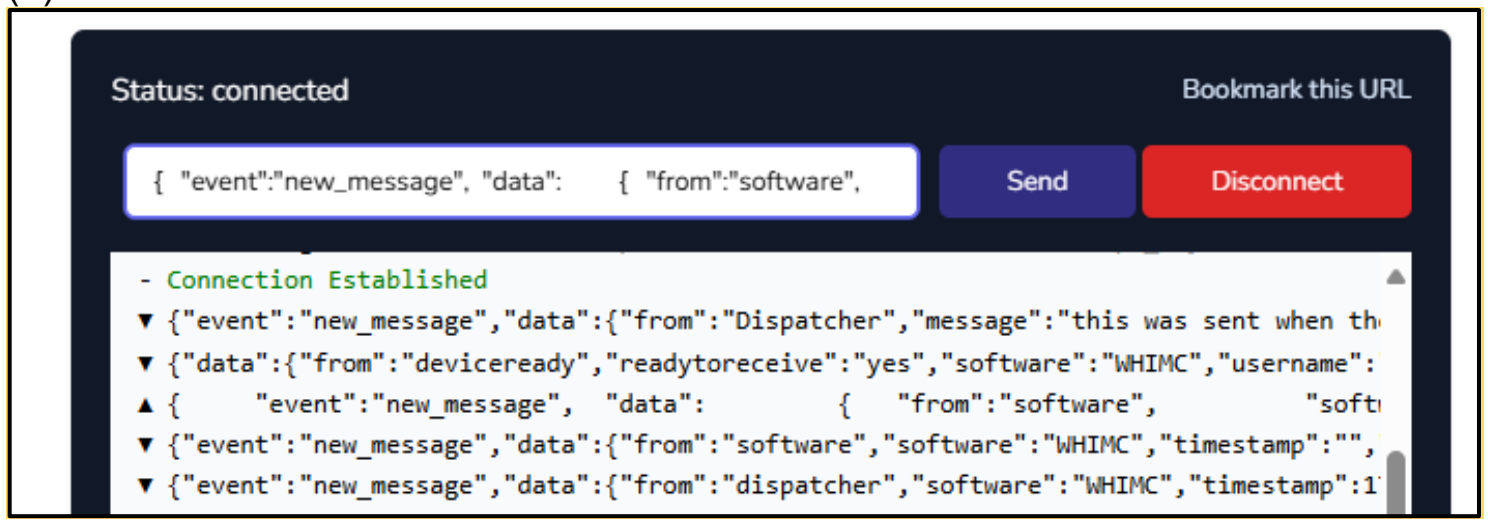

1. Reload your Websocket Tester: [https://piehost.com/websocket-tester?url=wss%3A%2F%2Ffree.blr2.piesocket.com%2Fv3%2F1%3Fapi_key%3D4TRTtRRXmvNwXCWUFIjgKLDdZJ0zwoKpzn5ydd7Y%26notify_self%3D1](https://piehost.com/websocket-tester?url=wss%3A%2F%2Ffree.blr2.piesocket.com%2Fv3%2F1%3Fapi_key%3D4TRTtRRXmvNwXCWUFIjgKLDdZJ0zwoKpzn5ydd7Y%26notify_self%3D1) and don't forget to indicate your Personal Websocket Channel in the link and click

(2)

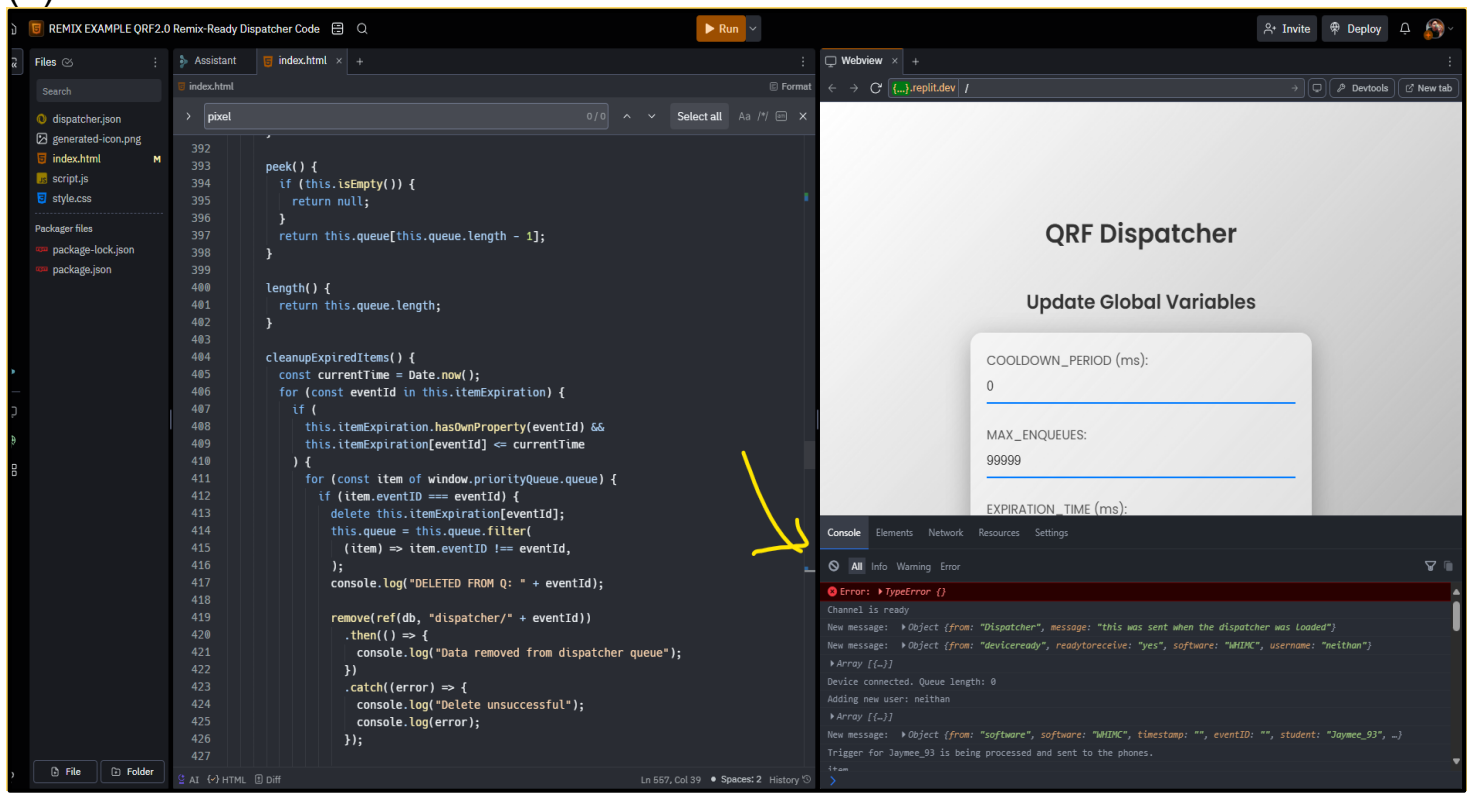

2. Re-run your dispatcher. It should display the same confirmation message that the dispatcher was successfully run and a callback should have been sent to the websocket tester.

(3 - 4)

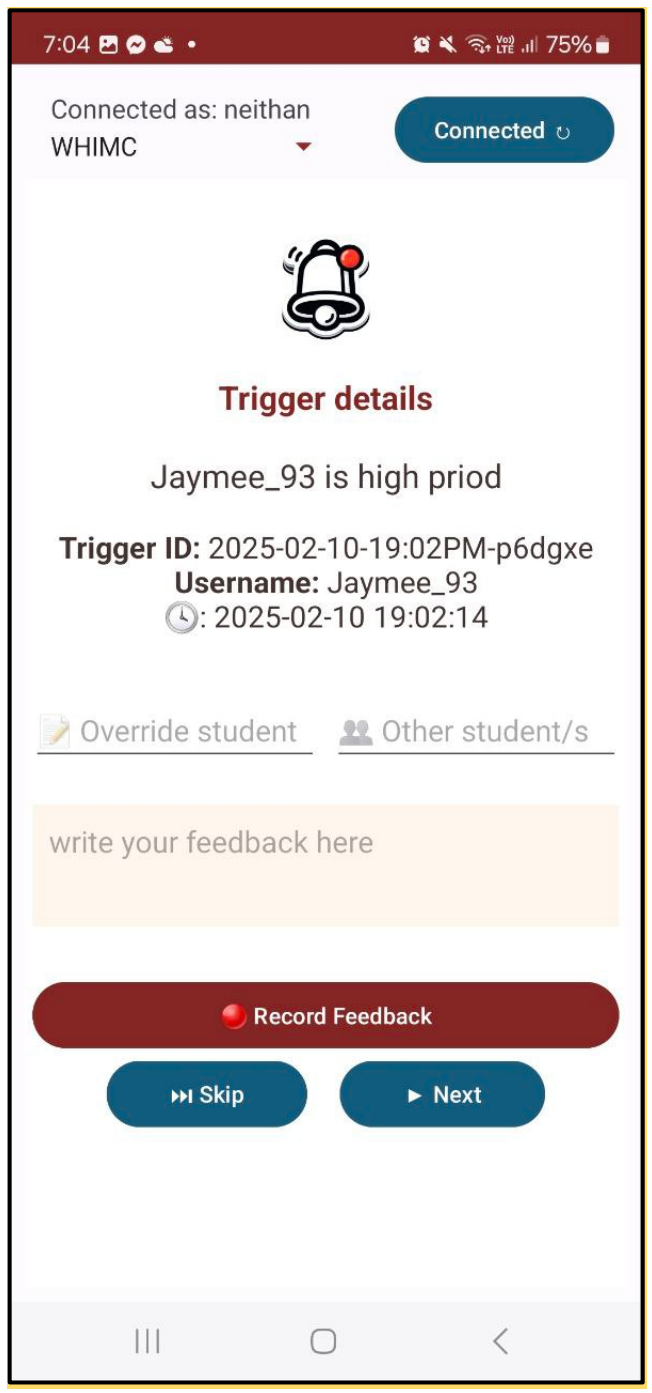

3. Login to your QRF Phone app via any username with the password "qrf2024". Once you login and tap on the "Ready" button check that a callback was sent to both the dispatcher and the websocket tester. If you can see those messages the whole system is now connected.
4. Try sending the json packet below (dark orange text) in the websocket tester (directly just copy it to the input text box and hit send). Your Phone app should be able to receive this immediately.

QRF JSON PACKET (paste this in the websocket tester to simulate a trigger)

```
{
    "event":"new_message",
    "data":
        {   "from":"software",
            "software":"WHIMC",
            "timestamp":"",
            "eventID":"",
            "student":"Jaymee_93",
            "trigger":"Jaymee_93 is high priod",
            "priority": 1,
            "masterlogs":
                {   "from":"software",
                    "software":"WHIMC",
                    "timestamp":"",
                    "eventID":"",
                    "student":"Jaymee_93",
                    "trigger":"Jaymee_93 is high prio",
                    "reviewer": "",
                    "end":"",
                    "feedbackTXT":"",
                    "feedbackREC":""
                }
        }
}
```

YOU ARE NOW DONE WITH STEP 4

# 5 Setup the QRF Dashboard in your Server

(1)

```
25  <script type="module">
26      import { initializeApp } from "https://www.gstatic.com/firebasejs/10.10.0/firebase-app.js";
27      import { getDatabase, ref, onValue, remove } from "https://www.gstatic.com/firebasejs/10.10.0/firebase-database.js";
28
29      const firebaseConfig = {
30          apiKey: "",
31          authDomain: "",
32          databaseURL: "",
33          projectId: "",
34          storageBucket: "",
35          messagingSenderId: "",
36          appId: "",
37          measurementId: ""
38      };
39
40      const app = initializeApp(firebaseConfig);
41      const db = getDatabase(app);
42
43      let showAllRecords = false;
44      let allEntries = [];
45
46      function fetchData() {
47          const dbRef = ref(db, 'masterlogs');
```

(2)

```
108
109             let output = `<center>
110
111                         <button onclick="downloadCSV()">${csvLabel}</button>
112                         <a href="https://www.pixelzero.solutions/minihoster/zipAndDownload.php" target="_blank">
113                             <button id="downloadAudioFiles">Download Audio Files (All)</button>
114                         </a>
115                         <button onclick="toggleRecords()">${seeAllLabel}</button>
116
117                     </center><br>
118                     <center><strong>${showAllRecords ? 'Displaying all records' : 'Displaying latest 50 records'}
                        </strong></center>
119                     <center>
120                         <input type="checkbox" id="filterTaken" onclick="applyFilter()" ${filterTakenElement &&
```

(3)

```
170                         <td>${feedbackTXTTrimmed === "" ? "No text feedback" : feedbackTXTTrimmed}</td>
171                         <td>${overrideStudent}</td>
172                         <td>${otherStudent}</td>
173                         <td>${!item.feedbackREC || item.feedbackREC === "skipped" || item.feedbackREC.trim() === "" |
                            item.feedbackREC.trim() === "No audio feedback" ? "No audio feedback" : `<audio style="width:
                            "audio_${item.key}" controls><source src=
                            "https://www.pixelzero.solutions/minihoster/qrf/${item.feedbackREC}" type="audio/mp4"></audio
174                         <td id="duration_${item.key}">N/A</td>
175                         <td id="start_${item.key}">N/A</td>
176                         <td id="end_${item.key}">${formattedEnd}</td>
177                         <td><button onclick="deleteRecord('${item.key}')">Delete</button></td>
178                     </tr>`;
179         });
180         output += `</table></div>`;
181
182         if (!showAllRecords && filteredEntries.length > 50) {
183             let label;
184             if (filterTakenElement && filterTakenElement.checked && filterUntakenElement && filterUntakenElement.checked) {
185                 label = "See All";
186             } else if (filterTakenElement && filterTakenElement.checked) {
187                 label = "See All Taken Triggers";
```

**Download the QRF Dashboard Code from this link:**

https://drive.google.com/file/d/1ee6jMRatRUvR-lpkvV5qJ51sCUxBkVG0/view?usp=sharing

Revise the following:

1. Line 29, use your const firebaseConfig code block
2. Line 112, use your own zipAndDownload.php webhook
3. Line 173, use your own path for the saved audio files

Deploy this code to your server, and it will host the QRF Dashboard. The dashboard enables real-time monitoring of triggers and provides options to download audio files and trigger logs in .csv format.

YOU ARE NOW DONE WITH STEP 5

# Appendix 3: Guidelines for DDCI Practices

**Interviewing techniques should be tailored to the learner population, using the following approach:**

**Guideline 1: Tell the students about the process before you conduct any interviews.** You should introduce yourself to the students and make it clear that they are not going to be in trouble for anything they tell you about their learning experience. Your goal is to understand how they are learning, and anything they tell you will be used to help other kids learn better.

**Guideline 2: Take a helpful but non-authoritative approach as an interviewer.** The essence of this guideline starts with using open-ended questions that that are academically neutral, but more specifically, we recommend that interviewers adopt a role that we are calling the **Big Sister Approach (BSA).** In the BSA, the interviewer is friendly and kind, but they also actively avoid taking on the role of a tutor, teacher, or other authority figure. This approach is important because students expect information and support from people with power and authority, and providing this information—while potentially helpful—does not help you understand why they are stuck. Your goal is to get them to explain their experience, and this cannot happen if you are answering their questions. Some students may also be less likely to talk to you—especially in an honest fashion—if they are afraid they will get in trouble for something they're doing. It is also important because as students get older, self-presentation effects will start to become more prominent. Students who think they are being judged will not provide you with as much information about how and why they are struggling.

**Guideline 3: Obtain assent for each interview.** This does not need to be a formal procedure, but an interviewer should ask each student if they are okay talking with right now. It is normal for students to get frustrated during learning, and finding out why is often an important part of educational research. However, if the student is *overly* upset, send a teacher to check on them, but do **not** proceed with an interview at that time. You are a guest in this classroom, and the students' autonomy and privacy should always be treated with respect.

**Guideline 4: Students may only provide information if you ask about. (Prepare for priming and salience effects).** As much as you want to let students lead these conversations, some students will not have the metacognitive skills to bring up certain parts of their experience. A child who is distressed about their current performance, for example, may struggle to tell you about past successes in a positive fashion. If you need to know how their current situation contrasts with those, you should work to prompt them with those specific questions as the interview (or the series of interviews) goes on.

**Guideline 5:** Individual questions can ask about students' struggle, but the DDCI protocol should take an asset-based approach. In line with the BSA, your goal is to identify what would improve students learning, and in line with research about good pedagogy (Ladson-Billings, 1995), that should include finding out what their strengths are.

# Appendix 4: Daily Research Concerns

**At the start of day one, the interviewers should:**

1. **Introduce yourself:** preferably as someone who is external to the software development team and not heavily invested in this version of the software being perfect as is. Feel free to say things like, "You can be completely honest with me about what you like and don't like in the game because I didn't design it so you can't offend me!" (see Big Sister Approach).
2. **Introduce your research project:** explain why you are conducting interviews in age-appropriate terms. This is important because students cannot assent if they do not have an age-appropriate understanding of the project.
3. **Demonstrate how the microphone works:** We recommend humorously demonstrating the use of a microphone with another adult in the classroom to acclimate the children.
4. **Define Assent:** Introduce the concept of assent to the students and explain that they can withdraw their assent at any time. Make sure that they know there is no penalty for not participating in your research.
5. **Determine Parental Consent:** Confirm which students have parental consent. Make sure that the interviewers know which students do not have parental consent so that they do not accidentally interview those students.

**At the start of each day, we recommend that interviewers double check:**

1. **Seating Chart/Student Tracking:**
    a. Do you have a seating chart (if needed) to help you keep track of which students are sitting where?
    b. Have you confirmed that students have gone back to the same seat?
    c. Do you know which students have consented/assented
    d. Does your support team know what additional measures need to be administered?
2. **Microphone Checks:**
    a. Are your microphones working?
    b. Are they supplied with new batteries?
    c. Have you turned on the USB Pre that connects the microphone to your phone (if necessary)?
3. **Interview Question Review:**
    a. Remind yourself to start with open ended questions
    b. Reflect on the kinds of interview questions most likely to help you to answer your research questions

**At the end of each fieldwork day, the research team should regroup to share observations, clarify context, and prepare for subsequent sessions:**

1. **Comparing Interviewer Notes:** Team members can briefly review and compare notes taken during the learning experience, or during specific interviews, to identify emerging patterns, discrepancies in interpretation, or gaps in coverage.
2. **Check Recordings:** There are many ways that recording equipment can fail. For example, even if you are using unidirectional microphones, the USB Pre (the device that allows you to connect these microphones to your phone) could be turned off or the batteries it requires could die. On the first day of fieldwork, we suggest checking a couple of interviews at the first opportunity you get. However, at the end of each day, it's a good idea to spot check your interviews to make sure nothing has gone wrong.
3. **Flagging Missing or Unusual Data:** The team can identify and record any disruptions (e.g., early student dismissals, fire drills), absences, or unexpected events that could influence the data but are not automatically captured in the QRF system.
4. **Discussing Trigger Performance:** If certain types of triggers are over- or under-firing (e.g., clustering around the same students repeatedly or not capturing enough variability), these trends should be discussed and documented.

5. **Updating Prioritization Strategies:** Based on the day's experiences, the team may decide to adjust how triggers are prioritized—either manually (e.g., deciding to skip certain triggers if they appear too often the next day), or by providing input for updates to the prioritization algorithm. Take notes about any changes you made that day and collect them for future use.
6. **Planning for Tomorrow:** Reviewing the next day's schedule (e.g., which classes or students will be observed), noting any missing data and making a plan to collect it, deciding which team members will focus on which activities, and pre-emptively noting any expected absences or logistical constraints.

To support this process, a shared daily debriefing template can be useful for capturing these reflections and decisions in a consistent, centralized format. These daily summaries can serve as a valuable audit trail for understanding how fieldwork decisions evolved over time and can be useful during data analysis and interpretation phases.

# Appendix 5: Overview of the Transcription to Coding Process

1. **Transcription & Line Segmentation.** As discussed in **Chapter 5**, following an established transcription process will help you to make consistent decisions about segmenting at the utterance level and for documenting speech overlap, pauses, or speaker turns that go uninterrupted for longer periods of time. Consistency is key, particularly when making decisions about segmenting your data, as this can influence subsequent analyses.

2. **Generating Codes.** As discussed in **Section 5.4**, most qualitative coding involves a mix of deductive analysis (to align with theoretical constructs) and inductive analysis (to ensure that you are taking full advantage of the ways in which constructs related to your research question might emerge within your unique data context). This stage is handled in different ways by different research teams and sometimes not reported on in the literature.

3. **Data Sampling.** Once data have been segmented into lines, researchers, we recommend that IRR should be checked by having two coders independently review a sample of the data before working independently to complete the full data set. Often researchers choose to sample 100 or 200 lines of data to check for IRR, but to some degree this will depend on the types of codes you are applying and the way your data has been segmented. We recommend randomly sampling representative interviews across different groups (e.g., different classes, different interviewers, different student demographics) until the appropriate sample size is reached.

4. **Independent Coding.** How you code will depend on your data, but binary codes (e.g., 1 for present/0 for absent) across each line of data are very common. We recommend each coder closely review the codebook to refresh their understanding of the constructs and definitions before they sit down to code. Selecting an IRR metric will depend on the kind of coding you are doing and how many raters you are using (see Table 5.8, **Section 5.6.2** for articles that discuss these issues).

5. **Interrater Reliability (IRR) Metrics.** After interdependent coding, IRR can be calculated to see how consistently each coder applied the codes. A common metric for binary coding with two coders is Cohen's Kappa, but as discussed **Section 5.6.2**, what you should use will depend on the parameters of your data. Often agreement at $\kappa \geq 0.8$ is considered sufficient, particularly if the consequences of being wrong are relatively benign. If your IRR metric is close to the threshold that is acceptable for your field (e.g., $\kappa \geq 0.76$), you may simply want to code more data, especially if the construct you were coding was relatively rare (see Table 5.8, **Section 5.6.2** for articles about Cohen's $\kappa$ and base rate issues). If your IRR metric is farther removed from an acceptable threshold (e.g., $\kappa \geq 0.49$), then you might want to go through a social moderation process before continuing with your coding process.

6. **Social Moderation of Disagreements.** One common practice for remediating a low IRR metric is to have your coders go through the process of social moderation (as discussed in **Section 5.6.1** and **Section 5.6.3**), where coders go through the codes that they have disagreed upon to determine the cause of the disagreement and/or to refine the codes. In general, we recommend that a second round of codes be collected independently after the social moderation process and checked separately to see if IRR improves. However, there may be some cases of infrequent codes for which there is agreement that the construct is qualitatively different from other items being coded but the amount of data present is insufficient for reaching IRR. In these cases, social moderation may be sufficient.